\pdfoutput=1% forces arxiv to use pdflatex
\documentclass[a4paper,12pt]{article}
\usepackage{ifpdf}
\ifpdf
\usepackage[pdftex]{graphicx}
\usepackage[usenames,dvipsnames]{color}
\else
\usepackage{graphicx}
\usepackage[usenames,dvips]{color}
\fi
\usepackage{amssymb,amsmath}
\usepackage[colorlinks=false]{hyperref}
\usepackage{caption}
\usepackage{paper}

\numberwithin{equation}{section}

\begin{document}

\title{\bfseries A Smooth Lattice construction of the Oppenheimer-Snyder spacetime.}
\author{%
Leo Brewin and Jules Kajtar\\[10pt]%
School of Mathematical Sciences\\%
Monash University, 3800\\%
Australia}
% \date{05-Jan-2009}% 1st version started
% \date{22-Feb-2009}% 1st version finished
% \date{22-Mar-2009}% 2nd version finished
\date{31-Mar-2009}% 3rd version finished
\reference{Preprint}

\maketitle

\begin{abstract}
\noindent
We present test results for the smooth lattice method using an Oppenheimer-Snyder spacetime.
The results are in excellent agreement with theory and numerical results from other authors.
\end{abstract}

% ============================================================================================
\section{Introduction}
\label{sec:Intro}

In recent times many numerical relativists have good reason to celebrate -- the long battle to
secure the holy grail \cite{shapiro:1986-02} is over (though some might prefer to redraw the battle
lines). The works of Pretorius \cite{arXiv:gr-qc/0507014,arXiv:0710.1338} and others
\cite{arXiv:gr-qc/0511048,arXiv:gr-qc/0511103} have opened a new era for computational general
relativity. This has spawned many new projects that directly address the needs of the gravitational
wave community. Many groups are now running detailed simulations of binary systems in full general
relativity as a matter of course. Does this mean that the development of computational methods for
general relativity is now over? The experience in other fields would suggest otherwise, look for
example at computational fluid dynamics where a multitude of techniques are commonly used,
including spectral methods, finite element methods, smooth particle hydrodynamics, high resolution
shock capture methods and the list goes on. The important point to note is that one method does not
solve all the problems and thus in numerical relativity it is wise, even in the face of the current
successes, to seek other methods to solve the Einstein equations. It is in that spirit that we have
been developing what we call the smooth lattice method
\cite{brewin:2002-01,brewin:1998-01,brewin:1998-02}. This is a fundamentally discrete approach to
general relativity based on a large collection of short geodesic segments connected to form a
lattice representation of spacetime. The Einstein equations are cast as evolution equations for the
leg-lengths with the Riemann and energy-momentum tensors acting as sources. Of course the Riemann
tensor must be computed from the leg-lengths and this can be done in a number of related ways, such
as by fitting a local Riemann normal coordinate expansion to a local cluster of legs or to use
the geodesic deviation equation, or, and with more generality, to use the second variation of
arc-length. Past applications of the method have included a full 3+1 simulation of the vacuum
Kasner cosmology \cite{brewin:1998-02} and a 1+1 maximally sliced Schwarzschild spacetime
\cite{brewin:2002-01}. In both cases the simulations were stable and showed excellent agreement
with the known solutions while showing no signs of instabilities (the maximally sliced
Schwarzschild solution ran for $t>1000m$ and was stopped only because there was no point in running
the code any longer).

In this paper we report on our recent work using the Oppenheimer-Snyder \cite{opp-snyder:1939-01}
spacetime as a benchmark for our smooth lattice method
\cite{brewin:2002-01,brewin:1998-01,brewin:1998-02}. We chose this spacetime for many reasons, it
has been cited by many authors
\cite{shapiro:1985-01,shapiro:1986-01,schinder:1988-01,gourg:1991-01,shapiro:1992-01,baumg:1995-01,romero:1996-01}
as a standard benchmark for numerical codes (and thus comparative results are available), the
analytic solution is known (in a number of time slicings), the equations are simple and there are
many simple diagnostics that can be used to check the accuracy of the results (as described in
sections \ref{sec:Diagnst}, \ref{sec:Results}).

In an impressive series of papers, Shapiro and Teukolsky
(\cite{shapiro:1985-01,shapiro:1986-01,shapiro:1992-01}) used the Oppenheimer-Snyder spacetime as
the first in a series of test cases. They were motivated by certain problems in relativistic
stellar dynamics (such as the formation of neutron stars and black holes from supernova) and they
developed a set of codes based on the standard ADM equations, adapted to spherical symmetry, in
both maximal and polar slicing and using an $N$-body particle simulation for the hydrodynamics.
They made limited use of the exact Schwarzschild solution to develop an outer boundary condition
for the lapse function while using both the Schwarzschild and FRW solutions to set the initial
data. Though their discussion on the size of their errors is brief (for the Oppenheimer-Snyder test
case), they did note that the errors were of the order of a percent or so (for a system with 240
grid points and 1180 dust particles). In a later work, Baumgarte \etal \cite{baumg:1995-01}
extended their work by expressing the metric and the equations in terms of an out-going null
coordinate. This leads to a slicing that covers all of the spacetime outside (and arbitrarily close
to) the event horizon. In this version of their code Baumgarte \etal \cite{baumg:1995-01} chose to
solve only the equations for the dust ball by using the Schwarzschild solution as an outer boundary
condition.

This idea, to replace the exterior equations with the known Schwarzschild solution, has been used
by Gourgoulhon \cite{gourg:1991-01}, Schinder \etal \cite{schinder:1988-01} and Romero \etal
\cite{romero:1996-01}. Gourgoulhon \cite{gourg:1991-01} used a radial gauge and polar slicing while
solving the equations using a spectral method and reported errors in the metric variables between
$10^{-7}$ to $10^{-5}$. However, with the onset of the Gibbs phenomena, the code
could only be run until the central lapse collapsed to around $2\times10^{-3}$. Schinder \etal
\cite{schinder:1988-01} used the same equations as Gourgoulhon \cite{gourg:1991-01} but with a
discretisation based on a standard finite difference scheme. They reported errors of order 1\% for
evolution times similar to those of Gourgoulhon \cite{gourg:1991-01}. The work of Romero \etal
\cite{romero:1996-01} differs from that of Schinder \etal \cite{schinder:1988-01} in that they used
high-resolution shock capture methods for the hydrodynamics. They report evolutions down to a
central lapse of $1.3\times10^{-10}$.

Our results compare very well against those given above with our errors being of the order
fractions of a per-cent for 1200 grid points. Our code runs, without any signs of instabilities,
for maximal slicing out to $t=500m$ where the central lapse has collapsed to $10^{-110}$ (see
Figure \ref{Max:ExpoLapse}). We make no use of the known solutions other than the conservation of
local rest energy (we use a particle like method to compute the rest density). We also provide
extensive comparisons of our results with the exact solution (see section \ref{sec:Results}).

In the following sections we will describe all aspects of our code, including the design of the
lattice (section \ref{sec:OSlattice}), the curvature and evolution equations (sections
\ref{sec:Riemann}, \ref{sec:EvolEqtns} and \ref{sec:EvolVnVz}), computing the density (section
\ref{sec:Rho}), the junction conditions (section \ref{sec:JctCond}), setting the initial data
(section \ref{sec:IniData}) and finally the results (section \ref{sec:Results}).

We will make frequent reference to two papers, our earlier paper on the Schwarzschild spacetime
\cite{brewin:1998-02} and a companion paper showing how the Einstein equation can be applied to the
lattice \cite{brewin:2009-04}. We will refer to
these as \PaperI\ and \PaperII\ respectively.

% ============================================================================================
\section{The Oppenheimer-Snyder lattice}
\label{sec:OSlattice}

\begin{figure}[t]
\centerline{\includegraphics[width=0.8\textwidth]{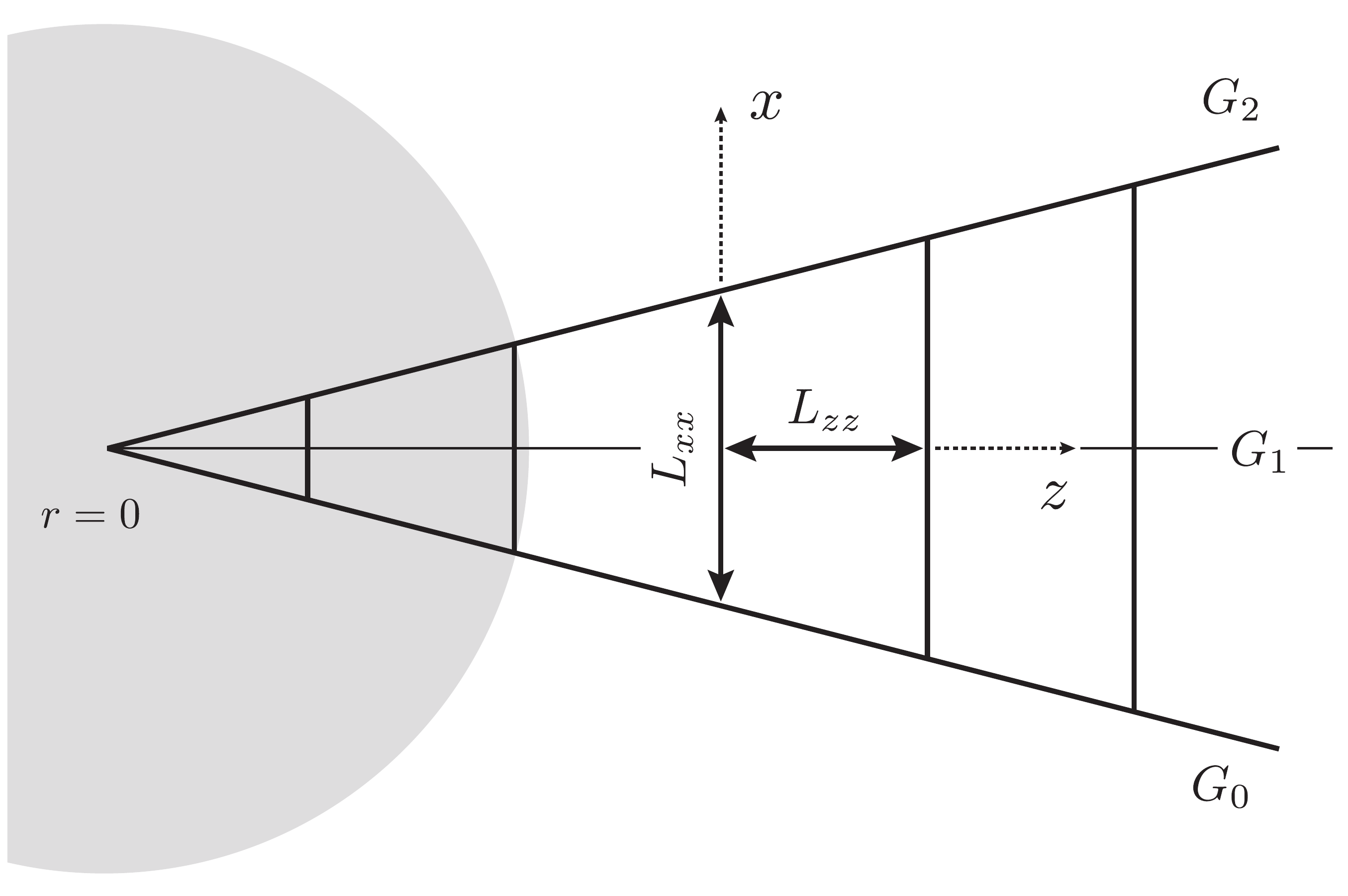}}
\vskip0.5cm
\caption{\normalfont In this figure we show how the lattice is constructed from
two radial geodesics $G_0$ and $G_2$ and the series of interconnecting legs $\Lxx$.
The third geodesic $G_1$ lies midway between $G_0$ and $G_2$ is used to define the
radial legs $\Lzz$. The grey patch to the left represents (part) of the dust ball.}
\label{fig:Lattice}
\end{figure}

What design should we choose for the lattice? We will take a minimalist approach -- build the
simplest lattice that captures the required symmetries while being sufficiently general to allow the
full dynamics to be expressed through the evolution of the lattice data. Here is a construction of
such a lattice. Take a single spacelike radial geodesic, in one Cauchy surface, extending from the
centre of the dust ball out to the distant asymptotically flat regions and sub-divide it into a
series of short legs with lengths denoted by $\Lzz$. We will refer to the end points of each leg as
the lattice nodes. Note that we are free to choose the $\Lzz$ as we see fit (in the same way that we
are free to choose the lapse function in an ADM evolution). Now construct a clone of this geodesic
by rotating it through any small angle (while remaining in the Cauchy surface). Finally connect the
corresponding nodes of the pair of geodesics by a second set of geodesic legs, with lengths denoted
this time by $\Lxx$ (see Figure \ref{fig:Lattice}). We now have a spacelike 3-dimensional lattice
contained within one Cauchy surface. From here on in we allow this lattice to vary smoothly with
time. 

Note that each leg of this lattice is a geodesic segment of the 3-metric of the Cauchy surface. We
could also connect the nodes of the lattice with geodesic segments of the full 4-dimensional
spacetime (much like constructing chords to arcs of a circle). This gives us two representations of
the lattice, both sharing the same node points with the first composed of short 3-geodesics and the
second composed of short 4-geodesics. Suppose that typical leg-lengths in the two representations
are ${}^3\Lij$ and ${}^4\Lij$ respectively. Then it is not hard to see that ${}^4\Lij={}^3\Lij +
{\cal O}{({}^3\Lij^3)}$. The upshot is that in all of our equation in this paper we are free
to use either representation (the differences being at least as small as the truncation errors).

The $\Lzz$ and $\Lxx$ are all that we need to describe the geometry of each Cauchy surface but we
also need some way to represent the dust ball on the lattice. Again, we shall take a minimalist
approach -- we know that the dust can be described as a set of particles travelling on timelike
geodesics with conserved rest mass. Thus we add a series of dust particles on the radial geodesic
with each particle carrying a conserved rest mass.

\begin{figure}[t]
\centerline{\includegraphics[width=0.8\textwidth]{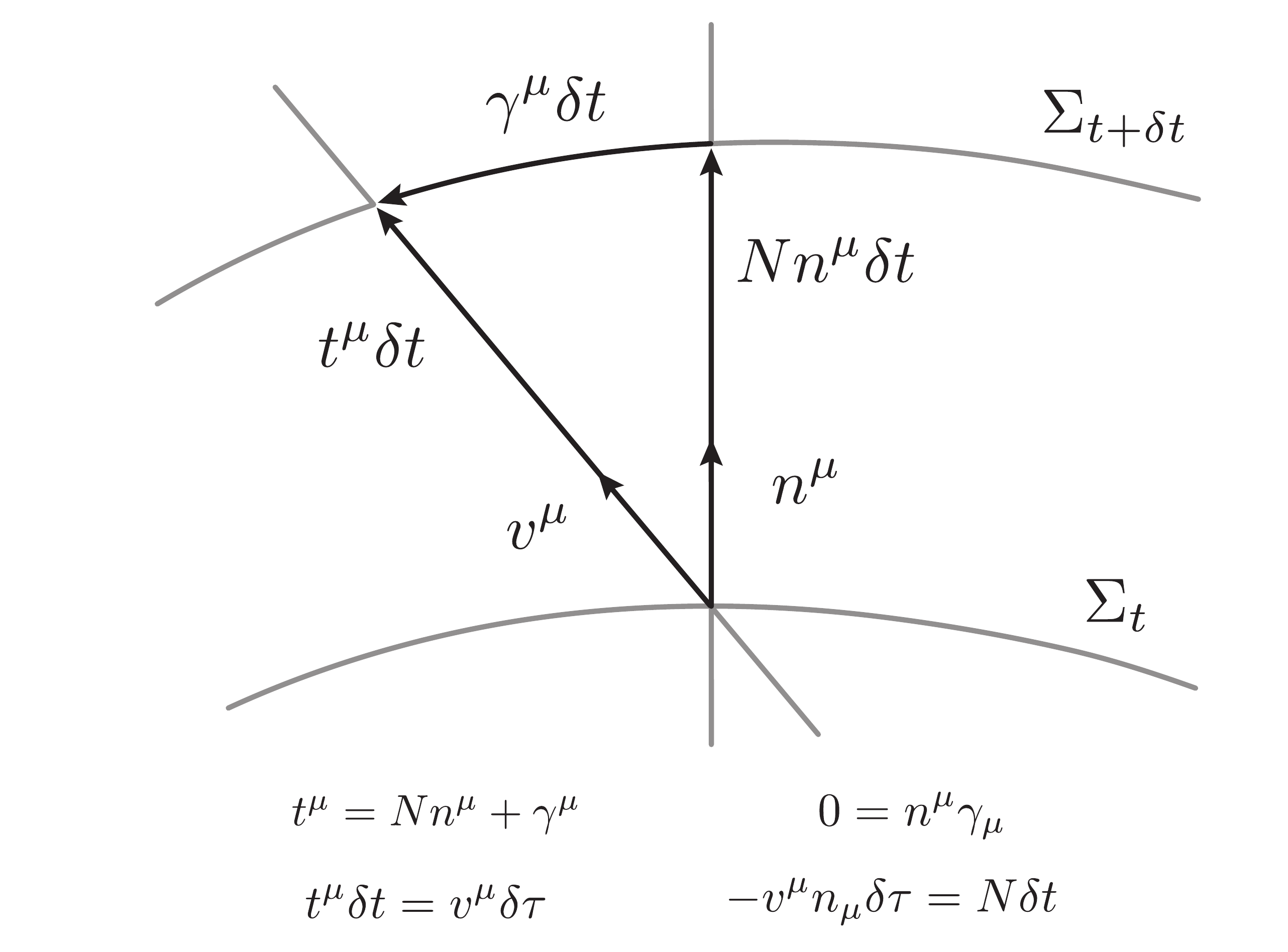}}
\vskip0.5cm
\caption{\normalfont In this figure we introduce most of the kinematical quantities on the lattice.
The dust particle's unit 4-velocity is $v^\mu$, while $\gamma^\mu$ is the drift vector, $n^\mu$ the
unit normal to the Cauchy surface $\Sigma_t$, $N$ is the lapse function and $\delta\tau$ is the
proper time measured along the dust particle's trajectory.}
\label{fig:Define}
\end{figure}

As noted above, we are free to distribute the lattice nodes as we see fit. How should we do this?
We know that the dust ball will collapse so it makes sense to tie the lattice nodes to the dust
particles, i.e.\ the lattice nodes follow in-falling timelike geodesics. But what of the nodes
outside the dust ball? Again, by appeal to simplicity, we demand that \emph{every} lattice node,
interior and exterior, follow the in-falling timelike geodesics. In this scheme the lattice nodes
do \emph{not} follow the trajectories normal to the Cauchy surface (in contrast to the scheme in
\PaperI). This introduces a drift vector $\gamma^\mu$ (see Figure \ref{fig:Define}) (which is
similar to but distinct from the shift vector, see \cite{brewin:1998-02}).

The lattice just described differs from the Schwarzschild lattice of \PaperI\ in a number of
important ways -- it contains an internal boundary (the edge of the dust ball), the lattice nodes
are not at rest in each Cauchy surface, the lattice carries a set of dust particles and at the
inner boundary $\Lxx=0$. Thus we will need to develop new boundary conditions (section
\ref{sec:JctCond}), new evolution equations for the nodes (i.e.\ adapt the geodesic equations to
the lattice, section \ref{sec:EvolVnVz}) and an algorithm to compute the rest energy density from
the rest masses carried by the dust particles (section \ref{sec:Rho}).

In \PaperI\ we employed Riemann normal coordinates as a stepping stone to develop the purely scalar
equations for the leg-lengths, time derivatives, constraints etc. We went on to speculate whether
or not these coordinates imbued the numerical scheme with any favourable properties (we argued that
they did not). One way to avoid this coordinate issue is simply to derive the equations without
reference to a coordinate system. In this paper we will represent tensors, such as the Riemann and
extrinsic curvatures, by their frame components. We will use an orthonormal frame built as follows.
We choose the first two basis vectors, $m^\mu_x$ and $m^\mu_z$, to be the unit tangent vectors to
$\Lxx$ and $\Lzz$ respectively at the mid-point of $\Lxx$, see Figure (\ref{fig:Lattice}). The
remaining two basis vectors ($m^\mu_y$ and $m^\mu_n$) can be chosen freely (subject to the
orthornormal condition, e.g.\ $m^\mu_n$ could be chosen as the unit normal to the Cauchy surface).
With this choice of basis a typical frame component for the extrinsic curvature could be written as
$K_{\mu\nu} m^\mu_a m^\mu_b$. Such notation quickly becomes tiresome so we will introduce the
abbreviation $\Kab$ to represent $K_{\mu\nu} m^\mu_a m^\mu_b$ with an obvious generalisation to
other tensors.

We will allow a slight variation to this notation. On occasions we will find it useful to refer to
a leg by its end points, such as $i$ and $j$. That leg will have its own unit tangent vector which
we denote by $m^\mu$. We will then take $\Kij$ to be $K_{\mu\nu}m^\mu m^\nu$. This small change
will only ever be used for the extrinsic curvature.

The dust particles follow future pointing timelike geodesics. We will use $v^\mu$ to denote the
velocity 4-vector of the dust particles and we will record the frame components as $v_n= - v_\mu
n^\mu$ and $v_z=v_\mu m^\mu_z$ where $n^\mu$ is the unit normal to the Cauchy surface.

The notation just introduced sits quite nicely with the notation used in \PaperI. In that paper we
wrote $K_{xx}$, for example, to denote the $x-x$ coordinate components of $K_{\mu\nu}$ in the local
Riemann normal frame. In that frame we chose the three metric $g_{\mu\nu}(x)$ at the origin to
equal $\diag(1,1,1)$ and the basis vectors $m^\mu_a$ to have values $\delta^\mu_a$. Thus $K_{xx} =
K_{\mu\nu}m^\mu_x m^\nu_x$. The upshot is that coordinate components of \PaperI\ have the same
numerical values as the frame components used in this paper. Thus we would reasonably expect that
the equations used in \PaperI\ should carry over to this paper with only minor changes to
accommodate the introduction of the dust. This indeed proves to be the case (which is reassuring).
The details will be presented in section \ref{sec:EvolEqtns} where we will use a formalism
developed in \PaperII\ to derive, from scratch, the evolution equations for the lattice.

% ============================================================================================
\section{The Riemann curvatures}
\label{sec:Riemann}

The question here is: How do we compute the Riemann curvatures, $\Rx$ and $\Rz$, from the $\Lxx$ and
$\Lzz$? In our previous paper, \PaperI, we computed two Riemann curvatures, $\Rx$ and $\Rz$, using
\begin{align}
0 &= \ddLxxdz + \Rz \Lxx
  &\text{geodesic deviation}\label{eq:GD}\\
0 &= \frac{d\left(\LLxx \Rx\right)}{dz} - \Rz\dLLxxdz
  &\text{Bianchi identity}\label{eq:BI}
\end{align}
These equations will be used as follows. First we use the geodesic deviation equation to compute
the $\Rz$ for each node across the lattice. This then allows us to integrate the Bianchi identity
for $\Rx$ from the centre to the outer boundary.

This scheme sounds simple but there a number of (obvious) complications. Firstly, the equations are
singular at the centre (where $\Lxx=0$) and secondly, $\Rz$ will not be continuous across the
junction. These complications are new to this investigation but we also inherit one further
complication from \PaperI: what boundary condition should we use at $z=0$ when integrating the
Bianchi identity? This last problem is rather easy to deal with. At the centre of the dust ball we
know that the metric must be isotropic and thus we can be certain that $\Rx = \Rz$ at $z=0$.

How do we handle the singularity at $z=0$? Again, by symmetry arguments we can assert that
$\Rzmu=0$ at $z=0$. Thus in a small neighbourhood of $z=0$ we must have $\Rz = A + Bz^2$
where $A$ and $B$ are independent of position and $z$ is the radial proper distance measured from
$z=0$. Thus it is not unreasonable to use a quadratic \emph{interpolation} of $\Rz$ to estimate
$\Rz$ at $z=0$. Our experience shows that this works very well but it does require some care (see
section \ref{sec:JctRxRz} for the full details). Before dealing with the junction issue we should
emphasise that this process is an interpolation rather than an extrapolation of the data to $z=0$.
To see this just imagine extending the radial geodesics of the lattice through $z=0$ so that we can
use $\Rz = A + B z^2$ with $z$ in a range $-z_0<z<z_0$ for some small $z_0$.

The frame components $\Rxx = R_{\mu\nu} m_x^\mu m_x^\nu$ and $\Rzz = R_{\mu\nu} m_z^\mu m_z^\nu$ of
the Ricci tensor are rather easy to construct from $\Rx$ and $\Rz$. Using the orthonormal frame
$m_x^\mu$, $m_y^\mu$ and $m_z^\mu$ we can easily deduce that
\begin{align}
\Rxx &= \Rx + \Rz\label{eq:Rxx}\\
\Ryy &= \Rx + \Rz\label{eq:Ryy}\\
\Rzz &= \Rz\label{eq:Rzz}
\end{align} 
with all other $\Rab = 0$. From these equations it is easy to verify that the scalar
curvature is given by
\begin{equation}
R = 2 \Rx + 4 \Rz\label{eq:R}
\end{equation}

The one remaining complication is the discontinuity in $\Rz$ at the junction. This will be
discussed in detail in section \ref{sec:JctCond}.

% ============================================================================================
\section{The evolution equations}
\label{sec:EvolEqtns}

The equations of \PaperI\ have served us well so far but now we must chart a new path. The reason
is that, unlike our approach in \PaperI, here we allow the lattice nodes to drift across the Cauchy
surfaces and this will introduce extra terms in the evolution equations. There is also the issue of
introducing the energy momentum sources but, as we shall see later, this is really very easy to do
(it amounts to little more than adding a term of the form $8\pi k T_{\mu\nu} m^\mu m^\nu$ to the
vacuum equations). So how do we develop evolution equations for a non-zero drift vector? In
\PaperII\ we showed how the standard 3+1 ADM equations with a zero shift vector can be recovered
from the equations for the second variation of arc-length. And as arc-lengths of geodesics are
central to our smooth lattice approach this new formalism is well suited to our current task.

We begin by recalling from \PaperII\ the equations for the first and second variation of the
geodesic segment that connects nodes $i$ and $j$
\begin{align}
\DLij &= \left[m_\mu t^\mu\right]_i^j = \int_i^j\>m_\mu m^\nu t^\mu{}_{;\nu}\>ds\label{eq:DLij}\\
\DDLij &= \left[ t^\alpha{}_{;\mu} t^\mu m_\alpha \right]_i^j
        - \int_i^j\> {}^4R_{\mu\alpha\nu\beta} m^\mu m^\nu t^\alpha t^\beta \>ds\notag\\
       &\>+\int_i^j\> \left(
            t_{\mu;\alpha}t^{\mu}{}_{;\nu}m^\alpha m^\nu
          - \left(m_\mu m^\nu t^\mu{}_{;\nu}\right)^2 \right) \>ds\label{eq:DKij}
\end{align}
It is tempting to jump in by setting $t^\mu = Nn^\mu + \gamma^\mu$ and to let the equations take us
where they will. Indeed this works well for the first variation. We start be making the said
substitution and massage the result as follows
\begin{align*}
\DLij &= \left[m^\mu t_\mu\right]_i^j\\
      &= \left[m^\mu (Nn_\mu)\right]_i^j + \left[m^\mu\gamma_\mu\right]_i^j\\
      &= \int_i^j\>m^\mu m^\nu (Nn_\mu)_{;\nu}\>ds + \left[m^\mu\gamma_\mu\right]_i^j\\
      &= \int_i^j\>m^\mu m^\nu (-NK_{\mu\nu})\>ds + \left[m^\mu\gamma_\mu\right]_i^j\\
      &= -N\Kij \Lij + \left[m^\mu\gamma_\mu\right]_i^j
\end{align*}
In the second last line we have used $N n_{\mu;\nu} = -\left(\bot N_{,\mu}\right) n_\nu - N
K_{\mu\nu}$ and $m^\mu n_\mu =0$ while in writing the last line we have assumed that the leg-length
is sufficiently short that the integrand can estimated by a simple quadrature, in this case the
mid-point rule (later in section \ref{sec:DbyDt} we will have reason to change this to a
Trapezoidal rule). A similar equation can be found in \PaperII\ (differing only in the absence of
the $\gamma$ terms). For the pair of legs $\Lxx$ and $\Lzz$ we thus obtain
\begin{align*}
\DLxx & = -N \Kxx \Lxx + [m_x^\mu \gamma_\mu]\\
\DLzz & = -N \Kzz \Lzz + [m_z^\mu \gamma_\mu]
\end{align*}
and to keep the notation a little less cluttered we have not written the end points on the
$[\cdots]$ terms.

\begin{figure}[t]
\centerline{\includegraphics[width=0.8\textwidth]{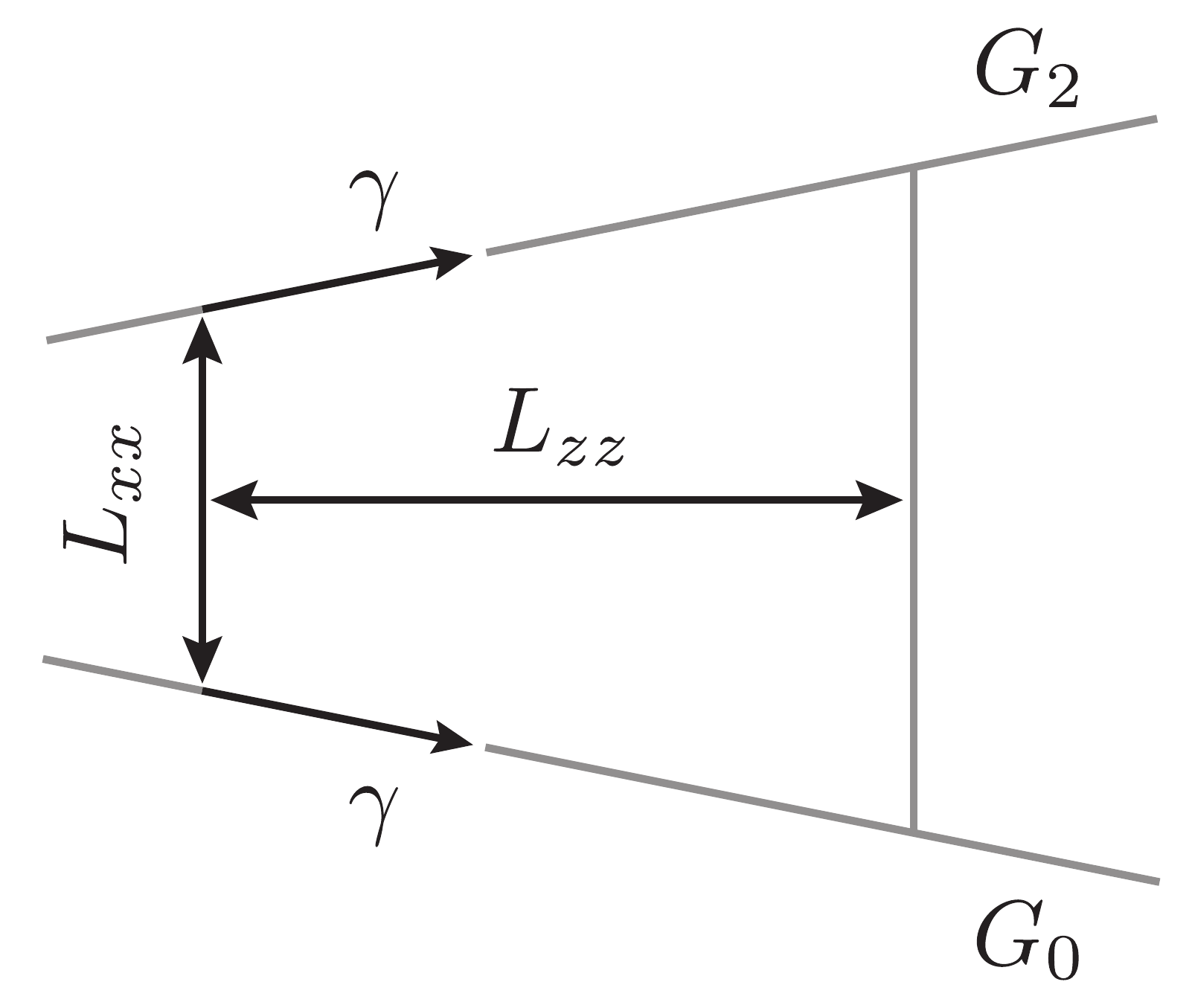}}
\vskip0.5cm
\caption{\normalfont In this diagram it is easy to see that 
$[m_{x\mu}{\tilde\gamma}^\mu] = d\Lxx/dz$ where $\tilde\gamma$ is the
unit vector parallel to $\gamma$. This result is used in section \ref{sec:EvolEqtns}
when deriving equation \EQref{eq:DLxx}.}
\label{fig:dLxxdz}
\end{figure}

What can we say about the $[m_\mu\gamma^\mu]$ terms? Let ${\tilde\gamma}^\mu$ be the unit vector
parallel to $\gamma^\mu$. Then we can immediately use the first variation equation once again (see
Figure \ref{fig:dLxxdz}) to deduce that $[m_{x\mu} {\tilde\gamma}^\mu]$ equals $d\Lxx/dz$ where, as
usual, $z$ is the radial proper distance measured along $G_1$. However, ${\tilde\gamma}^\mu =
\gamma^\mu/\gamma^z$ and by spherical symmetry we know that $\gamma^z$ does not change from one
radial geodesic to the next. Thus we deduce that $[m_{x\mu}\gamma^\mu] = \gamma^z d\Lxx/dz$. We now
turn to the other leg, $\Lzz$. In this case $m_z^\mu$ and $\gamma^\mu$ are parallel and thus we can
not invoke the first variation equation. But that is of no concern simply because
$m_{z\mu}\gamma^\mu = \gamma^z = Nv_z/v_n$. Thus we have $[m_{z\mu}\gamma^z] = [Nv_z/v_n]$. The
equations for the first time derivatives of $\Lxx$ and $\Lzz$ can now be written as
\begin{align}
\DLxx & = -N \Kxx \Lxx + \left(\gammaz\right)\dLxxdz\label{eq:DLxx}\\
\DLzz & = -N \Kzz \Lzz + \left[\gammaz\right]\label{eq:DLzz}
\end{align}
We turn our attention now to adapting the equations for the second variation to our simple lattice.

Our job would be greatly simplified if it happened that $\gamma^\mu=0$ but on the current lattice
that is not the case. So we now introduce a second lattice on which we set $\gamma^\mu=0$. The
nodes of the first lattice will follow the dust particles while those of the second lattice will
follow trajectories normal to their Cauchy surfaces (which will differ from those of the first
lattice). Note that the second lattice has been introduced solely to aid the exposition -- the
second lattice will never be needed nor used in our actual computer programs. To keep the
bookkeeping clear we will identify data on the second lattice by the addition of a dash. The second
lattice is created at some generic time, say $t=t_0$, and we choose to assign identical initial
data to both lattices, i.e.\ $\Lijdash=\Lij$, $\Kijdash=\Kij$ etc. on $\Sigma(t_0)$. We have no
reason to use distinct Cauchy surfaces for each lattice (we only want to set $\gamma^\mu=0$) so we
are free to set $N'=N$ and $dN'/dt=dN/dt$. It follows that we also have ${}^4R'_{\mu\nu\alpha\beta}
= {}^4 R_{\mu\nu\alpha\beta}$ across $\Sigma(t_0)$. Our task now is to adapt the equations for the
first and second variations to the second lattice. This has already been done in \PaperII\ where we
have shown that
\begin{spreadlines}{8pt}
\begin{align*}
\DLijdash  &= -NK'_{\mu\nu}m^\mu m^\nu \Lijdash\\
\DDLijdash &= \frac{1}{N}\frac{dN}{dt} \DLijdash
                - \frac{1}{\Lijdash} \left(\DLijdash\right)^2
                + N^2 K'_{\mu\alpha}K'^\mu{}_{\beta} m^\alpha m^\beta \Lijdash\\
           & + N N_{;\alpha\beta} m^\alpha m^\beta \Lijdash
             - N^2 \left({}^4R_{\mu\alpha\nu\beta}\right) m^\mu m^\nu n^\alpha n^\beta \Lijdash
\end{align*}
\end{spreadlines}
Where we head next depends upon what type of equations we wish to work with. We can develop either
a second-order set of equations involving both $d\Lijdash/dt$ and $d^2\Lijdash/dt^2$ or a first
order system involving $d\Lijdash/dt$ and $d\Kijdash/dt$. We will take the second approach for two
reasons, it mimics the standard ADM approach and, more importantly, it eliminates the $dN/dt$ term
(which would add undue complexity when using maximal slicing). Between this pair of equations we
can easily eliminate $d^2\Lijdash/dt^2$ with the following result
\begin{align*}
\DKijdash &= -N_{;\alpha\beta}m^\alpha m^\beta 
            + N\left({}^4R_{\mu\alpha\nu\beta}\right) m^\mu m^\nu n^\alpha n^\beta\\
          & + 2 N \left(\Kijdash\right)^2 - N K'_{\mu\alpha}K'^{\mu}{}_{\beta}m^\alpha m^\beta
\end{align*}
where $\Kijdash:=K'_{\mu\nu}m^\mu m^\nu$. This last equation controls the evolution of $\Kijdash$
for the second lattice. Can we use this information to deduce the evolution of $\Kij$ on the first
lattice? Yes, by simply splitting the evolution into a part parallel to the normal plus a part
parallel to the drift vector. Since $\Kij$ is a scalar function we can use a standard chain rule to
write
\begin{equation*}
\DKij = \DKijdash + \KijmuDash \gamma^\mu
\end{equation*}
and as $\Kij=\Kijdash$ on $\Sigma(t_0)$ we arrive at
\begin{align*}
\DKij &= -N_{;\alpha\beta}m^\alpha m^\beta 
        + N\left({}^4R_{\mu\alpha\nu\beta}\right) m^\mu m^\nu n^\alpha n^\beta\\
      & + 2 N \left(\Kij\right)^2 - N K_{\mu\alpha}K^{\mu}{}_{\beta}m^\alpha m^\beta 
        + \Kijmu \gamma^\mu
\end{align*}

In a moment we will apply this equation to $\Kxx=K_{\mu\nu} m_x^\mu m_x^\nu$ and $\Kzz = K_{\mu\nu}
m_z^\mu m_z^\nu$ but first we recall that both $\gamma^\mu$ and $m^\mu$ are tangent to $\Sigma$,
$T^{\mu\nu} = \rho v^\mu v^\nu$ and for our spherically symmetric lattice, $K_{\mu\nu}$ is diagonal
and $\gamma^z=Nv_z/v_n$ . We will also need the contracted Gauss equation, namely,
\begin{equation*}
{}^4R_{\mu\alpha\nu\beta} m^\mu m^\nu n^\alpha n^\beta 
 = \left(-\bot{}^4R_{\mu\nu} + R_{\mu\nu} 
         + K K_{\mu\nu} - K_{\alpha\mu}K^{\alpha}{}_{\nu}\right) m^\mu m^\nu
\end{equation*}
We then find that the above equation for $d\Kij/dt$ when applied to $\Kxx$ and $\Kzz$ leads to
\begin{align}
\DKxx &= -\dNdxx + N\left( \Rxx + K \Kxx - 4\pi k \rho \right)
         + \left(\gammaz\right) \KxxDz\label{eq:DKxx}\\
\DKzz &= -\dNdxx + N\left( \Rzz + K \Kzz + \left(4 - 8 v^2_n\right)\pi k \rho \right)
         + \left(\gammaz\right) \KzzDz\label{eq:DKzz}
\end{align}
This pair of equations coupled with (\ref{eq:DLxx},\ref{eq:DLzz}) are the evolution equations for
the lattice.

% ============================================================================================
\section{The constraints}
\label{sec:HamMom}

The general form of the Hamiltonian and momentum constraints are
\begin{gather*}
R + K^2 - K_{\mu\nu} K^{\mu\nu} = 16 \pi k T_{\mu\nu} n^\mu n^\nu\\
\bot\left(K_{|\nu} - K^\mu{}_{\nu|\mu}\right) = 8\pi k \bot\left(T_{\mu\nu} n^\mu\right)
\end{gather*}
where $K=K^\mu{}_\mu$. It is a simple matter to apply these equations to the Schwarzschild
spacetime, see \PaperI\ for details. For the present case we need to account for the non-zero
$T_{\mu\nu}$ in the interior of the dust ball. We can easily adapt the equations of \PaperI\ by
simply adding on the terms $8\pi k T_{\mu\nu} n^\mu n^\nu$ for the Hamiltonian and $8\pi k
T_{\mu\nu} n^\mu m^\nu_{z}$ for the momentum constraints (projections in the other two directions
$m^\mu_x$ and $m^\mu_y$ yield the trivial equation $0=0$). This leads to
\begin{gather}
0 = \Rx + 2\Rz + \Kxx^2 + 2\Kxx\Kzz - 8 \pi k\rho v^2_n\label{eq:Ham}\\
0 =\frac{1}{\Lxx}\left( \Kzz \dLxxdz - \dLKxxdz\right) - 4 \pi k\rho v_n v_z\label{eq:Mom}
\end{gather}
(for simplicity we have cleared a common factor of 2 from both equations).

% ============================================================================================
\section{The particle equations}
\label{sec:EvolVnVz}

Here we will derive the equations governing the evolution of the particle 4-velocities.

We will use the geodesic equation $0=v^\mu{}_{;\nu}v^\nu$ to obtain evolution equations for the
$v_n$ and $v_z$ components of the particle's 4-velocity.

The computation are simple but do entail a few steps. We begin by writing $dv_n/dt$ as a
directional derivative along $t^\mu = \lambda v^\mu$. The Leibniz rule is then applied which in
turn allows the geodesic equations $0=v^\mu{}_{;\nu}v^\nu$ to be imposed. Finally, we use
\begin{equation}
N n_{\mu;\nu} = -\bot N_{,\mu} n_\nu - N K_{\mu\nu}\label{eq:nDeriv}\\
\end{equation}
to re-write $n_{\mu;\nu}$ in terms of the lapse and extrinsic curvatures. The details are as
follows.
\begin{align*}
\Dvn &= v_{n;\nu}t^\nu = -\left(v^\mu n_\mu\right)_{;\nu}(\lambda v^\nu) 
 = -\lambda v^\mu v^\nu n_{\mu;\nu}\\
&= \lambda v^\mu v^\nu\left( \frac{1}{N}\left(\bot N_{,\mu}\right)n_\nu + K_{\mu\nu}\right)\\
&=-v_z v_n\frac{\lambda}{N}\dNdz +\frac{1}{\lambda}\gamma^\mu\gamma^\nu K_{\mu\nu}
\end{align*}
But we also know that $\lambda=N/v_n$ and $0=\gamma^x=\gamma^y$ while $\gamma^z = Nv_z/v_n$ so
this last equation may be further reduced to just
\begin{equation}
\Dvn = - v_z \dNdz + N \frac{v_z^2}{v_n} \Kzz\label{eq:DVn}
\end{equation}

The computations for $dv_z/dt$ are much the same,
\begin{align*}
\Dvz &= v_{z;\nu} t^\nu = (v^\mu m_{z\mu})_{;\nu} (\lambda t^\nu) 
      = \lambda v^\mu v^\nu m_{z\mu;\nu}\\
&=\lambda\left(v_n n^\mu + v_z m_z^\mu\right) v^\nu m_{z\mu;\nu}\\
&=\lambda v_n n^\mu v^\nu m_{z\mu;\nu}
\end{align*}
The term $n^\mu v^\nu m_{z\mu;\nu}$ can be computed by expanding $0=(m_{z\mu}
n^\mu)_{;\nu}v^\nu$ and then using \EQref{eq:nDeriv} to obtain
\begin{equation*}
0=m_{z\mu;\nu} n^\mu v^\nu + \frac{v_n}{N}\dNdz - K_{\mu\nu}m_z^\mu m_z^\nu
\end{equation*}
which, when substituted into the previous equation for $dv_z/dt$, leads to
\begin{equation}
\Dvz = - v_n \dNdz + N v_z \Kzz\label{eq:DVz}
\end{equation}

One simple check we can immediately apply to our equations is to ask: do they preserve the unit
normalisation of $v^\mu$? Since we have chosen $n^\mu$ and $m_z^\mu$ to be unit vectors the
question reduces to asking if $d(-v_n^2+v_z^2)/dt$ vanishes for all $t$. From the above equations
this is easily seen to be so.

% ============================================================================================
\section{The density}
\label{sec:Rho}

There are at least two ways to compute the density, either by solving the Hamiltonian constraint or
by integrating the equations of motion for the dust, namely, $0=(\rho v^\mu v^\nu)_{;\nu}$.

Recall that the Hamiltonian constraint is given by
\begin{equation}
\Rx + 2\Rz + \Kxx^2 + 2\Kxx\Kzz = 8 \pi k\rho v^2_n\label{eq:RhoH}
\end{equation}
This equation is trivial to solve for $\rho$ since on each Cauchy surface all of the other
quantities are known. Notice that $v^2_n= 1 + v^2_z$ and thus $v_n \ge 1 > 0$.

Using the Hamiltonian constraint is one of many tricks used in numerical relativity to coerce
better stability properties from the evolution equations. The merits of doing so have been debated
over the years and is not something we will delve into here. However as we are trying to establish
the limitations of the smooth lattice method it makes sense to explore other methods to compute the
density. So for our second method we turn to the energy-momentum equations. From $0=(\rho v^\mu
v^\nu)_{;\nu}$ we learn two things (i) the dust particles follow time like geodesics, $0=
v^\mu{}_{;\nu}v^\nu$ and (ii) the rest mass is conserved along the worldtube generated by the dust
particles $0=d/dt\int\>\rho dV'$ where $d/dt$ is the time derivative following the dust and $dV'$
is the proper volume in the dust's rest frame. We will need both equations to compute the density.

Recall that we have chosen to tie the dust particles to the nodes of the lattice. As the nodes
drift relative to the Cauchy surface there will be a non-zero boost between the rest frame of the
dust and that of the Cauchy surface. Thus, in terms of the volume element $dV$ on the Cauchy
surface we have
\begin{equation*}
\int_{C_0}\>\rho v_n \> dV = \int_{C_1}\>\rho v_n \> dV
\end{equation*}
where $C_0$, $C_1$ denote the intersections of a dust worldtube with a pair of Cauchy surfaces, one
at time $t_0$ and another at a later time $t_1$.

\begin{figure}[t]
\centerline{\includegraphics[width=0.8\textwidth]{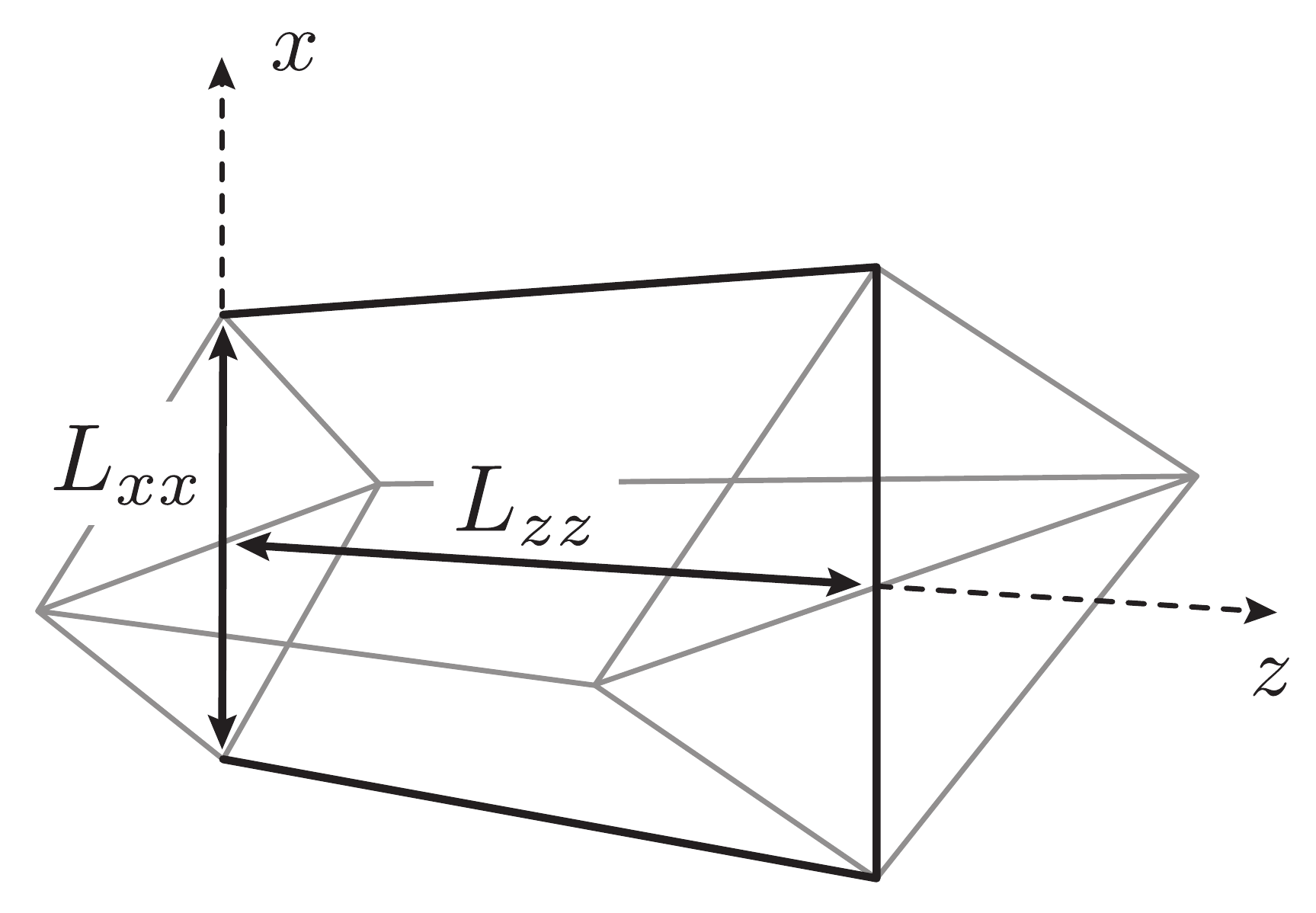}}
\vskip0.5cm
\caption{\normalfont Here we display the 3-dimensional cell which we use to compute
the energy density from the conservation equation \EQref{eq:RhoV}.}
\label{fig:Density}
\end{figure}

The question which arises now is: how do we construct the three dimensional cross-sections $C_0$
from the 2-dimensional lattice? The solution is depicted in Figure \ref{fig:Density} where we have
simply taken the original lattice and rotated it by $\pi/2$ about the central geodesic $G_1$. This
creates $C_0$ and $C_1$ as truncated pyramids with a square cross-section. In each of these we take
the density to be constant. The volume of $C_0$ and $C_1$ can be computed by elementary Euclidean
geometry (the dust is minimally coupled to the geometry and thus curvature corrections can be
ignored). This leads to
\begin{equation*}
\iV_i = \frac{1}{3}\iLzz_{i} 
        \left(\iLLxx_{i} + \iLxx_{i}\iLxx_{i+1} + \iLLxx_{i+1}\right)
\end{equation*}
where $\iLxx_{i}$ and $\iLxx_{i+1}$ are the values of $\Lxx$ at nodes $i$ and $i+1$ respectively.
The previous conservation equation can now be re-written as
\begin{equation}
3 m_i = \left(\rho v_n\right)_i 
        \iLzz_{i} \left(\iLLxx_{i} + \iLxx_{i}\iLxx_{i+1} + \iLLxx_{i+1}\right)\label{eq:RhoV}
\end{equation}
where $m_i$ is the conserved rest mass along the worldtube ($m_i$ is set as part of the initial
conditions). The $v_n$ are estimated at the centre of each cell by quadratic interpolation from the
neighbouring nodes (which will draw in nodes beyond this basic cell). This equation can then be
solved for $\rho$. We assign that $\rho$ to the centre of the cell and then use quadratic
interpolation to estimate $\rho$ at the lattice nodes.

% ============================================================================================
\section{Maximal slicing}
\label{sec:MaxN}

A maximally sliced spacetime is defined to be a spacetime for which $K=0$ everywhere. Such
spacetimes are often constructed by first setting $K=0$ on an initial Cauchy surface (e.g.\ on a
time symmetric initial slice) and then demanding that $dK/dt=0$ throughout the evolution. For our
lattice we have $K=2\Kxx+\Kzz$ and thus from the equations (\ref{eq:DKxx},\ref{eq:DKzz}) we see
that $dK/dt=0$ provided
\begin{equation*}
0 = 2\dNdxx + \dNdxx - N\left(R - 4\pi k\left(1+2v^2_n\right)\rho\right)
\end{equation*}
But in \PaperI\ we showed that under spherical symmetry 
\begin{equation*}
\dNdxx = \frac{1}{\Lxx}\dLxxdz \dNdz
\end{equation*}
which allows us to re-write the previous equation as
\begin{equation}
0 = \dNdzz + \frac{2}{\Lxx} \dLxxdz \dNdz 
  - N \left(R - 4\pi k\left(1+2v^2_n\right)\rho\right)\label{eq:MaxN}
\end{equation}
We treat this as an ordinary differential equation for $N$. The boundary conditions are simple, at
$z=0$ we require $dN/dz=0$ while at the outer boundary we require $1=\lim_{z\rightarrow\infty} N$.
Note also that the differential equation is singular at $z=0$ (due to the $1/\Lxx$ term). We deal
with this by appealing to the spherical symmetry of the solution at $z=0$ to deduce that
$\dNdxx=\dNdxx$ and thus our original differential equation for $N$ can be re-written as
\begin{equation} 
0 = 3\dNdzz - N \left(R - 4\pi k\left(1+2v^2_n\right)\rho\right)
   \quad\text{at }z=0\label{eq:MaxNIBC}
\end{equation} 
which is clearly non-singular. The same result can also be obtained by applying l'H\^{o}pital's
rule to $(1/\Lxx)(dN/dz)$ as $z\rightarrow0$. At the junction we know that $\rho$ and $R$ suffer a
jump discontinuity. Thus we expect a corresponding jump discontinuity in $d^2N/dz^2$ which in turn
forces both $N$ and $dN/dz$ to be continuous across the junction. This adds extra constraints to
the numerical solution of the above equation. We will cover this in more detail in section
\ref{sec:JctN} but for the moment we note that our method computes two separate solutions, one for
either side of the junction, which are then matched at the junction.

% ============================================================================================
\section{The junction conditions}
\label{sec:JctCond}

Darmois \cite{darmois:1927} and later Israel \cite{israel:1966-01} developed a very elegant
approach to handle discontinuities in a metric in General Relativity. However, their method
requires some work to push through so we defer the details to Appendix \ref{app:Darmois} preferring
instead to present here a direct approach.

By integrating the geodesic deviation equation \EQref{eq:GD} over a short interval $z\in
(-\eps,+\eps)$ we obtain
\begin{equation*}
0 = \left[\dLxxdz\right]^{+\eps}_{-\eps} + \int^{+\eps}_{-\eps}\> \Rz \Lxx\>dz
\end{equation*}
If we require $\Rz$ to be bounded on each Cauchy surface then we must have
\begin{equation*}
0 = \lim_{\eps\rightarrow0} \left[\dLxxdz\right]^{+\eps}_{-\eps}
\end{equation*}
and thus $d\Lxx/dz$ is continuous everywhere on the lattice and, most importantly, across the
junction. We also know that $\Lxx$ and $d\Lxx/dt$ must be continuous and thus from the evolution
equation \EQref{eq:DLxx} we see that $0=\lim_{\eps\rightarrow0}[\Kxx]^{+\eps}_{-\eps}$. From here
on we shall dispense with the limits on the square brackets and take $[\dots]$ to mean
$\lim_{\eps\rightarrow0}[\cdots]^{+\eps}_{-\eps}$.

Applying a similar integration to the Bianchi identity leads to
\begin{equation*}
0 = \left[\LLxx \Rx\right]
  - \lim_{\eps\rightarrow0}\int^{+\eps}_{-\eps}\> \Rz\dLLxxdz\>dz
\end{equation*}
and thus
\begin{equation}
0 = \left[\Rx\right]\label{eq:JRx}
\end{equation}
since $\Lxx$ must be continuous every where on the lattice. Thus we conclude that $\Rx$ is
continuous on the lattice. However, by inspection of the Hamiltonian constraint \EQref{eq:Ham}, we
see that the same can not be said for $\Rz$. Since we know that $0=[\Lxx]$ and $0=[\Kxx]$ we see
that continuity of the Hamiltonian requires
\begin{equation}
[\Rz] = [4\pi k\rho v_n^2 - \Kxx \Kzz]\label{eq:JRz}
\end{equation}
We also need suitable junction conditions for the lapse function when using maximal slicing. First
we demand that the clocks of a pair of observers travelling close to but on opposing sides of the
junction should remain synchronised throughout their journey. Thus we find that the lapse is
continuous across the junction, $0=[N]$. For the first derivative we follow the method outlined
above. Integrating the maximal slicing equation \EQref{eq:MaxN} over the short interval $z\in
(-\eps,+\eps)$ leads to
\begin{equation*}
0 = [\dNdz] + \lim_{\eps\rightarrow0} \int_{-\eps}^{+\eps}\left(
\frac{2}{\Lxx} \dLxxdz \dNdz 
  - N \left(R - 4\pi k\left(1+2v^2_n\right)\rho\right)\right)\>dz
\end{equation*}
and as we except all terms in the integral to be bounded (at worst) and $\Lxx>0$ we see that this
requires 
\begin{equation}
0 = [\dNdz]\label{eq:JNz}
\end{equation}
Equations \EQref{eq:JRx}, \EQref{eq:JRz} and \EQref{eq:JNz} constitute the full set of junction
conditions for our lattice. Other conditions such as $0=[\Lxx]$ and $0=[N]$ are trivially
implemented in the numerical code (they require no special care). However we have no freedom in our
data to guarantee $0=[d\Lxx/dz]$. The reason is that all of the $\Lxx$ leg lengths are subject to
the evolution equations and we have to live with what they dictate. Of course we expect the jump in
$d\Lxx/dz$ to be small and to vanish as the lattice is progressively refined.

% ============================================================================================
\section{Numerical methods}
\label{sec:NumMeth}

To obtain numerical solutions of our equations we turn once again to the techniques developed in
\PaperI. We use second order accurate finite differences (on a non-uniform grid) for all of the
spatial derivatives, such as $d\Lxx/dz$ and $d^2\N/dz^2$ (though with a two exceptions, as noted
below in section \ref{sec:DbyDz}, for the the three nodes centred on the junction). The time
integration employs a standard 4th-order Runge-Kutta method and the time step is chosen so that the
Courant factor for the smallest $\Lzz$ on the lattice is $1/2$ (the leg on which this occurs lies
on the surface of the dust ball).

The lattice and its attendant equations in this paper differ most notably from those of \PaperI\ by
the presence of the dust ball. This not only introduces new terms in the equations but it also
forces many of the variables, or their derivatives, to be discontinuous at the junction. Dealing
with these discontinuities requires some care. For the geodesic deviation equation \EQref{eq:GD},
the Bianchi identity \EQref{eq:BI} and the maximal lapse equation \ref{eq:MaxN} the general
approach is to solve those equations twice, once on either side of the junction, and then use the
junction conditions to match the solutions. The details are as follows.

% --------------------------------------------------------------------------------------------
\subsection{The Riemann curvatures}
\label{sec:JctRxRz}

The discretised forms of the geodesic deviation equation \EQref{eq:GD} and the Bianchi identity
\EQref{eq:BI} were given in \PaperI\ and, apart form some minor notational changes, are equivalent
to the following pair of equations
\begin{align}
\iLxx_i\iRz_i &= -\iddLxxdz_i\label{num:Rz}\\[10pt]
\begin{split}
2\iLLxx_{i}\iRx_i &= \iLLxx_{i}\big(\iRz_{i}+\iRz_{i-1}\big)\\
        & + \iLLxx_{i-1} \big(2\iRx_{i-1} - \iRz_{i} - \iRz_{i-1}\big)
\end{split}\label{num:Rx}
\end{align}
where the second derivatives of $\Lxx$ are computed using the second order non-uniform finite
differences (as described in section \ref{sec:DbyDz}).

Our plan is to use this pair of equations to calculate the Riemann curvatures on the lattice but we
immediately encounter two problems, the equations are singular at $z=0$ and, as previously noted,
the second derivatives of $\Lxx$ are not continuous across the junction. The first problem is
rather easy to deal with. We draw upon the required spherical symmetry at $z=0$ to deduce that
$d\Rz/dz=0$ at $z=0$ and thus $\Rz(z) = A + B z^2 +\BigO{z^3}$ near $z=0$. The coefficients $A$ and
$B$ are obtained by fitting $\Rz(z) = A + B z^2$ to two samples for $\iRz_i$ (typically $\iRz_{4}$
and $\iRz_{8}$ for $\njct=120$) and then setting $\iRz_i = A + B z^2_i$ for each node near $z=0$
(i.e.\ at $z=0,z_1,z_2$ and $z_3$). For $\Rz$ we again call on the spherical symmetry to assert
that $\iRx_0=\iRz_0$. Our numerical experiments show that we have no need to use the quadratic
interpolation scheme for $\iRx$ near $z=0$.

We turn now to the issue of the junction. As with the lapse function, we compute both Riemann
curvatures separately on each side of the junction. We first use the above equations to compute the
curvatures for all of the interior lattice nodes excluding the node at the junction. At the
junction we apply a series of interpolations in conjunction with the boundary conditions to set the
curvatures on the junction and one node point outside it. The details are as follows.

First we use cubic extrapolation to compute the one-sided limits $\lim_{z\uparrow\zjct}\Rx$ and
$\lim_{z\uparrow\zjct}\Rz$, which we abbreviate as $\iRx^\upM$ and $\iRz^\upM$. We then use the
junction condition \EQref{eq:JRx} and the Hamiltonian constraint \EQref{eq:Ham},
which we re-write as
\begin{align}
\iRx^\upP & = \iRx^\upM\label{num:JRx}\\
\iRz^\upP & = -\frac{1}{2}\left(\Rx + \Kxx^2 + \Kxx\Kzz\right)^\upP\label{num:JRz}
\end{align}
to step across the junction (with the $\upP$ super-script denoting the right hand one-sided limit).
We then return to the above discrete equations \EQref{num:Rz} and \EQref{num:Rx} to compute the
curvatures in the exterior region. This too requires some explanation. We first compute $\iRz_i$
from $i=\njct+2$ to $i=\nvac-1$ (i.e.\ we skip the first exterior node and stop one node in from
the outer boundary). We then return to the node we skipped over (i.e.\ $i=\njct+1$) and use cubic
interpolation (using the the nodes $\njct,\njct+2,\njct+3$ and $\njct+4$) to estimate $\iRz$ at
that node. The Bianchi identity can then be applied to all the exterior nodes (except the node on
the outer boundary). Finally, we use cubic extrapolation to compute the curvatures on the boundary
nodes. This completes the computation of the curvatures.

% --------------------------------------------------------------------------------------------
\subsection{Maximal slicing}
\label{sec:JctN}

The discrete form of the maximal lapse equation \EQref{eq:MaxN} is of the form
\begin{equation}
0 = a_i \iN_{i+1} + b_i \iN_{i} + c_i \iN_{i-1}\label{num:TriN}
\end{equation}
for some set of coefficients $a_i$, $b_i$ and $c_i$ (see Appendix \ref{app:MaxNeqtns} for the
details). We wish to solve this set of equations subject to the following conditions
\begin{alignat}{2}
0 &= \dNdz&\quad&\text{at }z=0\label{num:BCNz}\\
0 &= \left[\N\right]&&\text{at }z=\zjct\label{num:JN}\\
0 &= \left[\dNdz\right]&&\text{at }z=\zjct\label{num:JNz}\\
1 &= \lim_{z\rightarrow\infty}\N\label{num:BCN}
\end{alignat}
By reflection symmetry at $z=0$ we can easily extend the lattice to $z<0$. Thus a discrete version
of \EQref{num:BCNz} would be $\N_{-1} = \N_{+1}$. Continuity at $z=\zjct$ allows us to use one
value of $\N$ at $\zjct$, which we denote by $\Njct$. However, the continuity of $\dNdz$ is not
something we can prescribe but must be obtained by an iterative process (to be described below). We
denote the left and right hand limits for $\dNdz$ at $z=\zjct$ by $\idNdz^{\upM}$ and
$\idNdz^{\upP}$ respectively. We compute these one-sided limits, for a given set of $\iN_i$, by a
cubic extrapolation of $\idNdz_i$. This too requires some explanation. We start with the four nodes
nearest to but excluding the junction. We then use cubic extrapolation of the $\iN_i$ on these
nodes to extend the $\iN_i$ to the junction and two nodes beyond (we store this generated data in a
separate array so as not to overwrite the data already defined on those nodes). Finally we use the
standard non-uniform second order finite differences to estimate the first derivative at the
junction. This computation is done twice, once for each one-sided limit. For the outer boundary we
simply set $1=\iNvac$.

The discrete equations for $\iN_i$ are solved in three iterations with each iteration involving
separate solutions for $\iN_i$ on each side of the junction. The algorithm requires two guesses for
$\N$, one for $\N$ at $z=0$ and one for $\N$ at $z=\zjct$ which we denote by $\gNleft$ and $\gNjct$
respectively. With given values for these guesses we use a Thomas algorithm to solve the
tri-diagonal system \EQref{num:TriN} for $0\leq z\leq\zjct$ and again for $\zjct\leq z\leq\zvac$.
Our guesses are unlikely to be correct (at first) so we record the errors in the boundary
conditions by $\Eleft=\iN_{+1}-\iN_{-1}$ and $\Ejct=\idNdz^{\upP}-\idNdz^{\upM}$. Our aim is to
choose the two guesses so that $0=\Eleft$ and $0=\Ejct$. We chose three pairs of guesses $(0,0)$,
$(0,1/2)$ and $(1,1)$ for $(\gNleft,\gNjct)$ and we recorded the corresponding errors as
$\Eleft^{(j)}$ and $\Ejct^{(j)}$ for $j=1,2,3$. Since the discrete equations are linear and
homogeneous in $\iN_i$ we can form a linear combination such as
\begin{equation}
\N_i = \alpha_1 \N^{(1)}_i + \alpha_2 \N^{(2)}_i + \alpha_3 \N^{(3)}_i\label{num:MaxNSol}
\end{equation}
to satisfy the boundary and junction by an appropriate choice of constants $\alpha_1$, $\alpha_2$
and $\alpha_3$. The result is a 3 by 3 system of equations
\begin{alignat*}{2}
1 &=&\> \Nvac  &= \alpha_1 + \alpha_2 + \alpha_3\\
0 &=& \Eleft &= \alpha_1 \Eleft^{(1)} + \alpha_2 \Eleft^{(2)} + \alpha_3 \Eleft^{(3)}\\
0 &=& \Ejct  &= \alpha_1 \Ejct^{(1)} + \alpha_2 \Ejct^{(2)} + \alpha_3 \Ejct^{(3)}
\end{alignat*}
which is easily solved for the three weights $\alpha_i$ which in turn allows the final (correct)
solution for the maximal lapse to be computed from \EQref{num:MaxNSol}.

% --------------------------------------------------------------------------------------------
\subsection{The time derivatives}
\label{sec:DbyDt}

Spatial derivatives are calculated at each node using data from the surrounding nodes, and in cases
where this might draw in data from across the junction, we first use cubic extrapolation to extend
the data across the junction (which we store separately so as not to overwrite exiting data).

With the exception of the junction node there is no ambiguity in applying the evolution equations
to the nodes of the lattice. However, the discontinuities at the junction demand, once again, that
we tread carefully near and at the junction. Consider $\Kzz$ which in geodesic slicing will be
multiple valued at the junction. How do we handle this situation? We have already exhausted our
supply of junction conditions in forming the two jump conditions \EQref{num:JRx} and
\EQref{num:JRz} for the Riemann curvatures. So in the absence of any further information about
$\Kxx$ we have no choice but to consider its left and right hand limits as independent of each
other (despite the loose coupling afforded by the evolution equations). Each term could be evolved
by evaluating time derivatives built from one-sided limits of the source terms. There is however an
easier approach which we found to work quite well. The idea is to re-interpret the junction node
not as node on which to apply the evolution equations but rather as a convenient staging post to
impose the junction conditions. In this view we do \emph{not} evolve the data on the junction node.
Rather we treat that data as kinematical which we compute by one-sided extrapolations of the
surrounding data (which are evolved via the normal evolution equations).

So in our code we use \EQref{eq:DLxx}, \EQref{eq:DKxx}, \EQref{eq:DKzz}, \EQref{eq:DVn} and
\EQref{eq:DVz} (subject to a minor change noted below) to evolve $\Lxx$, $\Kxx$, $\Kzz$, $\Vn$ and
$\Vz$ on the nodes $i=0,1,2,\cdots \njct-1$ and $i=\njct+1, \njct+2, \njct+3, \cdots\nvac-1$. We
use \EQref{eq:DLzz} (again, see below) to evolve the $\Lzz$ for all legs not connected to the
junction. For the two legs attached to junction we use one-sided cubic extrapolation of $d\Lzz/dt$
to compute their time derivatives. At the outer boundary we impose static boundary conditions for
all of the data.

There is one exception to this simple algorithm. We use a one-sided extrapolation to set $d\Lxx/dt$
at the node $\njct-1$. This proved to be essential for long term stability with maximal slicing
(but made no difference in geodesic slicing). We can offer no reasonable explanation as to why this
works other than the following admittedly vague rationalisation. By extrapolating the time
derivatives outwards from the interior of the dust ball to the node $\njct-1$ we might be halting
or minimising the inward propagation of any errors that arise at the junction. Delving deeper
into this mystery is best left for another time.

There is one remaining subtlety that we must address. The careful reader may have noticed that in
the present context we are treating the $\Kxx$ and $\Kzz$ as being defined on the nodes whereas the
extrinsic curvatures arose in section \ref{sec:EvolEqtns} by approximating the integrals by a
mid-point rule. Thus if we wish to use node based values for $\Kxx$ and $\Kzz$ we should use a
Trapezoidal rule to estimate the integrals. This is a minor change and leads to the following
node-based equations
\begin{align}
\DLxx & = -\left\langle\N \Kxx\right\rangle \Lxx 
          +\left\langle\gammaz\right\rangle\dLxxdz\label{eq:DLxxAlt}\\
\DLzz & = -\left\langle\N \Kzz\right\rangle \Lzz + \left[\gammaz\right]\label{eq:DLzzAlt}
\end{align}
where the angle-brackets denotes an average of that quantity over the leg while the square-brackets
continues to denote the change across a leg. In fact for the $\Lxx$ equation the angle-brackets are
redundant (the end points carry identical values) but were retained simply for emphasis. Since the
Riemann curvatures are already node-based we see that no such averaging is required for the
extrinsic curvature equations \EQref{num:Rz} and \EQref{num:Rx}. Note also that the spatial
derivatives are also node based (by suitable choice of the finite difference operators).

% --------------------------------------------------------------------------------------------
\subsection{The spatial derivatives}
\label{sec:DbyDz}

The evolution equations (\ref{eq:DLxx}--\ref{eq:DKzz}) and the momentum constraint \EQref{eq:Mom}
require spatial derivatives of the $\Lxx$, $\Lzz$, $\Kxx$ and $\Kzz$. For all but the two nodes
either side of the junction (i.e.\ at nodes $\njct-1$ and $\njct+1$), and the junction itself, we
employ second order non-uniform spatial derivatives as described in \PaperI. On the two nodes
either side of the junction we use one-sided quadratic extrapolation. This is the only point in the
code where we used quadratic approximations and we do so because both linear and (interestingly)
cubic interpolation lead to instabilities forming at the junction (at around $t\approx13$ for cubic
extrapolation and only for one of our models with $\njct=240$ and $\nvac=1200$). The derivatives at
the junction are computed last using one-sided cubic extrapolation.

The only other spatial derivatives that need to be computed are the first and second derivatives of
the lapse function (for use in the maximal slicing equation \EQref{eq:MaxN} and in the particle
equations (\ref{eq:DVn},\ref{eq:DVz})). Once again we use the second order non-uniform spatial
derivatives from \PaperI\ for all of the nodes with the exception of the five nodes centred on the
junction. For nodes $\njct\pm2$ and $\njct\pm1$ we use cubic extrapolation to build an extended set
data. This introduces some temporary and artificial nodes which we chose to be symmetric to the
real nodes (e.g.\ when extending the data for node $\njct-1$ we create new nodes $\njct$, $\njct+1$
that are the mirror images (in $\njct-1$) of $\njct-2$ and $\njct-3$). The derivatives on nodes
$\njct\pm2$ and $\njct\pm1$ are then computed on this extended data set using the standard
non-uniform centred differences while the derivatives on the junction are computed using one-sided
cubic extrapolation.

Once the maximal slicing equation has been solved we do have the option of using that equation as
an alternative way to calculate the second derivatives of the lapse. We chose not to do so because
we did not want to give the smooth lattice method a helping hand -- we want to test the method
under conditions closer (albeit in 1+1 form) to what we would expect for other spacetimes (i.e.\
for a true 3+1 evolution).

% --------------------------------------------------------------------------------------------
\subsection{The initial data}
\label{sec:IniData}

We require two things of our initial data, first they must satisfy the constraints \EQref{eq:Ham}
and \EQref{eq:Mom}, and second they must describe a time-symmetric initial slice. This last
condition is readily satisfied upon setting $\Kxx=0$, $\Kzz=0$ and $\Vn=1$, $\Vz=0$ which in turn
ensures that the momentum constraint is also satisfied. What we are left with is the Hamiltonian
constraint, the leg lengths, $\Lxx$, $\Lzz$ and the density $\rho$, i.e.\ we have one constraint
for three (sets) of data. Clearly there are a range of options here, so what should we do? We turn
once again to the scheme developed in \PaperI. There we chose to set the $\Lzz$ and then use the
Hamiltonian constraint to set the $\Lxx$. But here we also need the density.

Keep in mind that our aim is neither to discover nor explore the Oppenheimer-Snyder solution but
rather to use it as a test of the smooth lattice method. Thus it is not unreasonable to borrow some
information from the exact solution to set some of the data on the lattice, in particular the
density. We recall here some basic equations from the exact solutions for the Oppenheimer-Snyder
spacetime (see \cite{opp-snyder:1939-01,mtw:1973-01,petrich:1985-01,hajicek:2008-01}).

There are two free parameters in the solution, the ADM mass $m$ and the Schwarzschild areal radius
$R_0$ of the dust ball. From these we can compute the proper radius of the dust ball $\zjct$, the
FRW parameters $a_m$ and $\chi_0>0$ and the density $\rho$ using
\begin{align}
\sin^2\chi_0 &=\frac{2m}{R_0}\label{eq:FRW1}\\
a_m &= \frac{2m}{\sin^3\chi_0}\label{eq:FRW2}\\
\zjct & = a_m\chi_0\label{eq:FRW3}\\
8\pi k \rho &= \frac{3}{a^2_m}\label{eq:FRW4}
\end{align}
Clearly, we also have $\rho=0$ in the Schwarzschild exterior.

We used these equations to set $\zjct$ and $\rho$, for a given $m$ and $R_0$. To this we added
choices for the total length of the lattice $\zvac$, the number of interior nodes $\njct$ and the
total number of nodes $\nvac$ on the lattice.

Note that we still have the freedom to distribute the nodes along the $z$-axis (this amounts to
setting the $\iLzz_i$). We know that some of the spatial gradients are zero at $z=0$, that they
rise to a maximum near the junction and then settle down in the distant asymptotically flat regions
of the lattice. Thus it makes sense to concentrate the nodes around the junction. With this in mind
we chose to start at the junction and use a geometric progression to set the $\iLzz_i$ in both the
interior and exterior regions. We chose the same geometric ratio in both regions while also
requiring $\iLzz_{\njct-1} = \iLzz_{\njct}$. From here it is simple matter to compute all of the
$\iLzz_i$ across the lattice.

We now turn to the problem of setting $\Lxx$ and the Riemann curvatures. By reworking the
Hamiltonian constraint, geodesic deviation and Bianchi identity we find that across the lattice
\begin{equation}
\begin{split}
\iLxx_{i} &= \iLxx_{i-1} + \frac{\iLzz_{i-1}}{\iLzz_{i-2}}
                           \left(\iLxx_{i-1}-\iLxx_{i-2}\right)\\
          &\quad  - \frac{1}{2}\iLzz_{i-1}\left(\iLzz_{i-1}+\iLzz_{i-2}\right)
                    \left(\Lxx\Rz\right)_{i-1}
\end{split}\label{ini:Lxx}
\end{equation}
while the curvatures in the dust-ball are constant and are given by
\begin{equation}
\iRz_i = \iRx_i = \frac{8\pi k\rho}{3}\label{ini:RiemIn}
\end{equation}
and finally, in the Schwarzschild region, we find
\begin{align}
\iRz_i &= \iRz_{i-1}\left(\frac{5\iLLxx_{i-1}
         -\iLLxx_{i}}{5\iLLxx_{i}
         -\iLLxx_{i-1}}\right)\label{ini:RzOut}\\
\iRx_i &= -2\iRz_i\label{ini:RxOut}
\end{align}
These equations can be used to set the $\Lxx$, $\Rx$ and $\Rz$ across the lattice (a process that
will require the junction conditions for the curvatures). But to start the ball rolling we must
make some choice for $\iLxx_0$, and $\iLxx_1$. Clearly $\iLxx_0=0$ but for $\iLxx_1$ we are free to
make any choice we like (we chose $\iLxx_1= 0.001 \iLzz_0$ so that $d\Lzz/dz=0.001$ at $z=0$, as
discussed below in section \ref{sec:Results}).

% --------------------------------------------------------------------------------------------
\subsection{Density}
\label{sec:RhoNumeric}

In section \ref{sec:Rho} we noted that the density can be computed using either the Hamiltonian
constraint, in the form \EQref{eq:RhoH}, or by the conservation equation \EQref{eq:RhoV}. We find
that, for long term stability when using the second method, we are forced to use the Hamiltonian
constraint at exactly the two nodes just inside the junction (i.e.\ at nodes $\njct-2$ and
$\njct-1$). This was found by pure numerical experimentation. Why this should be so is unclear to
us but it is probably tied to the same mechanism noted above (with regard to halting the inward
propagation of errors from the junction by imposing ``correct'' values near the junction).

% ============================================================================================
\section{Diagnostics}
\label{sec:Diagnst}

From the known solution for the Oppenheimer-Snyder spacetime a number of useful diagnostics can be
drawn. Here we will discuss those diagnostics which, in the following section, we will apply to our
numerical results.

For geodesic slicing it is rather easy to show \cite{mtw:1973-01} that the proper radius of the
dust-ball $\zjct$ varies with proper time $t$ according to
\begin{equation}
\zjct(t) = \frac{a_m\chi_0}{2}\left(1 + \cos\eta(t)\right)\label{tst:ZJ}
\end{equation}
where $a_m = 2m/\sin^3\chi_0$ and $\eta(t)$ is the solution of $0= -2t + a_m(\eta + \sin\eta)$ with
$\eta >0$ (notice that Petrich \etal use $\eta$ where we use $\eta-\pi$).

Another simple diagnostics arises from the central density which is given by
\begin{equation}
\rho(t) = 24 a_m^2 \left(1+\cos\eta(t)\right)^{-3}\label{tst:Rho}
\end{equation}
This is singular when $\eta=\pi$ at which point the proper radius is zero and the dust ball has
collapsed onto the singularity. This will occur after a proper time of
\begin{equation}
\TimeSgeo = \pi m \left(\frac{R_0}{2m}\right)^{3/2}\label{tst:TSg}
\end{equation}
and at this moment, or a short time before, we expect our code to crash.

As the dust-ball collapses an outer apparent horizon will form and this too provides useful checks
on our numerics. It is known that when the outer most apparent horizon forms it does so at the
surface of the dust-ball. In our numerical code we locate the horizon by noting where on the radial
axis the quantity $d\Lxx/dz - \Kxx\Lxx$ vanishes. The root of this equation is the location of the
apparent horizon (this follows from the condition that $0=\partial A/\partial u$ where $A$ is the
area of a 2-sphere and $\partial/\partial u$ is the outward pointing null vector to the 2-sphere,
see \PaperI\ for more details). The time at which the horizon forms is also well known and this
affords yet another check on our numerical results. For geodesic slicing it can be shown that the
time, $\TimeHgeo$, and location $\LocHgeo$, of the apparent horizon are given by
\begin{align}
\TimeHgeo &= \frac{m}{\sin^3\chi_0}\left(\pi - 2\chi_0 + \sin(2\chi_0)\right)\label{tst:THg}\\
\LocHgeo &=  \frac{m}{\sin^3\chi_0}\left(1 - \cos(2\chi_0)\right)\label{tst:ZHg}
\end{align}
Note that in geodesic slicing the nodes are at rest relative to the Cauchy surfaces and thus this
time $\TimeHgeo$ equals the proper time measured by the observer following that junction as it
falls inwards and eventually meets the outward expanding event horizon. The quantity $\LocHgeo$
measures the proper distance out from the centre of the dust-ball to the junction.

Hawking's area theorem can also be used as a diagnostic. The theorem requires that the area of the
event horizon should be constant once all of the dust has fallen within the event horizon. For our
lattice this would require that the $\Lxx$ on the event horizon should be constant for the
remainder of the evolution. This is easily checked (by interpolating the values of $\Lxx$ from the
nodes onto the event horizon).

Equations for the time and location of the horizon, as well as the density and radius diagnostics,
are also available for maximal slicing but with one drawback -- the equations as given by Petrich
\etal require a numerical integration of some elliptic integrals. This introduces its own set of
numerical issues and we found that our implementation of the Petrich equations could only be
reliably used for $t\lesssim32$ (for $m=1$ and $R_0=5$). Even so, this was sufficient time to allow
for a useful comparison to be made.

We also have one extra diagnostic for the case of maximal slicing. There it is known that the lapse
function will, after an initial period, settle into an exponential decay. Petrich \etal show that
$N(t,0) \sim A \exp(\beta t)$ where $A$ is a constant and $\beta = - (2/3)^{(3/2)} \approx
-0.5443311$. We can use this to test our code by measuring the slope of the $\log N$ versus $t$.

There are of course two other diagnostics -- the Hamiltonian and momentum constraints.

In summary we have the following set of diagnostics.

\begin{itemize}
\item The constraints.
\item The history of the junction.
\item The history of the central density.
\item The crash time for geodesic slicing.
\item The Petrich solution for maximal slicing.
\item The exponential collapse of the central lapse.
\item The time and location of the first apparent horizon.
\item The constancy of the area of the event horizon in the vacuum region.
\end{itemize}

Clearly we have a raft of diagnostics and it is now time to turn to the actual results.

% ============================================================================================
\section{Results}
\label{sec:Results}

% --- include data generated by the code ---
% written by programs/geodesic/lattice-evol:evolve
\def\TimeCrashGeo{12.41793}
\def\TimeCrashExact{12.41824}
\def\TimeStepCrashGeo{8.19\times10^{-6}}

% written by programs/petrich/horizon
\def\APHtimeMaximal{17.02246}
\def\APHtimeGeodesic{10.87804}
\def\APHlocationGeo{2.16527}
\def\APHlocationMax{2.37971}

% written by programs/geodesic/lattice-evol:evolve
\def\APHformedTimeGeo{10.87837}
\def\APHformedLocationGeo{2.16534}

% written by programs/maximal/lattice-evol:evolve
\def\APHformedTimeMax{16.98238}
\def\APHformedLocationMax{2.38015}

% written by programs/maximal/post-process/least-square
\def\LapseSlope{-0.54424}
\def\LapseSlopeExact{-0.54433}
\def\LapseSlopeBeg{25.0}
\def\LapseSlopeEnd{35.0}

% written by programs/geodesic/lattice-evol:evolve
\def\TimeStepBegin{5.38\times10^{-3}}

% ------------------------------------------

Our aim was to write a code that used as few assumptions as needed to obtain reliable results. In
the end we have split the computation of the lapse from the rest of the code. The evolution of the
code takes as input (at each time step) the values of the lapse across the lattice. We do not use
the Hamiltonian or momentum constraints apart from the two exceptions noted in sections
\ref{sec:JctRxRz} and \ref{sec:RhoNumeric}. We employ no artificial smoothing such as artificial
viscosity nor do we add on any constraint preserving terms. Our time integrations are conducted
using a 4th-order Runge-Kutta routine and our time step was updated after every time step by
setting it equal to $1/2$ the shortest $\Lzz$ on the grid (which usually is the leg on or just
inside the junction). This choice sets the Courant factor to $1/2$ for legs near the junction (with
smaller values for legs away from the junction).

We set our initial data using $8\pi k=1$, $m=1$, $R_0=5$, $\zvac=400$ and $d\Lxx/dz=0.001$ at
$z=0$. We ran the code for three separate models, with $(\njct,\nvac)=(60,300)$, $(120,600)$ and
$(240,1200)$ for both the geodesic and maximal slicing and one further model with
$(\njct,\nvac)=(240,2400)$ for maximal slicing. The results for a selection of quantities are
displayed in Figures (\ref{Geo:LxxLzz}--\ref{Max:ErrHorizon}). The first point to note is that the
results are well behaved with no apparent instabilities even through to very late in the evolution.
The junction remains sharp without any noticeable smoothing and the constraints, though not zero,
do not show the exponential growth often associated with unstable evolutions.

We ran the geodesic code until it crashed at time $\TimeS=\TimeCrashGeo$ which compares well with
the exact time $\TimeSgeo=\TimeCrashExact$ (note that the time step at the crash was
$\TimeStepCrashGeo$ which is considerably smaller than the initial time step of $\TimeStepBegin$).

For geodesic slicing we found the apparent horizon formed at $\TimeH=\APHformedTimeGeo$ and
$\LocH=\APHformedLocationGeo$ while the exact values are $\TimeHgeo=\APHtimeGeodesic$ and
$\LocHgeo=\APHlocationGeo$. While for maximal slicing the numerical values were
$\TimeH=\APHformedTimeMax$, $\LocH=\APHformedLocationMax$ compared with the exact values
$\TimeHmax=\APHtimeMaximal$, $\LocHmax=\APHlocationMax$.

For maximal slicing and the collapse of the lapse diagnostic we estimated the slope over the
interval $\LapseSlopeBeg\leq t\leq\LapseSlopeEnd$ and obtained $\beta=\LapseSlope$ compared with
the exact value of $\LapseSlopeExact$.

In Figures (\ref{Geo:CnvrgRadRho},\ref{Max:CnvrgRadRho}) we have plotted the fractional errors in
the radius and the central density for the first three models (as described above). For geodesic
slicing the errors are very small. For maximal slicing the errors do decrease with increasing
number of nodes however it would appear that the errors are not converging to zero. The simple
explanation is that we set $N=1$ on a finite outer boundary and this clearly incurs an error. To
test this we re-ran our code with different choices for the location of the outer boundary (while
retaining the same number of nodes). This showed that the peaks in Figures (\ref{Max:CnvrgRadRho})
varied inversely with the distance to the outer boundary $\zvac$. Incidentally, the broad peaks in
those figures correspond to the formation of the apparent horizon.

For maximal slicing we have taken a snapshot of the numerical data at a fixed time, Figure
(\ref{Max:CnvrgNRho}), to compare the density and the lapse with their exact values (from the
Petrich code) across the lattice. Once again we see an initial convergence from coarse to fine
resolutions but then the convergence appears to falter. This is also due to the use a of finite
outer boundary, the peaks in the errors being proportional to $1/\zvac$. Similar considerations
apply to the snapshots of the Hamiltonian and momentum constraints, see Figure
(\ref{Max:CnvrgHamMom}). The corresponding snapshot for geodesic slicing is shown in Figure
(\ref{Geo:CnvrgHamMom}). In this case the errors are not limited by $\zvac$ but instead depend only
on $\njct$ and $\nvac$ and with the limited data available (only three models) it appears that the
peaks in these figures reduce by a factor of about 4 for each doubling of $(\njct,\nvac)$.

The fractional changes in the horizon $\Lxx$ are shown in Figure (\ref{Max:ErrHorizon}). This shows
that for $t<32$ the horizon area varied by no more than $5\times10^{-2}$ percent for the coarsest
model improving to less than $1\times10^{-3}$ percent for the finest model. By $t\approx500$ the
error had grown to less than 2 percent for the finest model.

We also ran our code using the Hamiltonian constraint to set the density and found results very
similar to those just given.

% ============================================================================================
\section{Discussion}
\label{sec:Discuss}

The results just presented are very encouraging. They are consistent with our previous
investigations of the smooth lattice method \cite{brewin:1998-01,brewin:1998-02,brewin:2002-01}
yielding excellent results with only minor demands on computational resources. This gives us
confidence that the method is viable but further tests are certainly required in particular an
example in full 3+1 dimensions, without symmetries, is imperative. This is a work in progress and
we hope to report on this soon.

One striking feature of the results for maximal slicing which we have so far ignored is the
wave-like behaviour displayed in many of the plots (and similar behaviour was also noted in
\PaperI). This is certainly not a gravitational wave (the spacetime is spherically symmetric). Can
this behaviour be understood from the evolution equations? Without delving too far into the
analysis we note that the first order equations \EQref{eq:DLxx} and \EQref{eq:DKxx} can be recast
as a single second order equation for $\Lxx$. This will involve $d^2\Lxx/dt^2$ and the Riemann
curvatures. But in these late times, where the waves are apparent, we see that
$\vert\Rz\vert\ll\vert\Rx\vert$ and thus the curvatures are dominated by $\Rx$ which, through the
geodesic deviation equation, \EQref{eq:GD}, introduces $d^2\Lxx/dz^2$ into the second order
evolution equation for $\Lxx$. Thus we have in the one equation the two key elements of the
one-dimensional wave equation for $\Lxx$ and so wave-like behaviour is not surprising. Of course
this is a very loose argument and there are many more terms to contend with before it can be said
that the wave-like behaviour can be understood in standard terms. We will pursue this matter in a
later paper.

% ============================================================================================
\appendix

% ============================================================================================
\section{The Darmois-Israel junction conditions}
\label{app:Darmois}

Consider a spacetime $(g,{\cal M})$ and let ${\cal S}$ be some 3-dimensional time like surface in
${\cal M}$. This surface will divide ${\cal M}$ into two parts; one part, ${\cal M^L}$, to the left
of ${\cal S}$ and another part, ${\cal M^R}$, to the right. In the absence of surface layers (e.g.
infinitesimally thin shells of dust with non-zero energy) the Darmois-Israel junction conditions
\cite{darmois:1927,israel:1966-01} ensure that $g$ is a solution of Einstein's equations
everywhere in ${\cal M}$ provided it is a solution in ${\cal M/S}$, and most importantly, that the
first and second fundamental forms on ${\cal S}$ are continuous across ${\cal S}$.

Suppose we denote the first and second fundamental forms on ${\cal S}$ by ${\tilde h}$ and ${\tilde
K}$ respectively. Then each of these quantities can be calculated from the embedding of ${\cal S}$
in either ${\cal M^L}$ or in ${\cal M^R}$. The junction conditions requires that both computations
yield identical results, that is $0=[{\tilde h}]$ and $0=[{\tilde K}]$.

In our case we take $(g,{\cal M})$ to be the Oppenheimer-Snyder spacetime and ${\cal S}$ to be the
surface generated by the evolution of the surface of the dust. We will use a \~\ symbol to denote
quantities that live on $\cal S$, for example, ${\tilde h}$ and ${\tilde K}$ will represent the
3-metric and extrinsic curvatures respectively on ${\cal S}$. We extend this notation slightly to
allow ${\tilde n}$ to be unit (space like) normal to ${\cal S}$ in ${\cal M}$.

Our first task will be to express the junction conditions in terms of data on $\Sigma$.

We know that $\Lxx$ lies in ${\cal S}$ and thus the junction condition $0=[{\tilde h}]$ requires
both $0=[\Lxx]$ and $0=[d\Lxx/dt]$ while $0=[{\tilde K}]$ requires $0=[d\Lxx/dz]$ (note that $d/dz$
is not normal to ${\cal S}$ but it can be resolved into pieces parallel and normal to ${\cal S}$
and the result follows). Looking back at the evolution equation \EQref{eq:DLxx} we see that this
series of observations leads to the simple condition that $0 = [\Kxx]$. We will make use of this
result in the following discussions on the Riemann curvatures.
Consider the Gauss equation for ${\cal S}$, namely,
\begin{equation*}
{\tilde\bot} \left({}^4R_{\mu\alpha\nu\beta}\right) = {\tilde R}_{\mu\alpha\nu\beta}
-{\tilde K}_{\mu\nu}{\tilde K}_{\alpha\beta} + {\tilde K}_{\mu\beta}{\tilde K}_{\alpha\nu}
\end{equation*}
where ${\tilde\bot}$ is the projection operator for ${\cal S}$ i.e.\ ${\tilde\bot}^\mu{}_\nu =
\delta^\mu{}_\nu - {\tilde n}^\mu{\tilde n}_\nu$. Since the vectors $m_x^\mu$, $m_y^\mu$ are both
tangent to ${\cal S}$ and since $0=[{\tilde K}_{\mu\nu}]$ we have
\begin{equation*}
0 = [{}^4R_{\mu\alpha\nu\beta} m_x^\mu m_x^\nu m_y^\alpha m_y^\beta]
\end{equation*}
We can apply the Gauss equation once again, but this time for $\Sigma$ rather than ${\cal S}$, that
is
\begin{equation*}
{\bot} \left({}^4R_{\mu\alpha\nu\beta}\right) = 
    {R}_{\mu\alpha\nu\beta} + {K}_{\mu\nu}{K}_{\alpha\beta} - {K}_{\mu\beta}{K}_{\alpha\nu}
\end{equation*}
This leads to the simple equation
\begin{equation}
0 = [\Rx]\label{DI:JRx}
\end{equation}
where we have used $0=[\Kxx]$ and the fact that $K_{\mu\nu}$ is diagonal. This is one of our two
junction conditions for the Riemann curvature. The second condition will apply to $\Rz$ and as we
shall soon see amounts to no more than requiring continuity of the Hamiltonian constraint across
the junction (as we would expect).

We repeat the above procedure this time using the vectors $m_x^\mu$ and $t^\mu$ and after the first
Gauss equation we find
\begin{equation*}
0 = [{}^4R_{\mu\alpha\nu\beta} m_x^\mu m_y^\nu t^\alpha t^\beta]
\end{equation*}
Now $t^\mu$ is spanned by $n^\mu$ and $m_z^\mu$, that is $t^\mu = v_n n^\mu + v_z m_z^\mu$, and
thus we have
\begin{align*}
0 &= [   v_n^2 {\bot} \left(R_{\mu\alpha\nu\beta} n^\mu n^\nu\right) m_z^\alpha m_z^\beta
      + 2v_n v_z {\bot} \left(R_{\mu\nu\alpha\beta} n^\mu\right) m_z^\nu m_x^\alpha m_x^\beta\\
  &   + v_z^2 {\bot} \left(R_{\mu\nu\alpha\beta}\right) m_z^\mu m_z^\nu m_x^\alpha m_x^\beta ]
\end{align*}
where we have also included the projection operator $\bot{}$ for $\Sigma$ (since $m_x^\mu$ and
$m_z^\mu$ are both tangent to $\Sigma$) in preparation for the second application of the Gauss
equation. This time we will need the Gauss equation and its contractions with $n^\mu$, that is
\begin{align*}
{\bot} \left({}^4R_{\mu\alpha\nu\beta}\right) &= 
    {R}_{\mu\alpha\nu\beta} + {K}_{\mu\nu}{K}_{\alpha\beta} - {K}_{\mu\beta}{K}_{\alpha\nu}\\
{\bot} \left({}^4R_{\mu\alpha\nu\beta} n^\mu \right) &= 
    {K}_{\alpha\beta|\nu} - {K}_{\alpha\nu|\beta}\\
{\bot} \left({}^4R_{\mu\alpha\nu\beta} n^\mu n^\nu \right) &= 
   - {\bot} \left({}^4R_{\alpha\beta}\right)
   + R_{\alpha\beta}
   + K K_{\alpha\beta} - K_{\alpha\mu} K^{\mu}{}_{\beta}
\end{align*}
Using the Einstein equations, ${}^4R_{\alpha\beta} = 8\pi k(T_{\alpha\beta} - (1/2)g_{\alpha\beta}
T)$, the constraint equation $K_{|\mu}-K_{\mu}{}^\nu{}_{|\nu} = 8\pi k\bot(T_{\mu\nu} n^\nu)$ and
the diagonal character of $K_{\mu\nu}$ we find that
\begin{align*}
{\bot} \left(R_{\mu\nu\alpha\beta}\right) m_z^\mu m_z^\nu m_x^\alpha m_x^\beta &=
   \Rz + \Kxx\Kzz\\
{\bot} \left(R_{\mu\nu\alpha\beta} n^\mu\right) m_z^\nu m_x^\alpha m_x^\beta &= 
   -4\pi k\rho v_n v_m\\
{\bot} \left(R_{\mu\alpha\nu\beta} n^\mu n^\nu\right) m_z^\alpha m_z^\beta &=
   -4\pi k \rho + \Rx + \Rz + \Kxx\Kzz + \Kxx^2
\end{align*}
and thus our junction condition can be reduced to
\begin{equation}
0 = \left[ \frac{1}{2}\rho v_n^2 - \Rz - \Kxx \Kzz \right]\label{DI:JRz}
\end{equation}
where we have used $v_z^2 = v_n^2 - 1$ to eliminate $v_z$. Looking back at our constraint equations
\EQref{eq:Ham} we see that this last equation, along with $0=[\Rx]$ and $0=[\Kxx]$, shows that the
Hamiltonian constraint must be conserved across the junction (as expected).

% ============================================================================================
\section{The maximal lapse equation}
\label{app:MaxNeqtns}

Let $\iN_i$ be the node values of the lapse function across the lattice. Then using second order
accurate finite differences (on a non-uniform grid) we obtain the following discrete equations
\begin{equation}
0 = a_i \iN_{i+1} + b_i \iN_{i} + c_i \iN_{i-1}
\end{equation}
with
\begin{align}
a_i &= \frac{2}{\iLzz_{i}}\left(\frac{\iLzz_{i-1}}{\iLxx_{i}}\idLxxdz_{i}+1\right)\\
c_i &= \frac{-2}{\iLzz_{i-1}}\left(\frac{\iLzz_{i}}{\iLxx_{i}}\idLxxdz_{i}-1\right)\\
b_i &= \frac{4\iLbar_i}{\iLzz_{i-1}\iLzz_{i}}
       \left(\frac{\iDeltaL_{i}}{\iLxx_{i}}\idLxxdz_{i} -1 \right)\notag\\
    &\quad - 2\iLbar_i\left(R + \pi k\rho(8v^2_n + 4)\right)
\end{align}
for $z>0$ and
\begin{gather}
a_i = \frac{6}{\iLzz_{i}}\qquad
c_i = \frac{6}{\iLzz_{i-1}}\\
b_i = \frac{-4\iLbar_i}{\iLzz_{i-1}\iLzz_{i}}
    - 2\iLbar_i\left(R + \pi k\rho(8v^2_n + 4)\right)
\end{gather}
for $z=0$.  

In the above equations we have introduced $2\iLbar_i=\iLzz_{i}+\iLzz_{i-1}$ and
$\iDeltaL_i=\iLzz_{i}-\iLzz_{i-1}$.

% ============================================================================================
\clearpage

\captionsetup{margin=0pt,font=small,labelfont=bf}

\def\FigOne#1{%
\centerline{%
\includegraphics[width=0.95\textwidth]{#1}}}

\def\FigPair#1#2{%
\centerline{%
\includegraphics[width=0.6\textwidth]{#1}\hfill%
\includegraphics[width=0.6\textwidth]{#2}}}

\def\FigQuad#1#2#3#4{%
\centerline{%
\includegraphics[width=0.6\textwidth]{#1}\hfill%
\includegraphics[width=0.6\textwidth]{#2}}%
\centerline{%
\includegraphics[width=0.6\textwidth]{#3}\hfill%
\includegraphics[width=0.6\textwidth]{#4}}}

% === figures =================================================================

\clearpage

% === Geodesic slicing ========================================================

\begin{figure}[t]
\FigOne{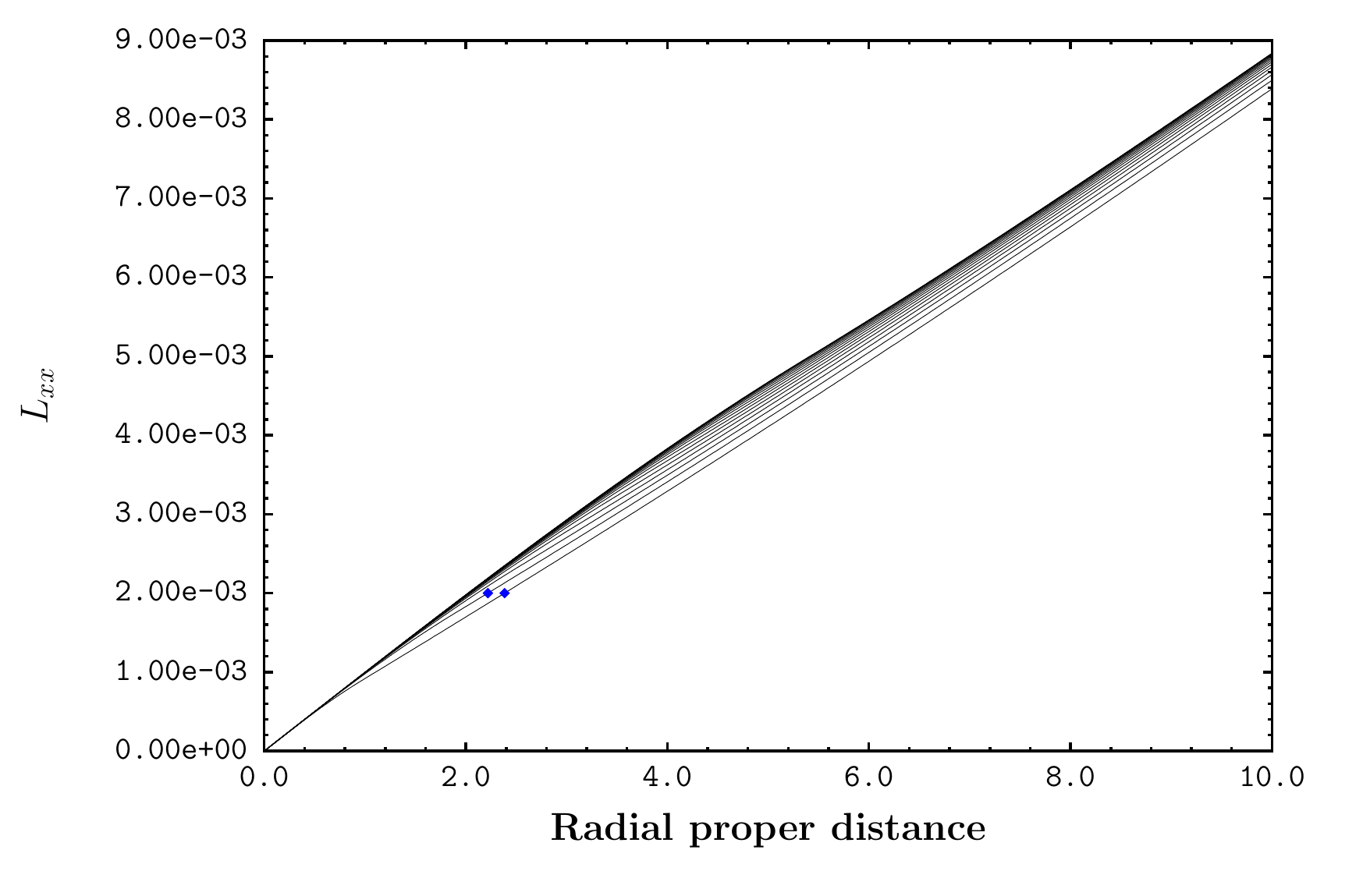}
\FigOne{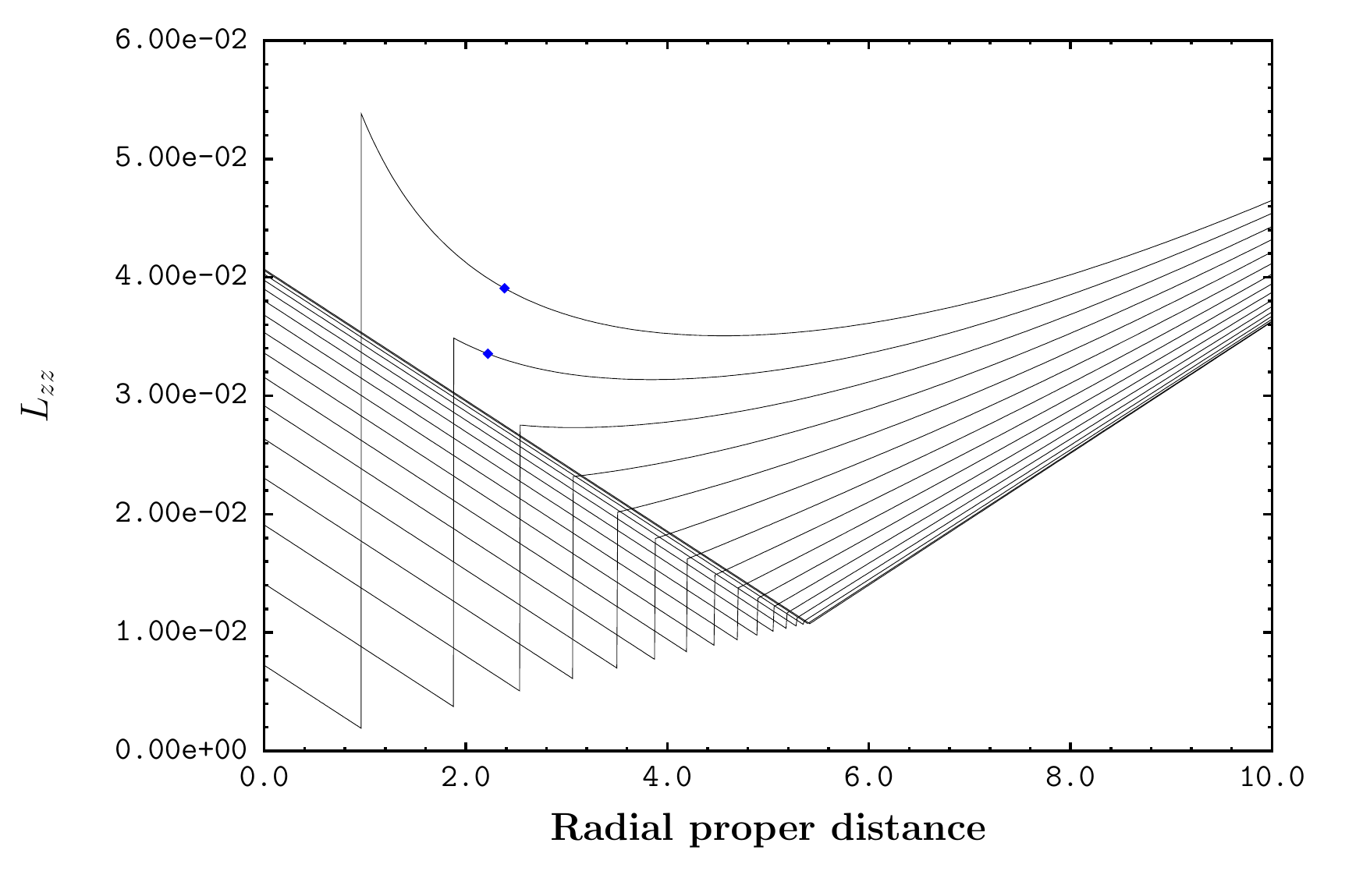}
\caption{\normalfont The $\Lxx$ and $\Lzz$ leg lengths plotted from $t=0$ to $t=12$ in steps of
0.8. The small dots denote the lattice node points. The larger diamonds denote the location of the
apparent horizon. This occurs late in the evolution and appears only on the last two curves. The
inward motion of the junction is also clearly evident in this plot.}
\label{Geo:LxxLzz}
\end{figure}

\clearpage

\begin{figure}[t]
\FigOne{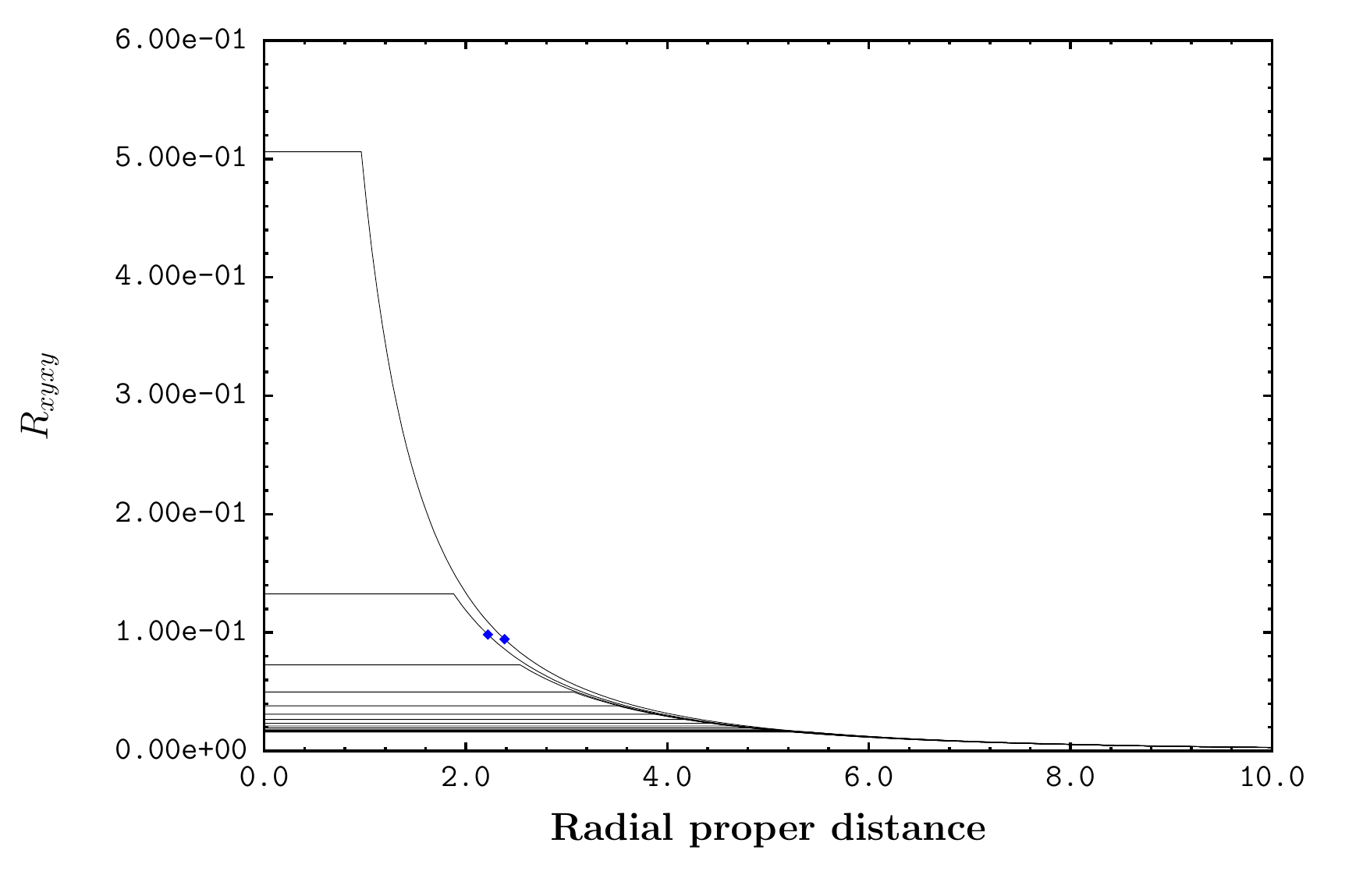}
\FigOne{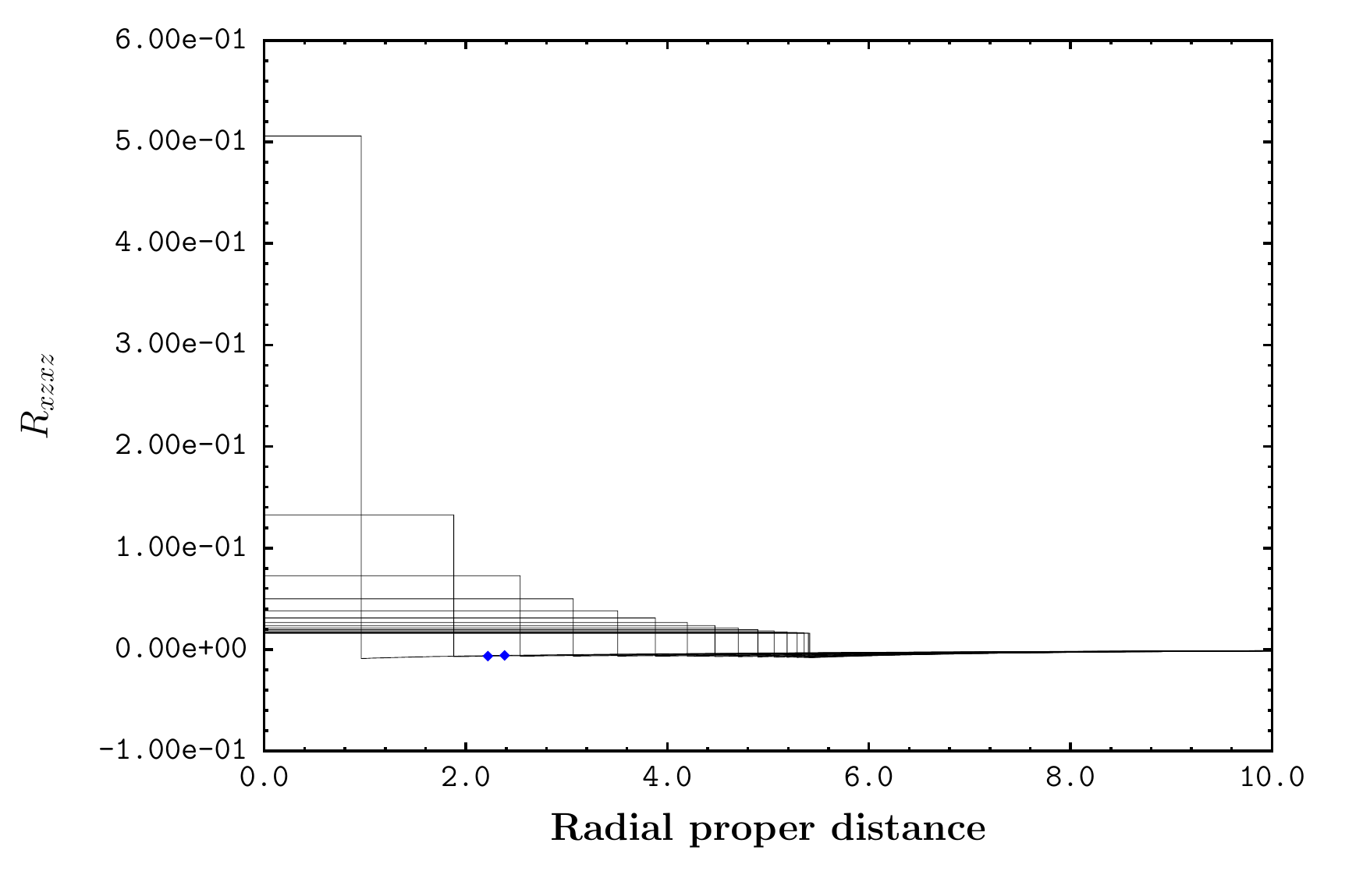}
\caption{\normalfont The Riemann curvatures, $\Rx$ top and $\Rz$ bottom. Notice the flat profiles
inside the dust ball. This feature can be seen in many of the following figures.}
\label{Geo:RxRz}
\end{figure}

\clearpage

\begin{figure}[t]
\FigOne{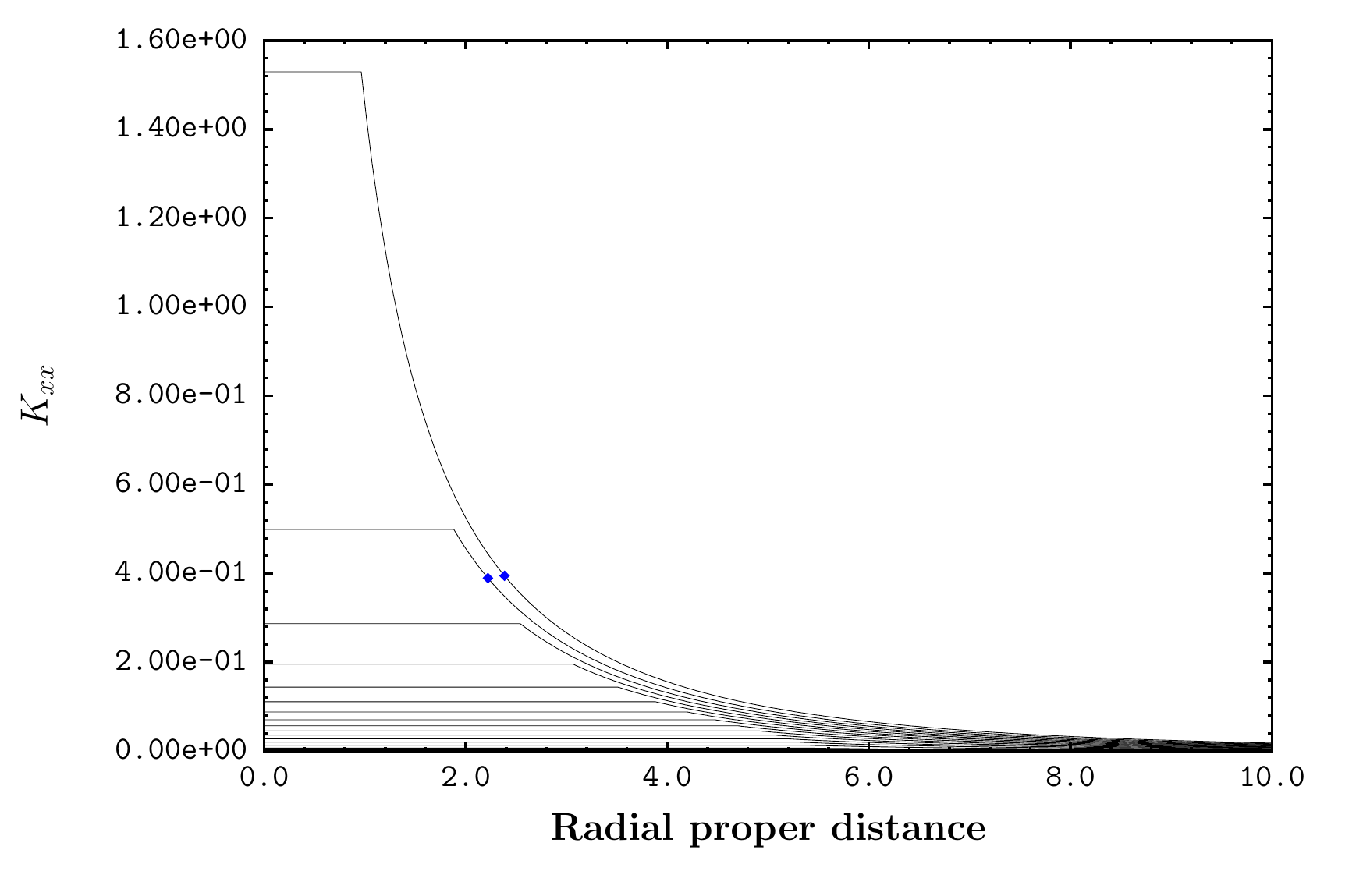}
\FigOne{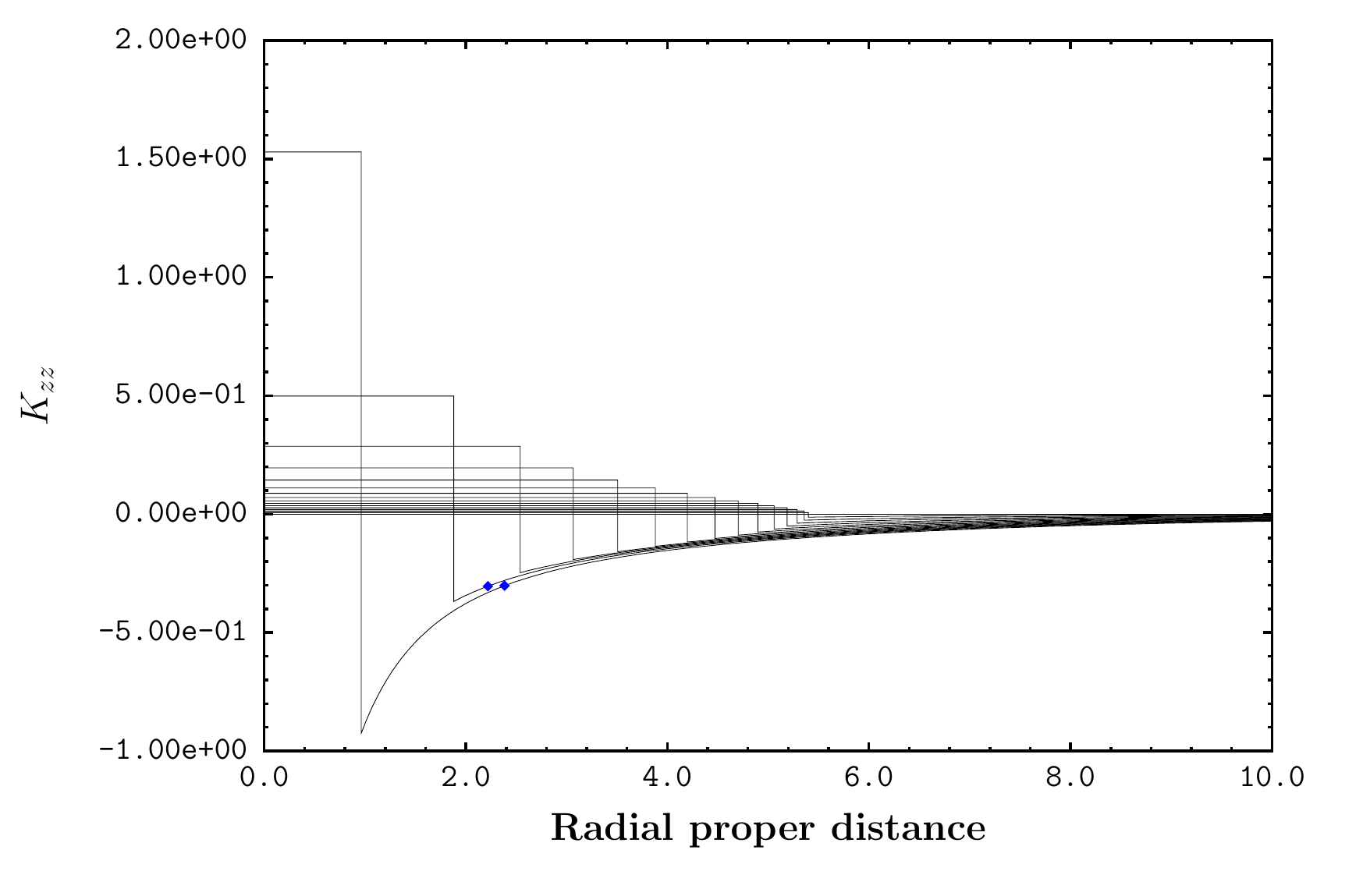}
\caption{\normalfont The extrinsic curvatures, $\Kxx$ top and $\Kzz$ bottom.}
\label{Geo:KxKz}
\end{figure}

\clearpage

\begin{figure}[t]
\FigOne{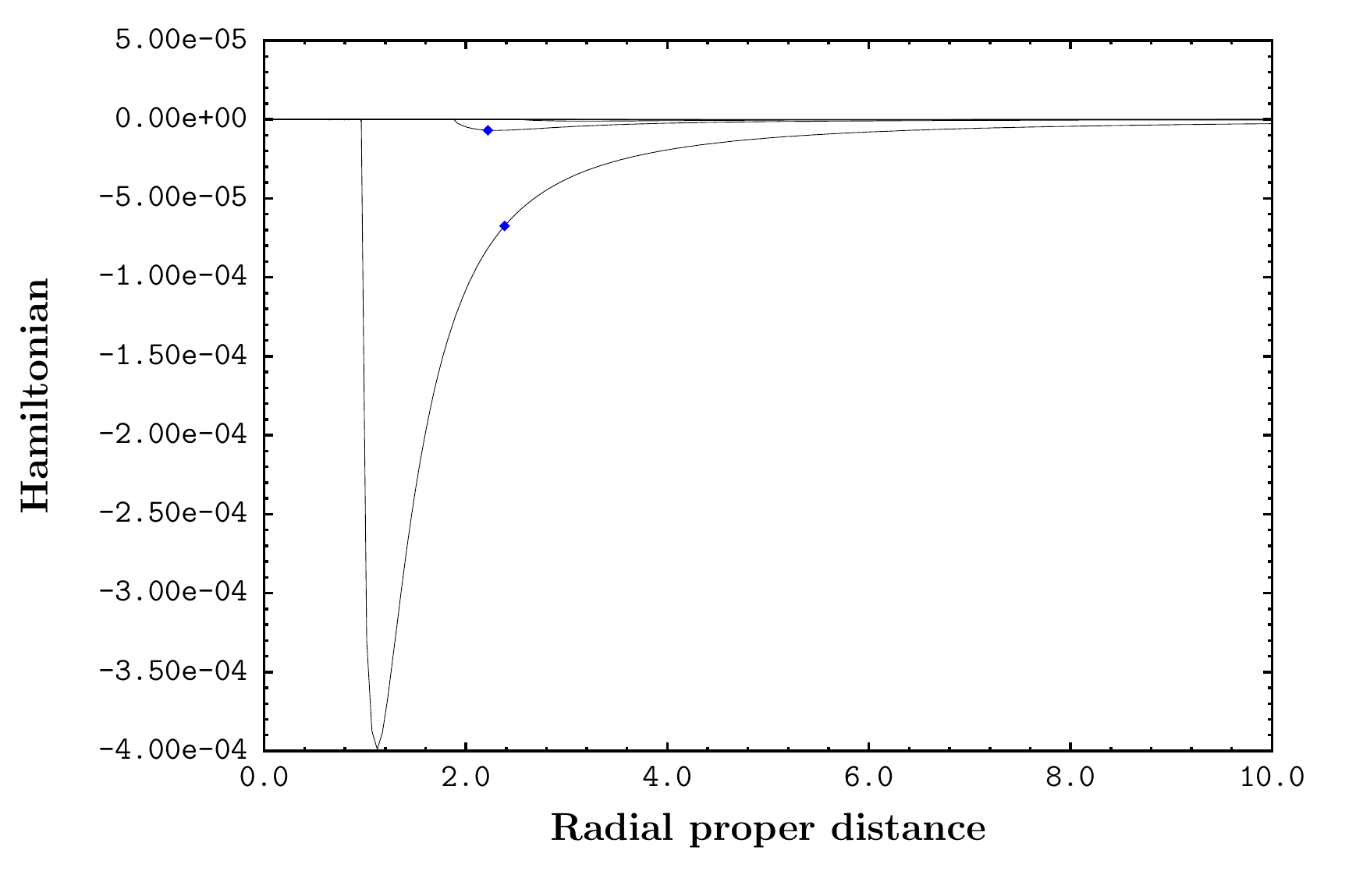}
\FigOne{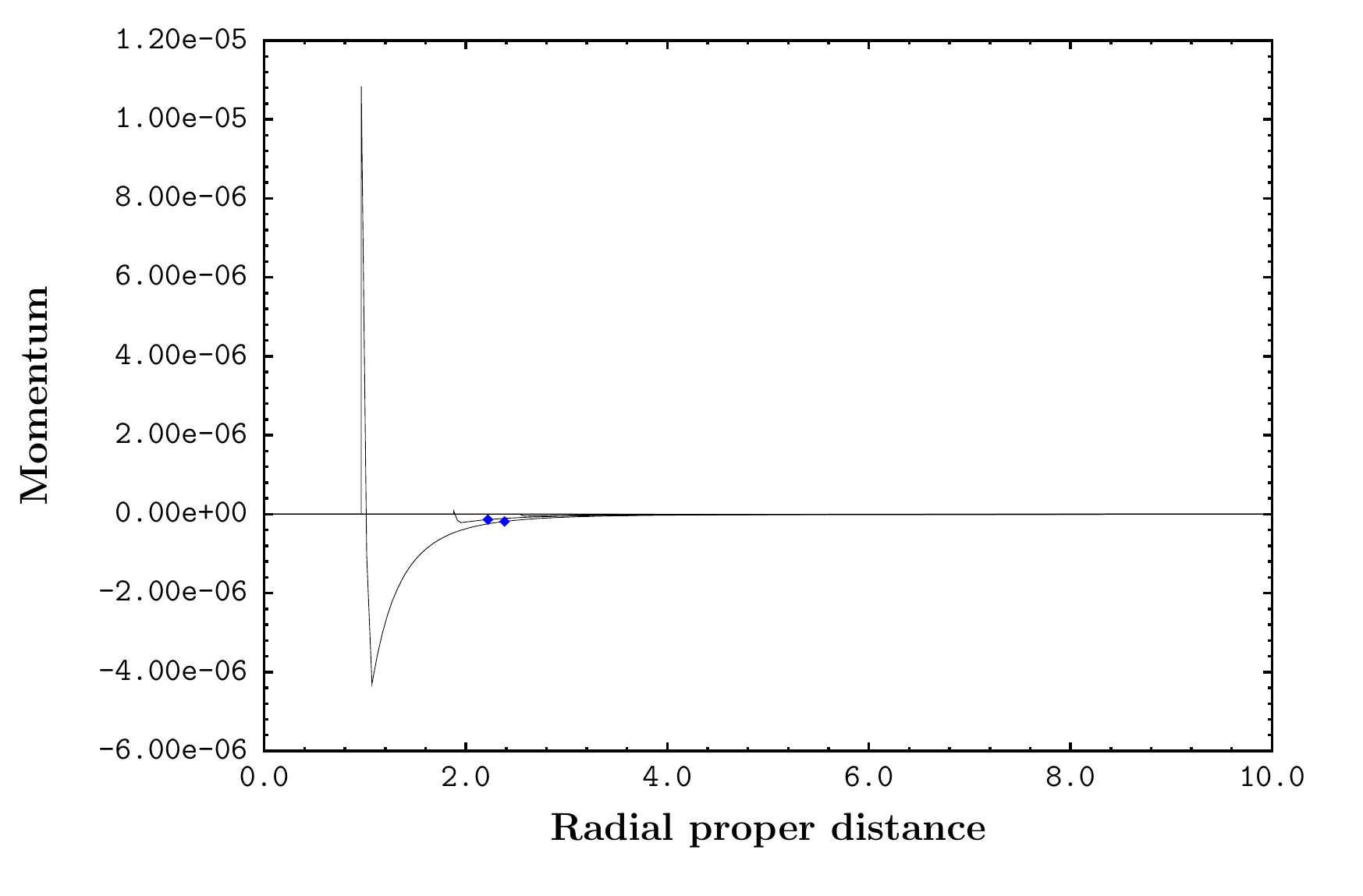}
\caption{\normalfont The constraints. These grow rapidly as the singularity is approached and this
causes the first 10 curves to be too small to be seen on this scale.}
\label{Geo:HamMom}
\end{figure}

\clearpage

% === Maximal slicing =========================================================

\begin{figure}[t]
\FigOne{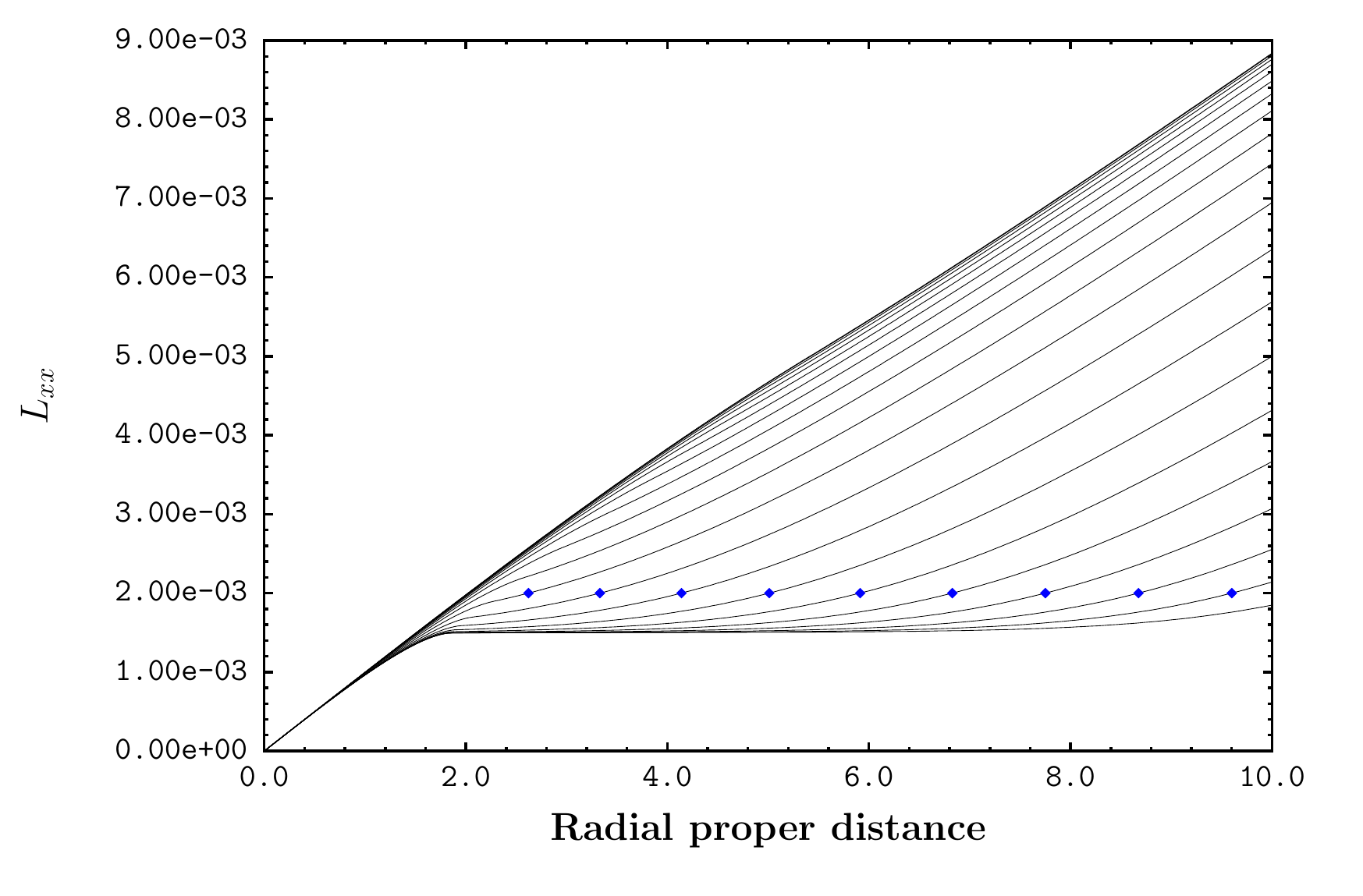}
\FigOne{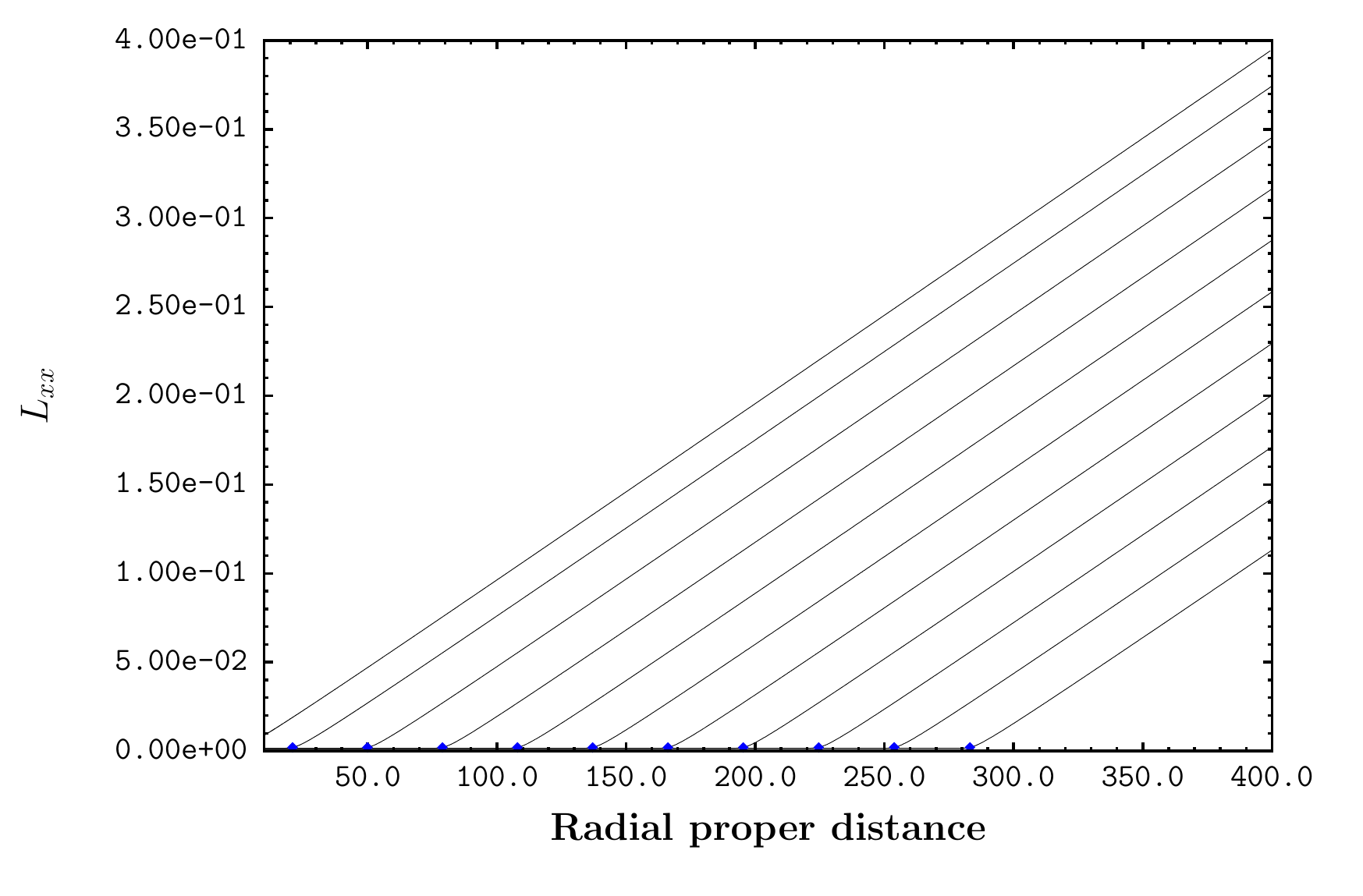}
\caption{\normalfont The $\Lxx$ leg lengths for $t=0$ to $t=32$ in 20 steps (top, with $0<z<10$)
and $t=0$ to $t=500$ in 10 steps (bottom, with $10<z<400$). This time the motion of the apparent
horizon is much more noticeable than for the case of geodesic slicing.}
\label{Max:Lxx}
\end{figure}

\clearpage

\begin{figure}[t]
\FigOne{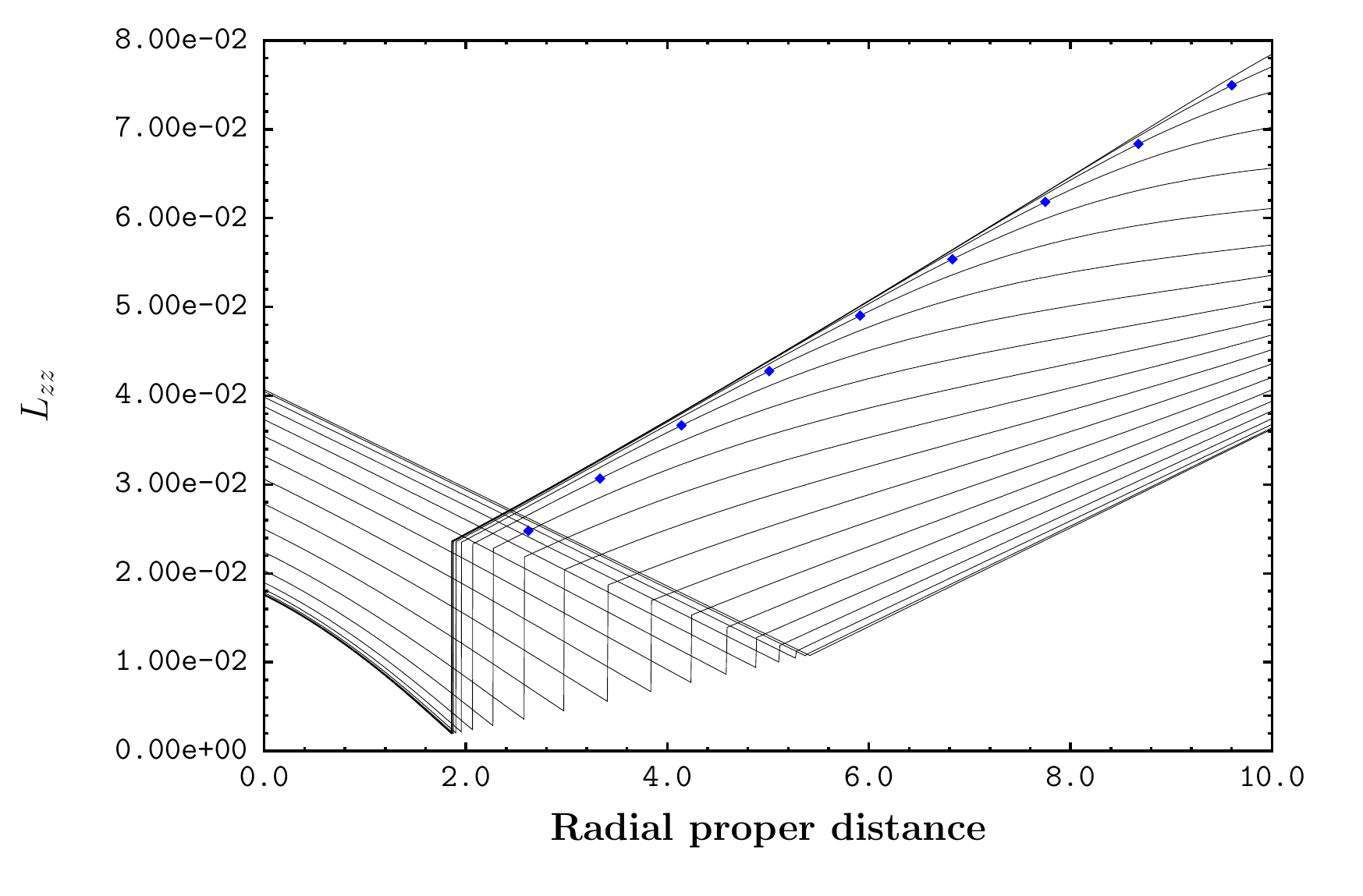}
\FigOne{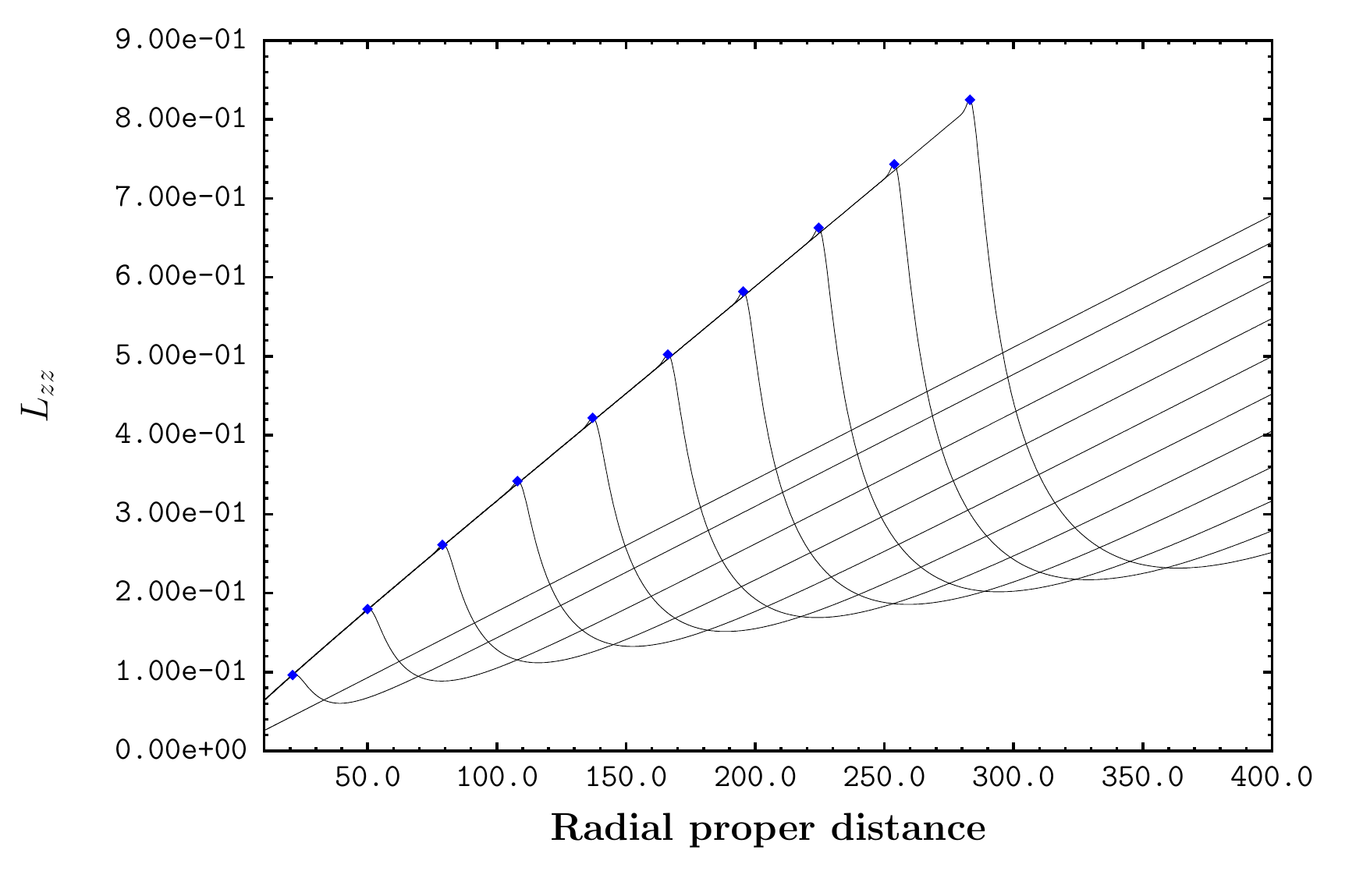}
\caption{\normalfont The $\Lzz$ leg lengths for $t=0$ to $t=32$ in 20 steps (top, with $0<z<10$)
and $t=0$ to $t=500$ in 10 steps (bottom, with $10<z<400$). Notice the extreme change in $\Lzz$ at
the junction. The curves bunch together late in the evolution due to the exponential collapse of
the lapse (see Figure \ref{Max:N}).}
\label{Max:Lzz}
\end{figure}

\clearpage

\begin{figure}[t]
\FigOne{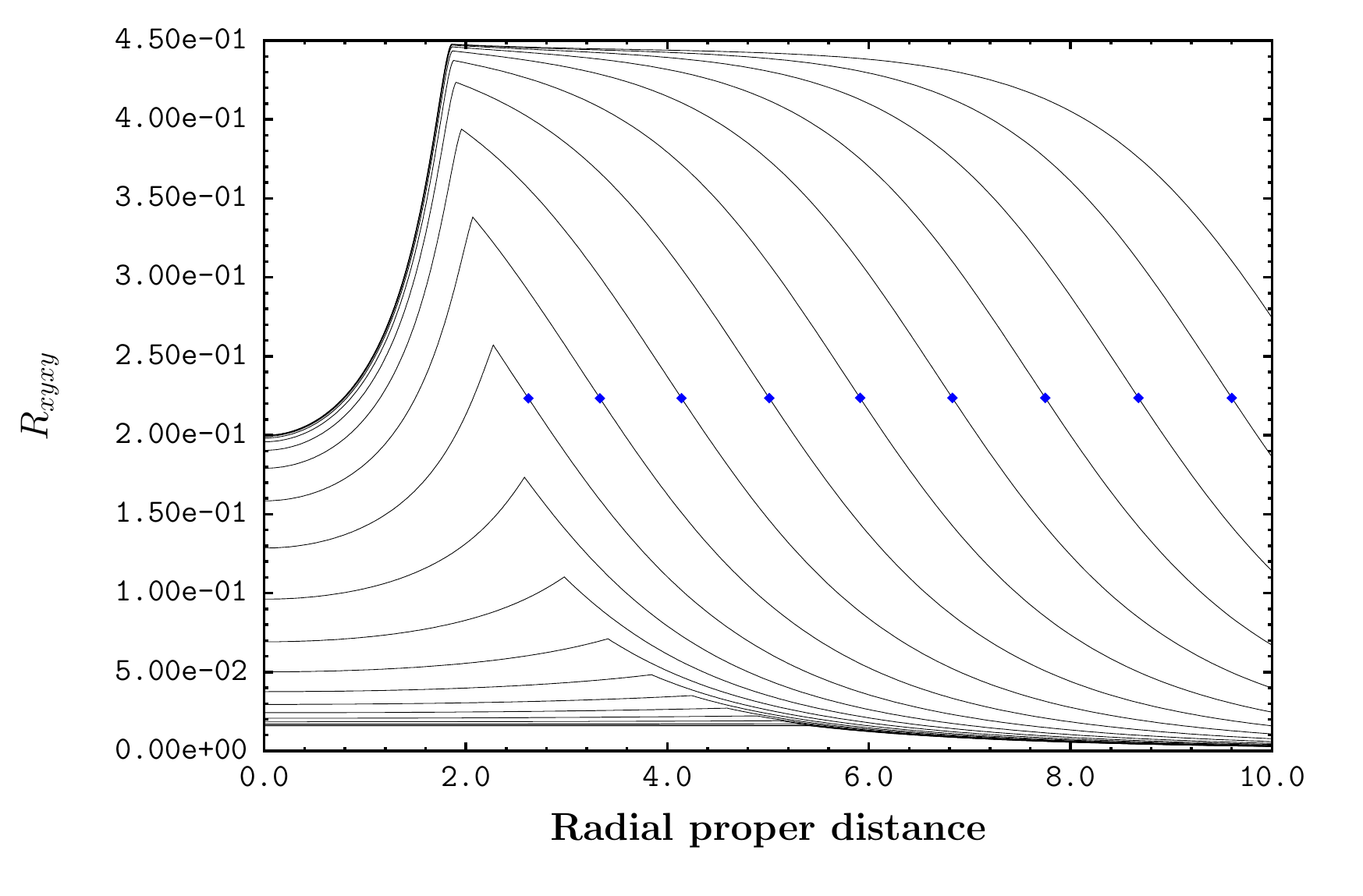}
\FigOne{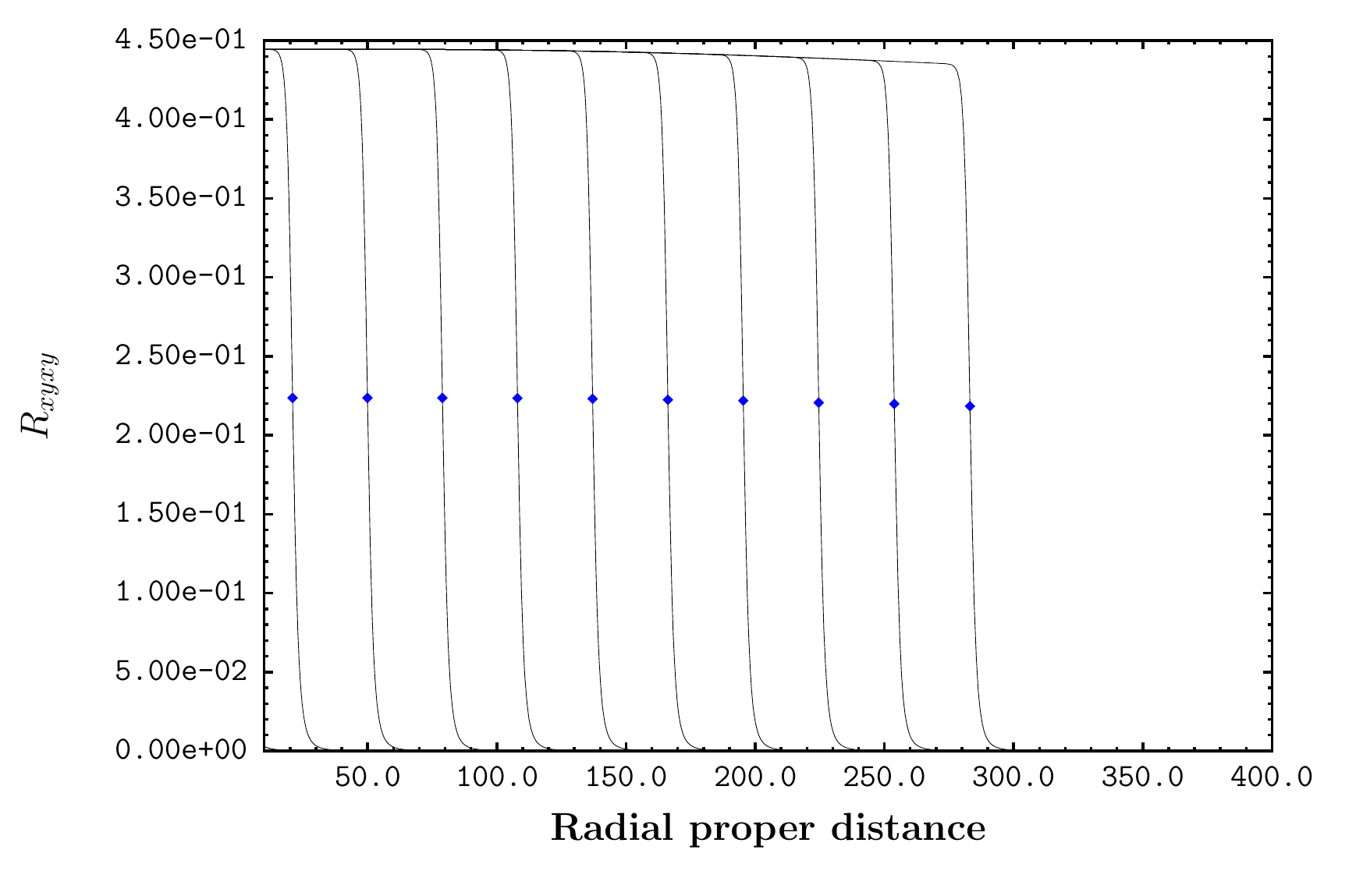}
\caption{\normalfont The Riemann curvature $\Rx$.}
\label{Max:Rx}
\end{figure}

\clearpage

\begin{figure}[t]
\FigOne{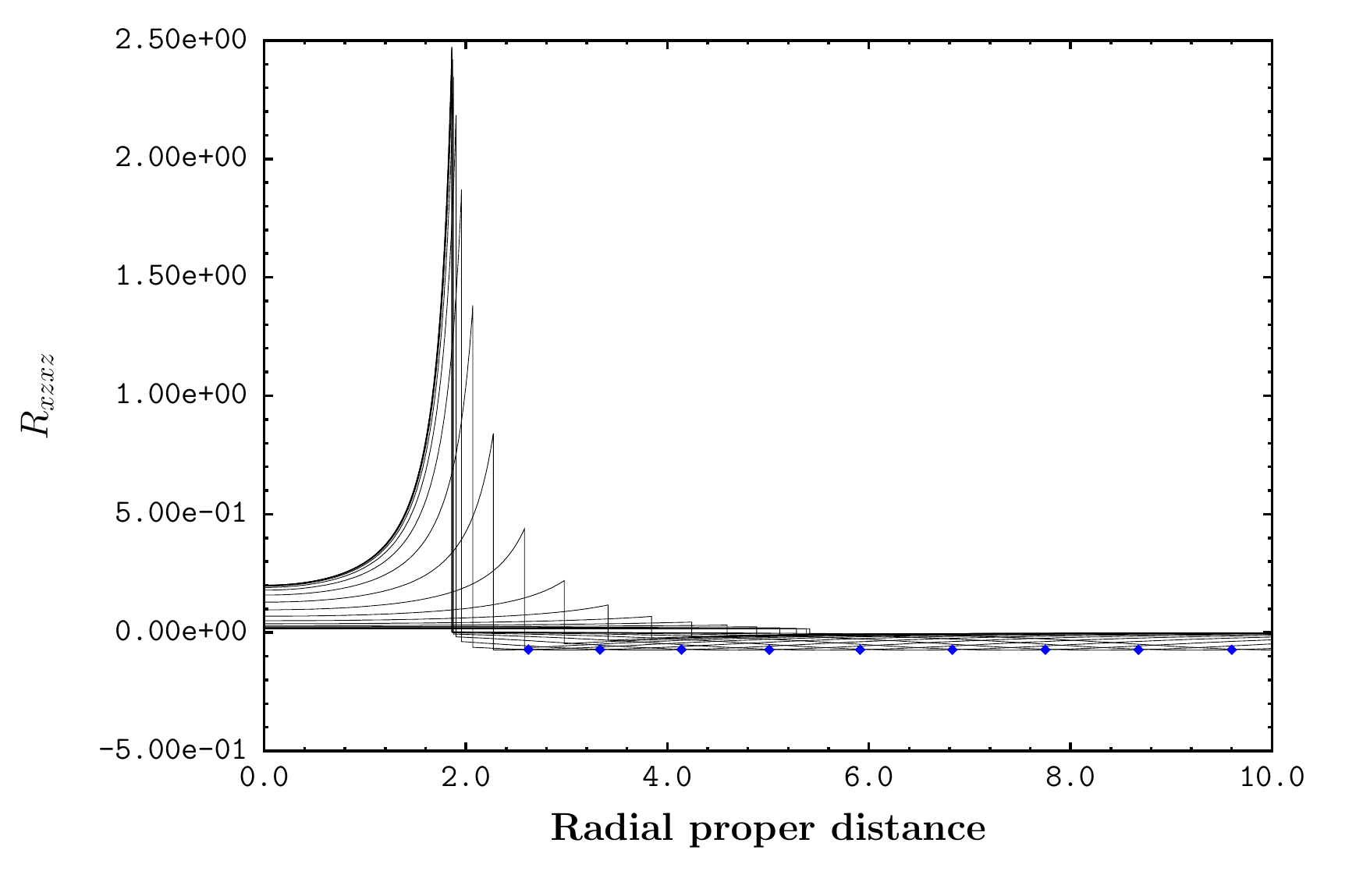}
\FigOne{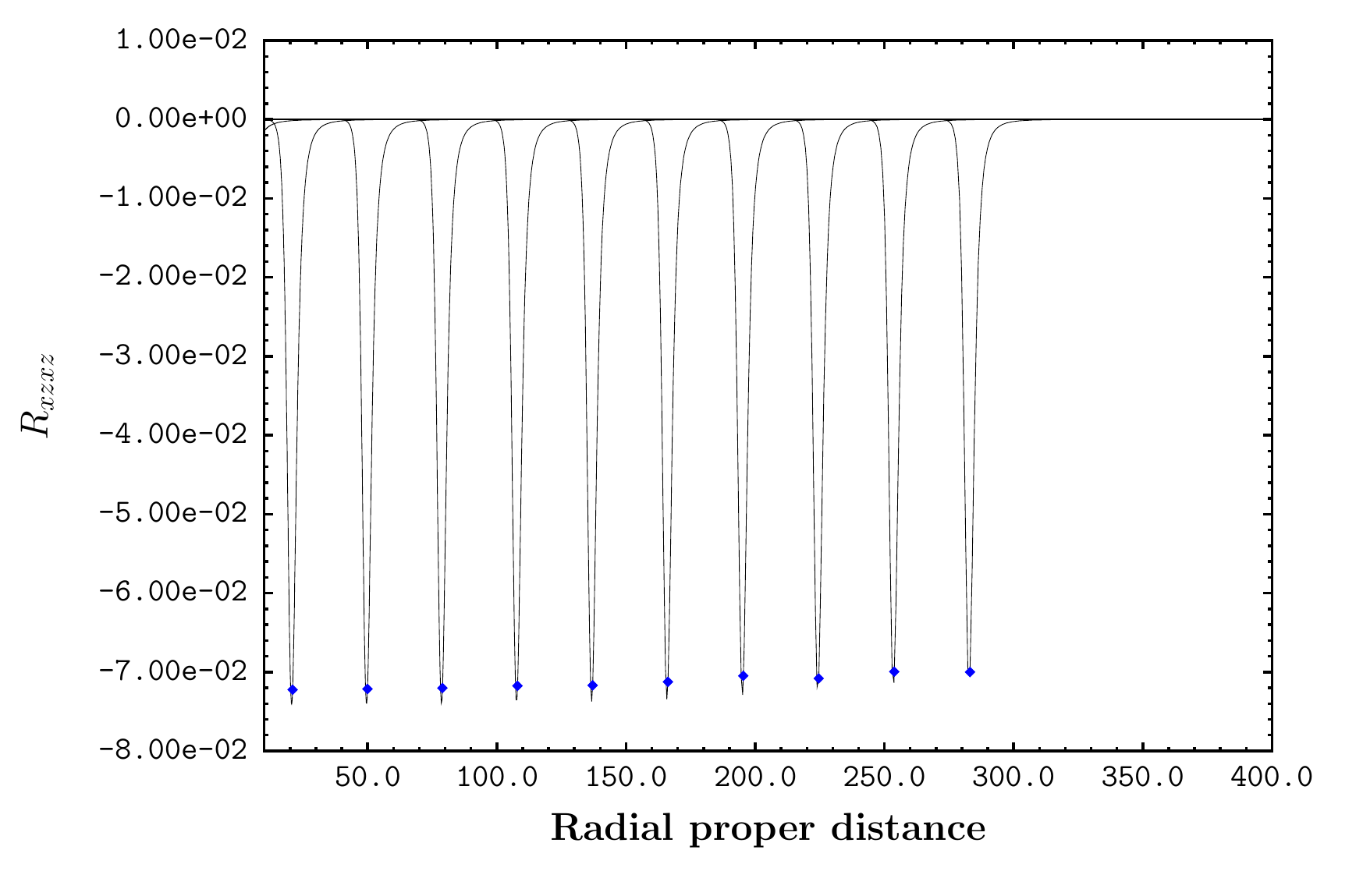}
\caption{\normalfont The Riemann curvature $\Rz$.}
\label{Max:Rz}
\end{figure}

\clearpage

\begin{figure}[t]
\FigOne{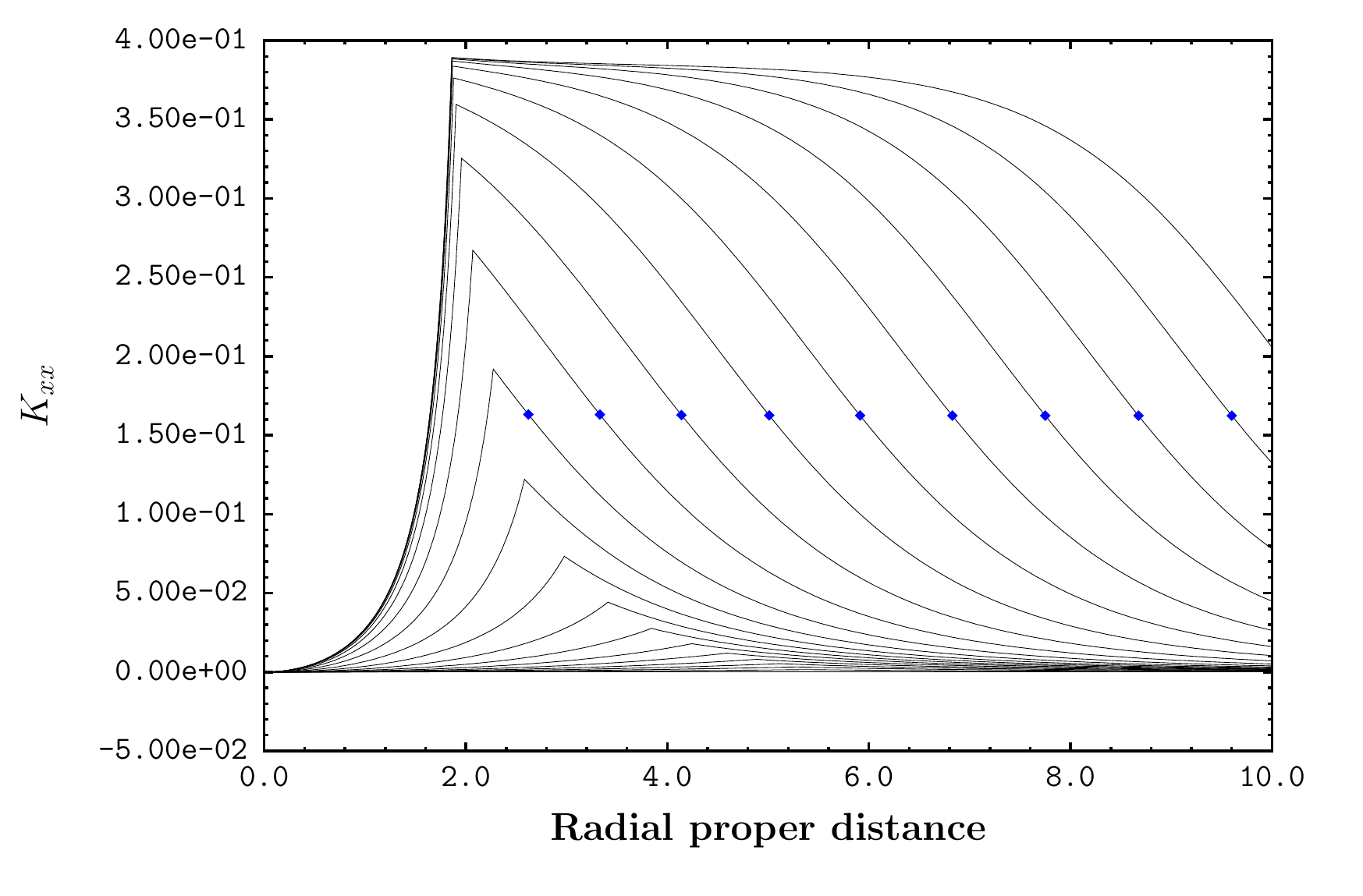}
\FigOne{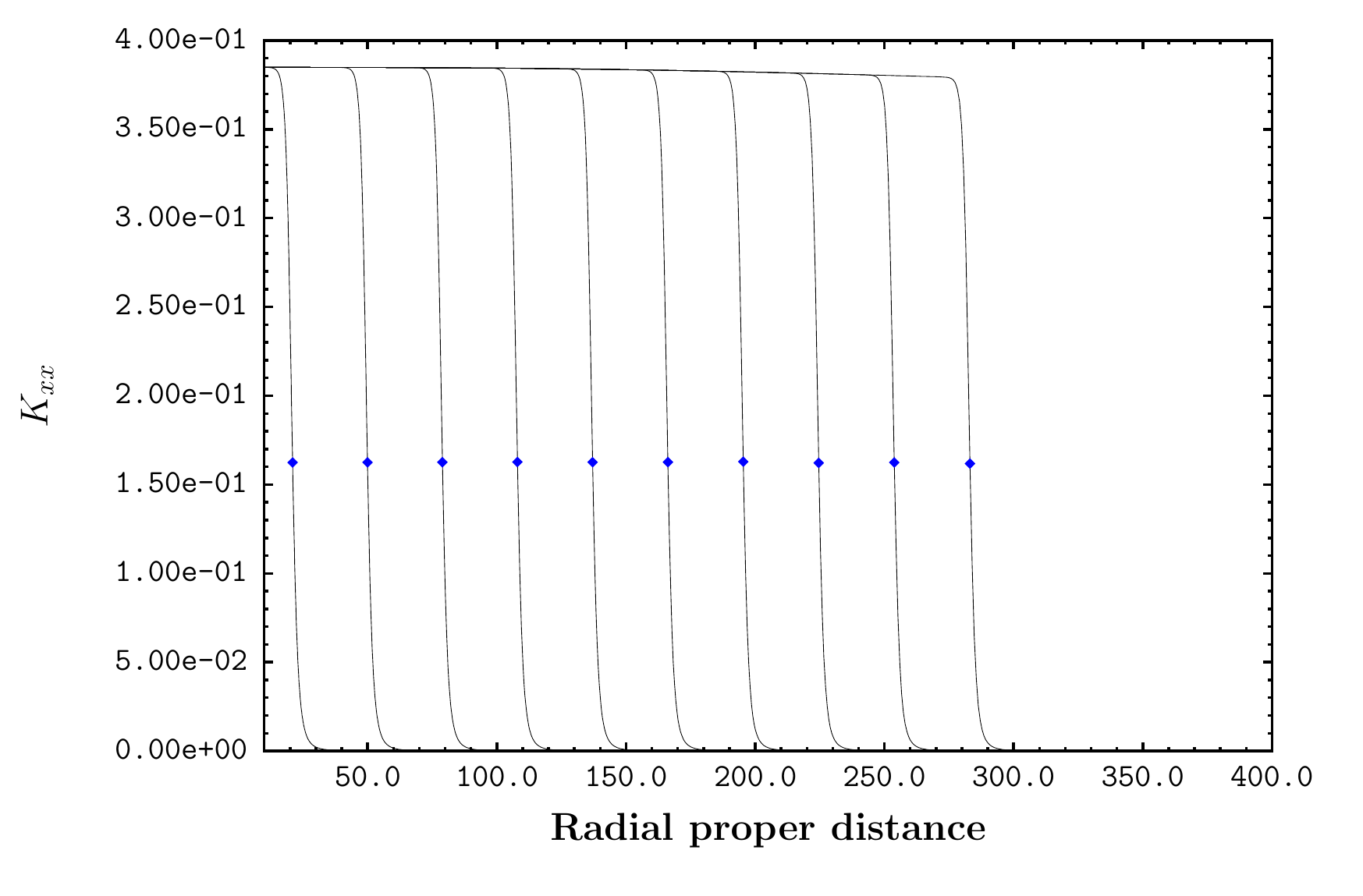}
\caption{\normalfont The extrinsic curvature $\Kxx$.}
\label{Max:Kx}
\end{figure}

\clearpage

\begin{figure}[t]
\FigOne{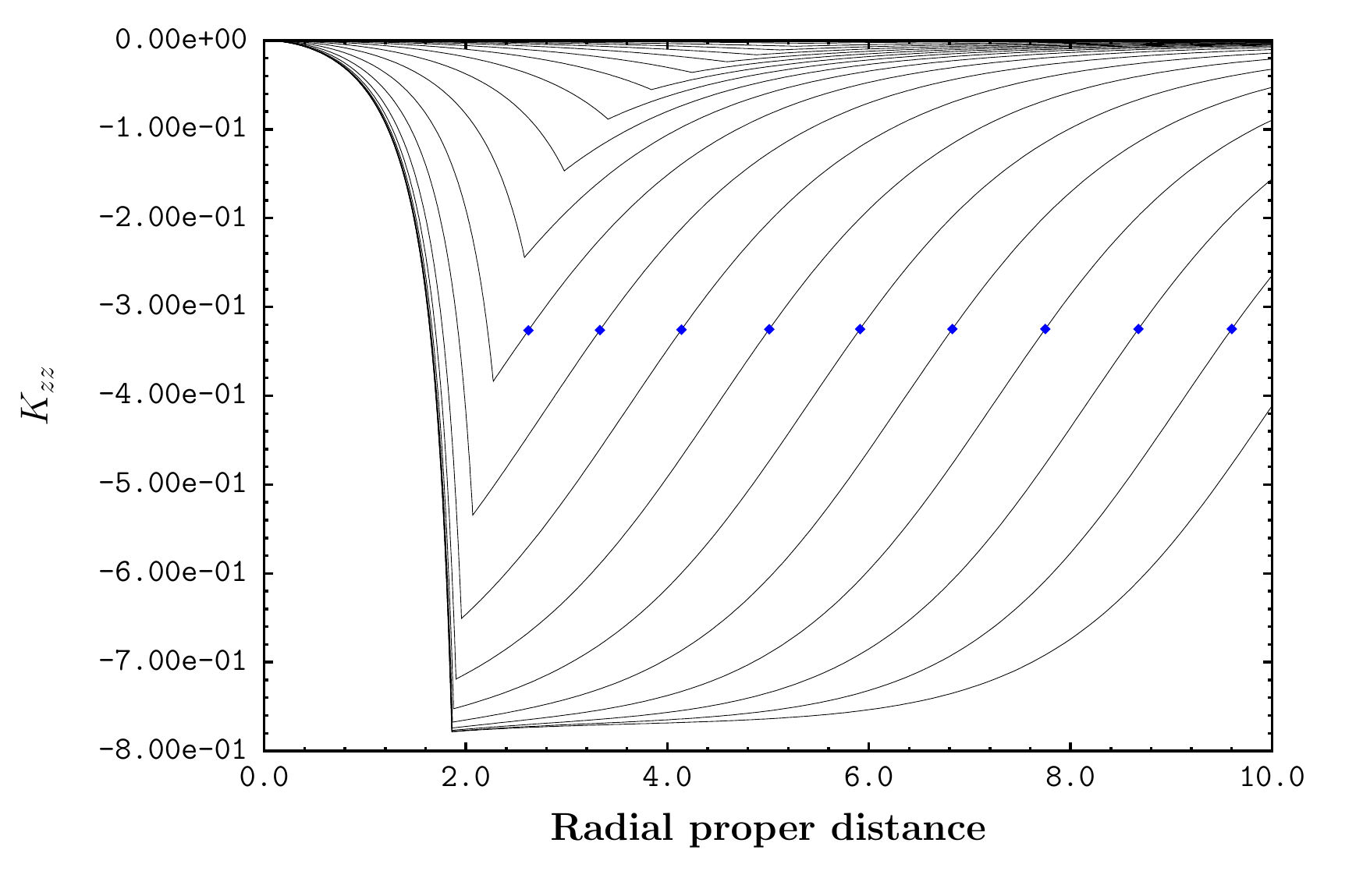}
\FigOne{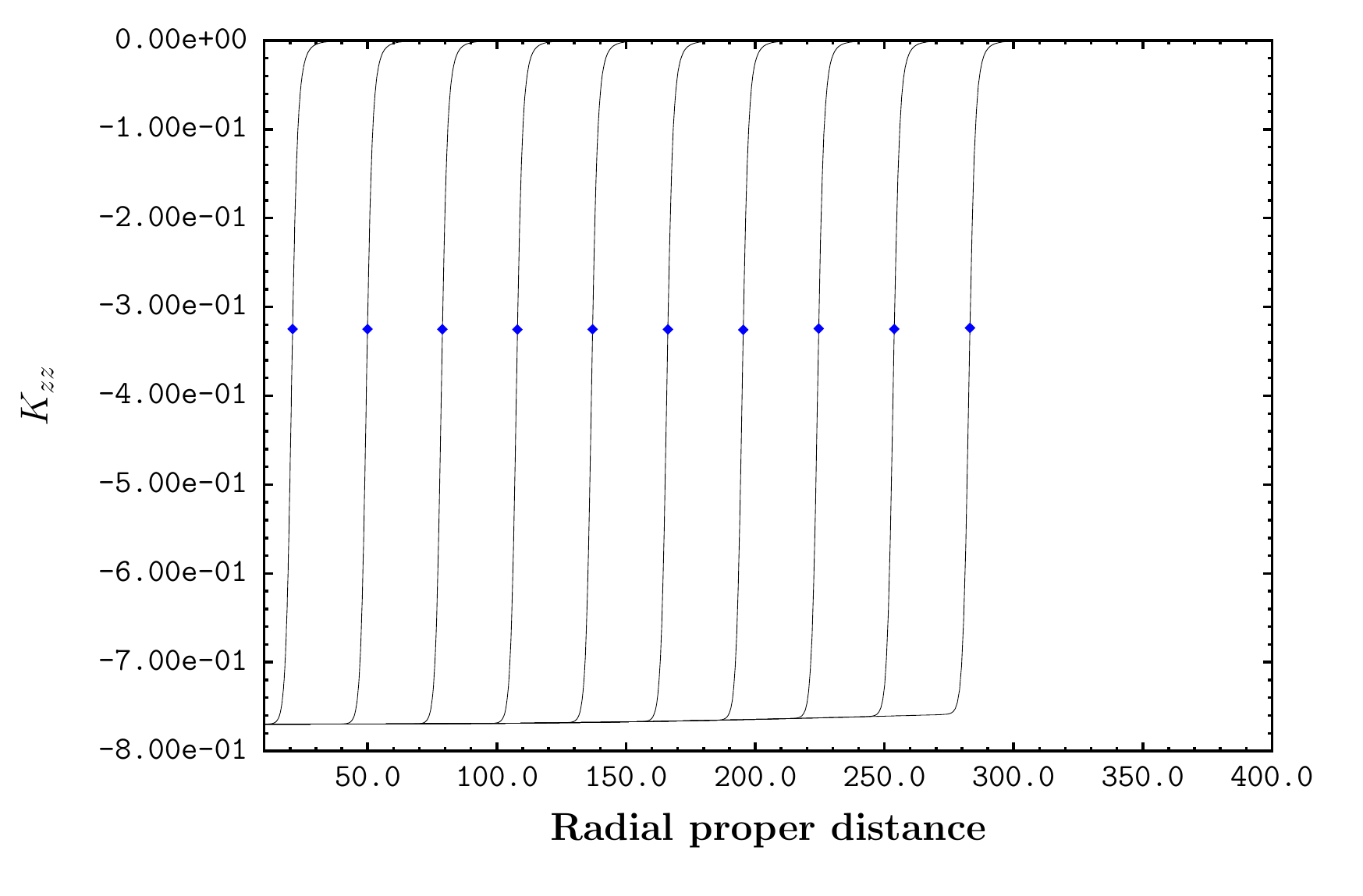}
\caption{\normalfont The extrinsic curvature $\Kzz$.}
\label{Max:Kz}
\end{figure}

\clearpage

\begin{figure}[t]
\FigOne{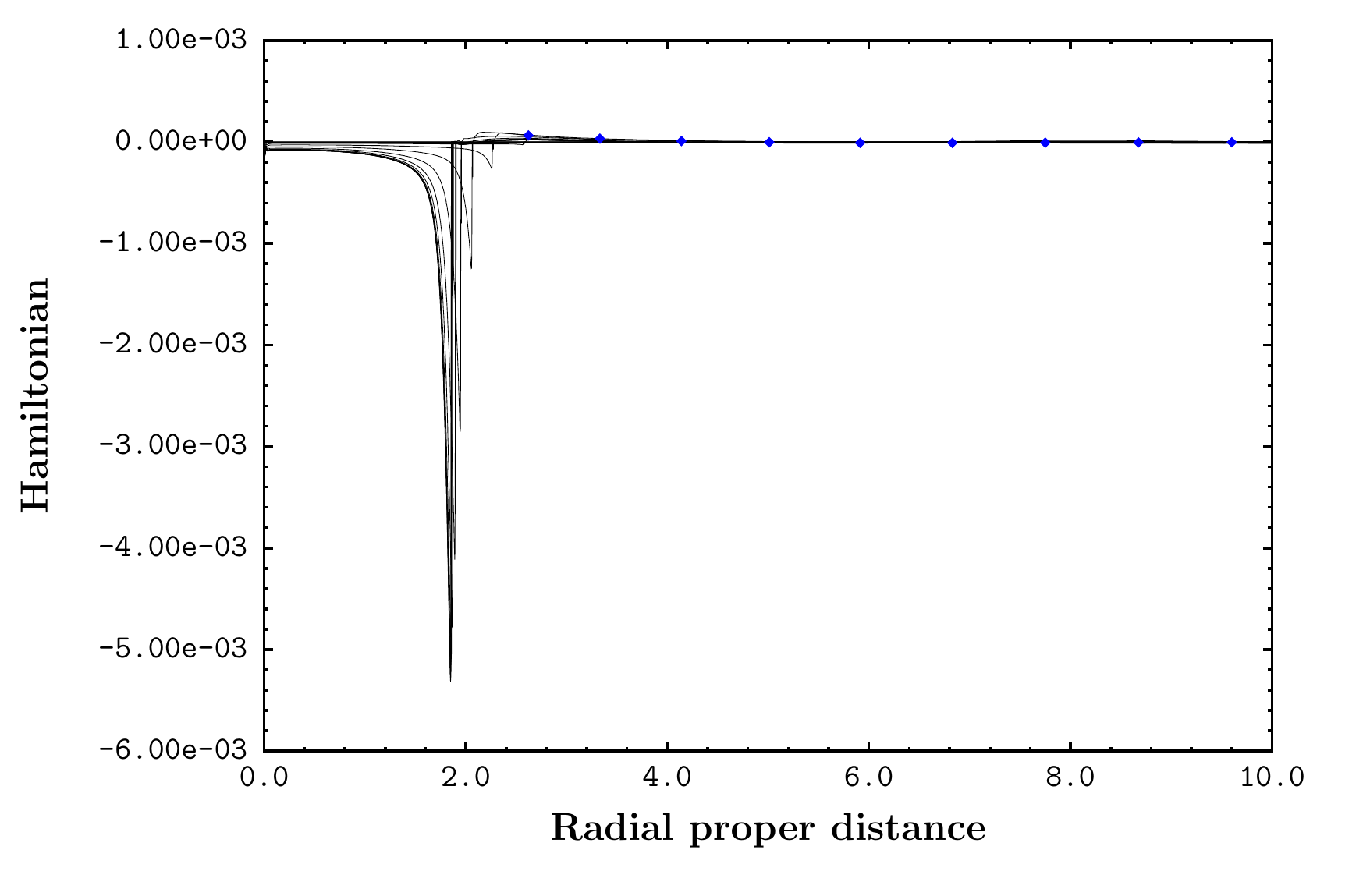}
\FigOne{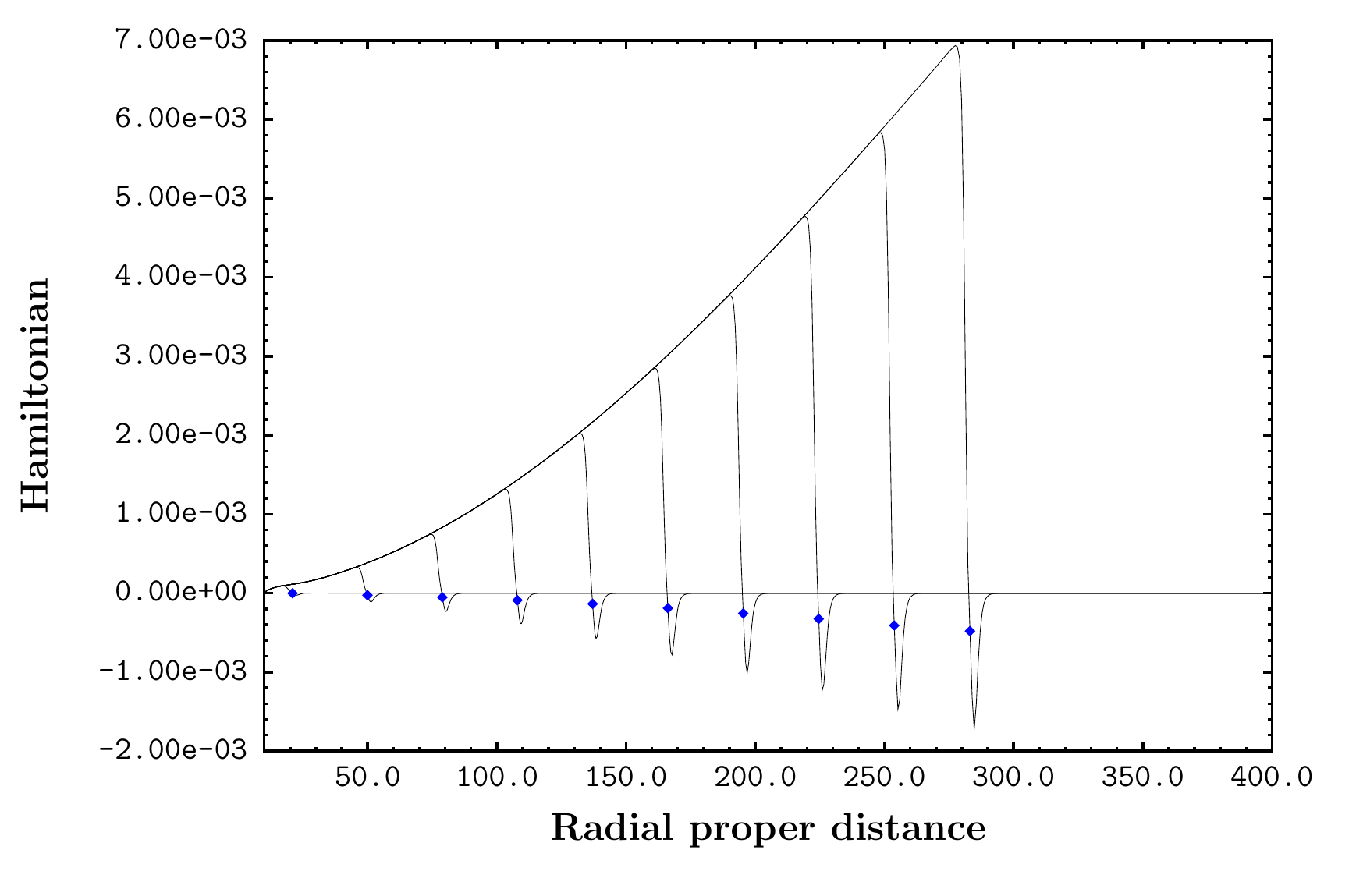}
\caption{\normalfont The Hamiltonian constraint. This shows a slowly growing error for later
times.}
\label{Max:Ham}
\end{figure}

\clearpage

\begin{figure}[t]
\FigOne{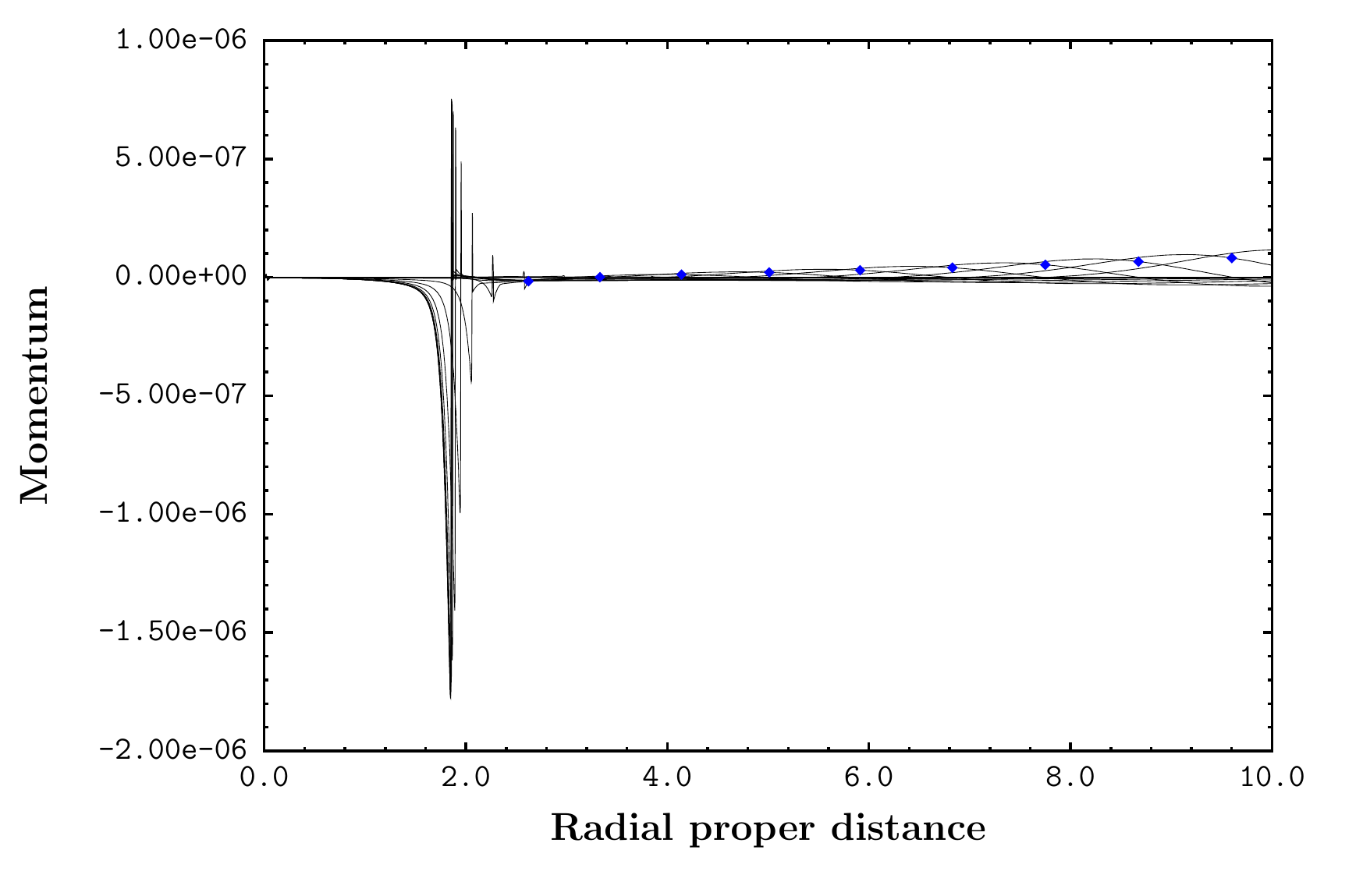}
\FigOne{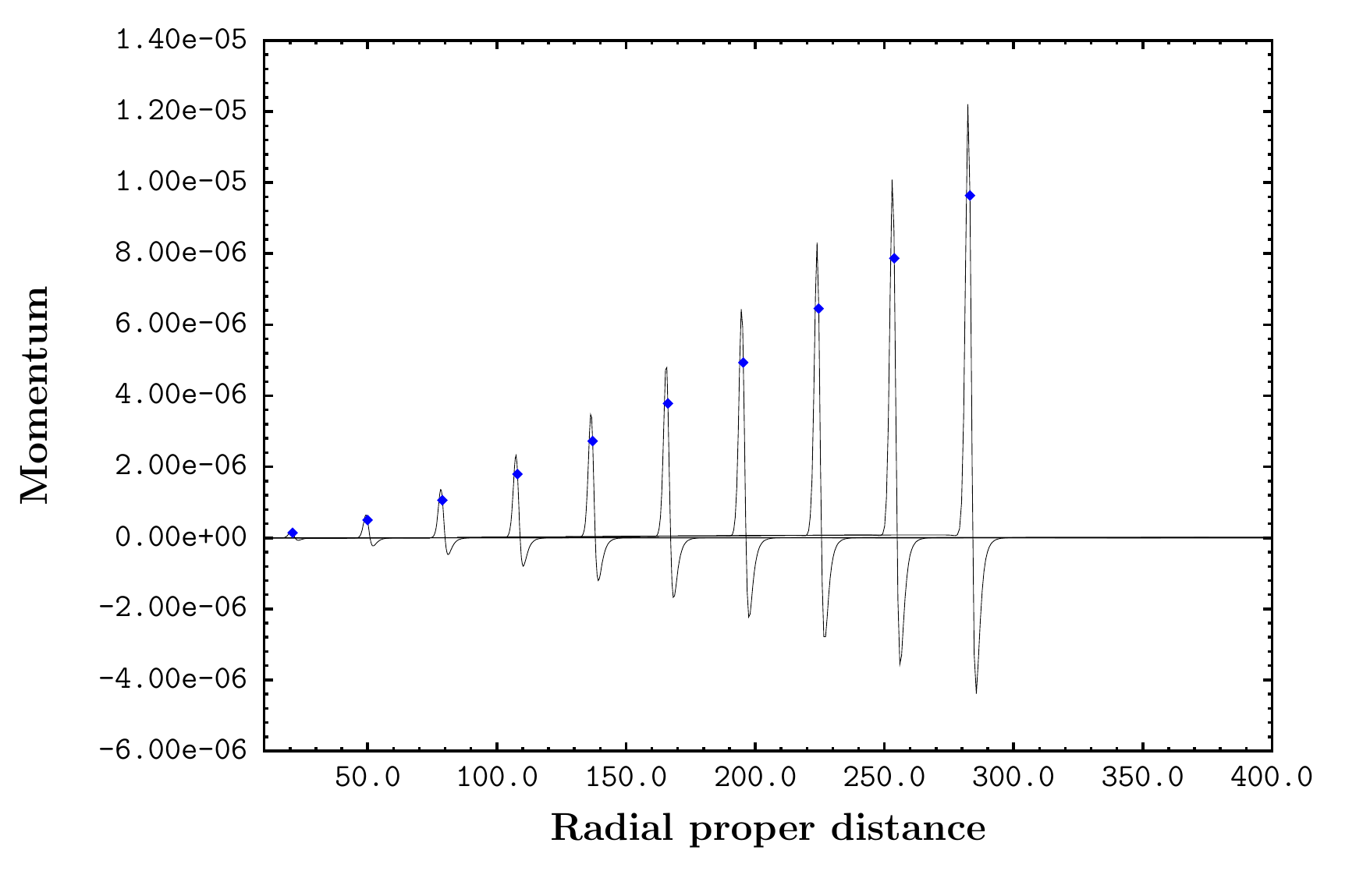}
\caption{\normalfont The momentum constraint. This shows a similar slow growing peak as seen in the
Hamiltonian constraint.}
\label{Max:Mom}
\end{figure}

\clearpage

\begin{figure}[t]
\FigOne{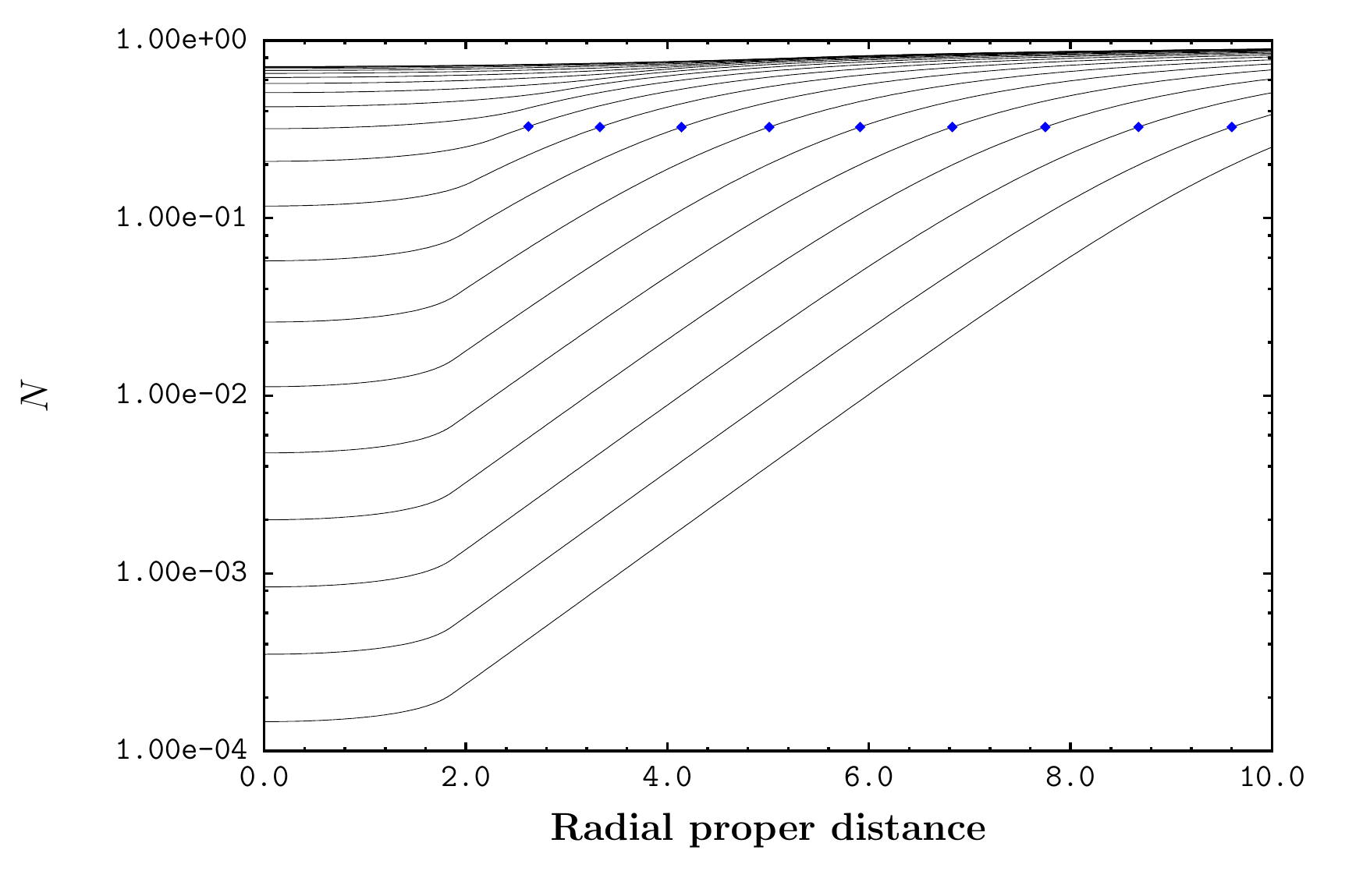}
\FigOne{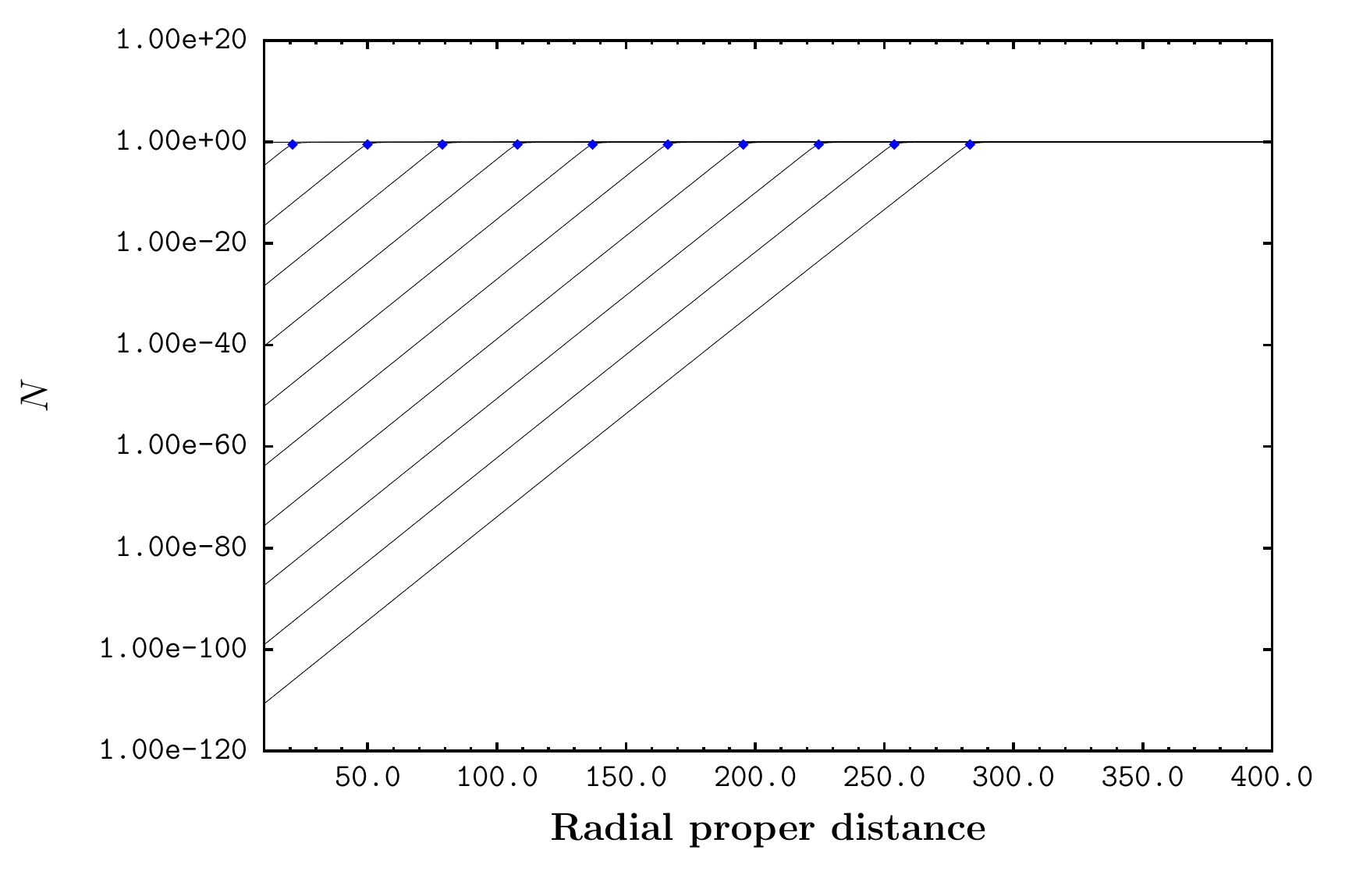}
\caption{\normalfont A logarithmic plot of the lapse. The even gaps between the curves shows
clearly that the collapse is exponential in time. Note the extreme value of the lapse at the origin
for late times, of order $10^{-110}$.}
\label{Max:N}
\end{figure}

\clearpage

\begin{figure}[t]
\FigOne{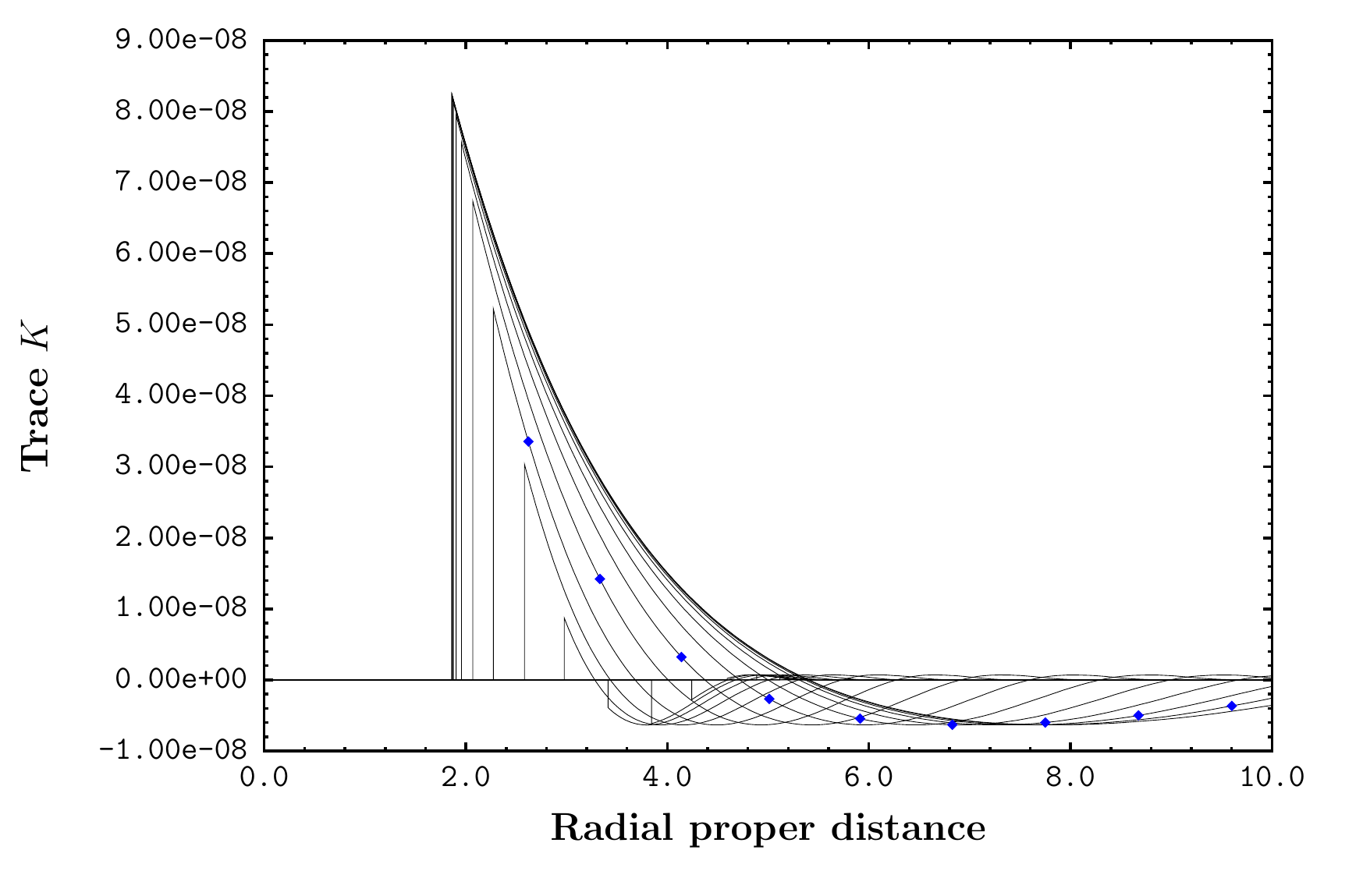}
\FigOne{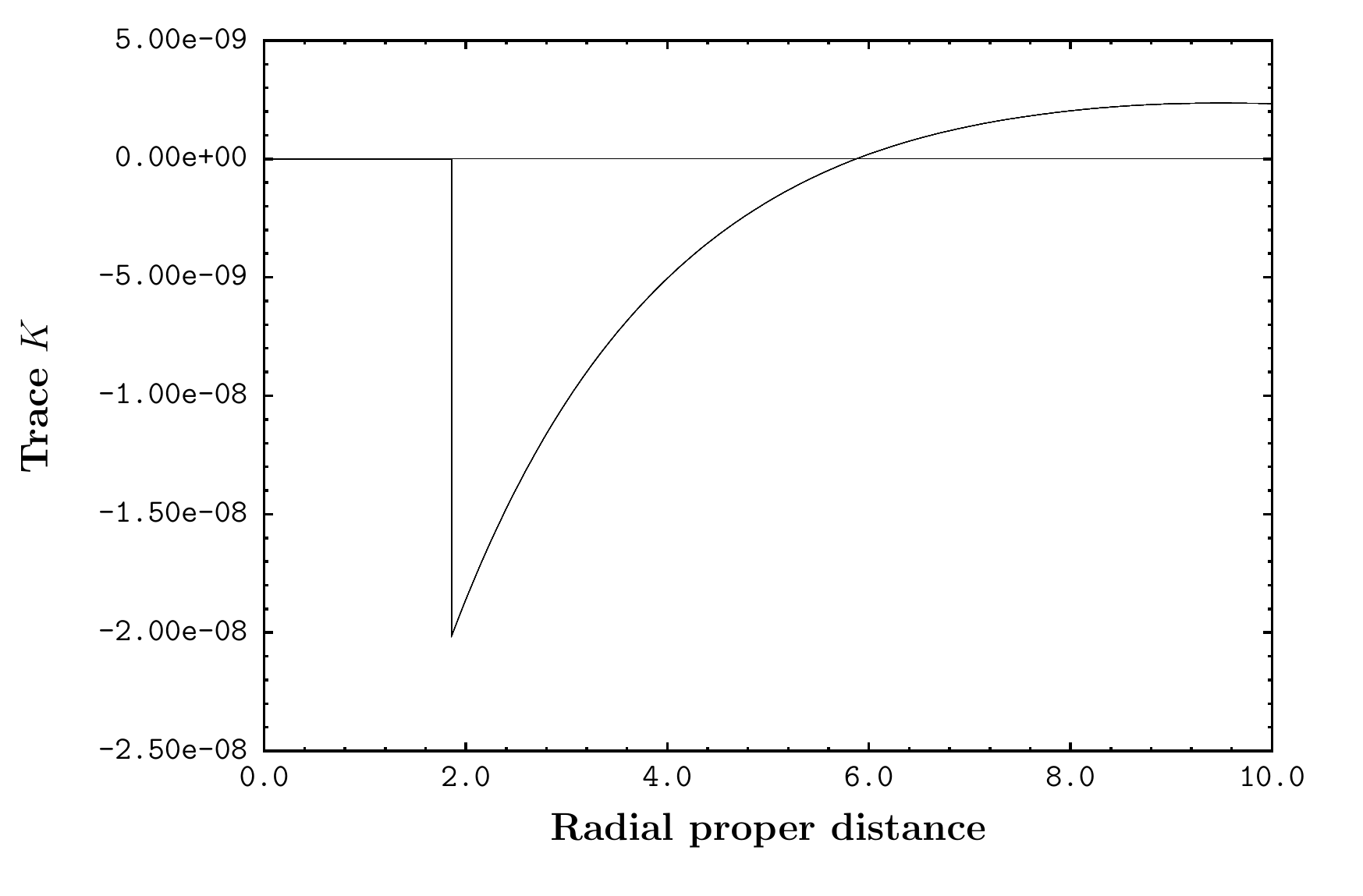}
\caption{\normalfont The trace of the extrinsic curvature $K$. This should be zero for all time. In
the lower plot the lapse has collapsed in that part of the lattice and thus there is no apparent
evolution in $K$. In the upper plot $(\njct,\nvac)=(240,1200)$ while for lower plot we used
$(\njct,\nvac)=(240,2400)$.}
\label{Max:K}
\end{figure}

\clearpage

\begin{figure}[t]
\FigOne{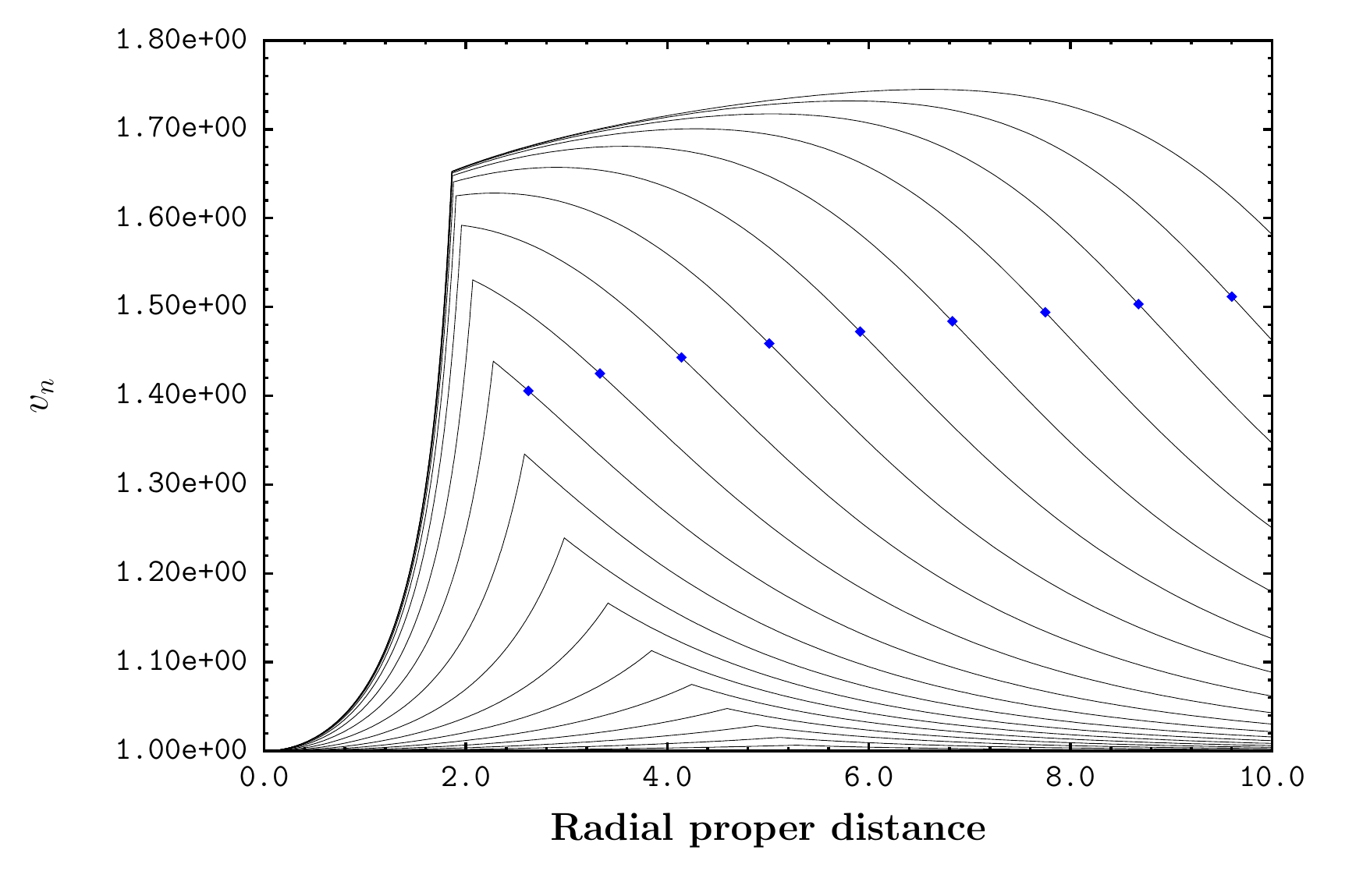}
\FigOne{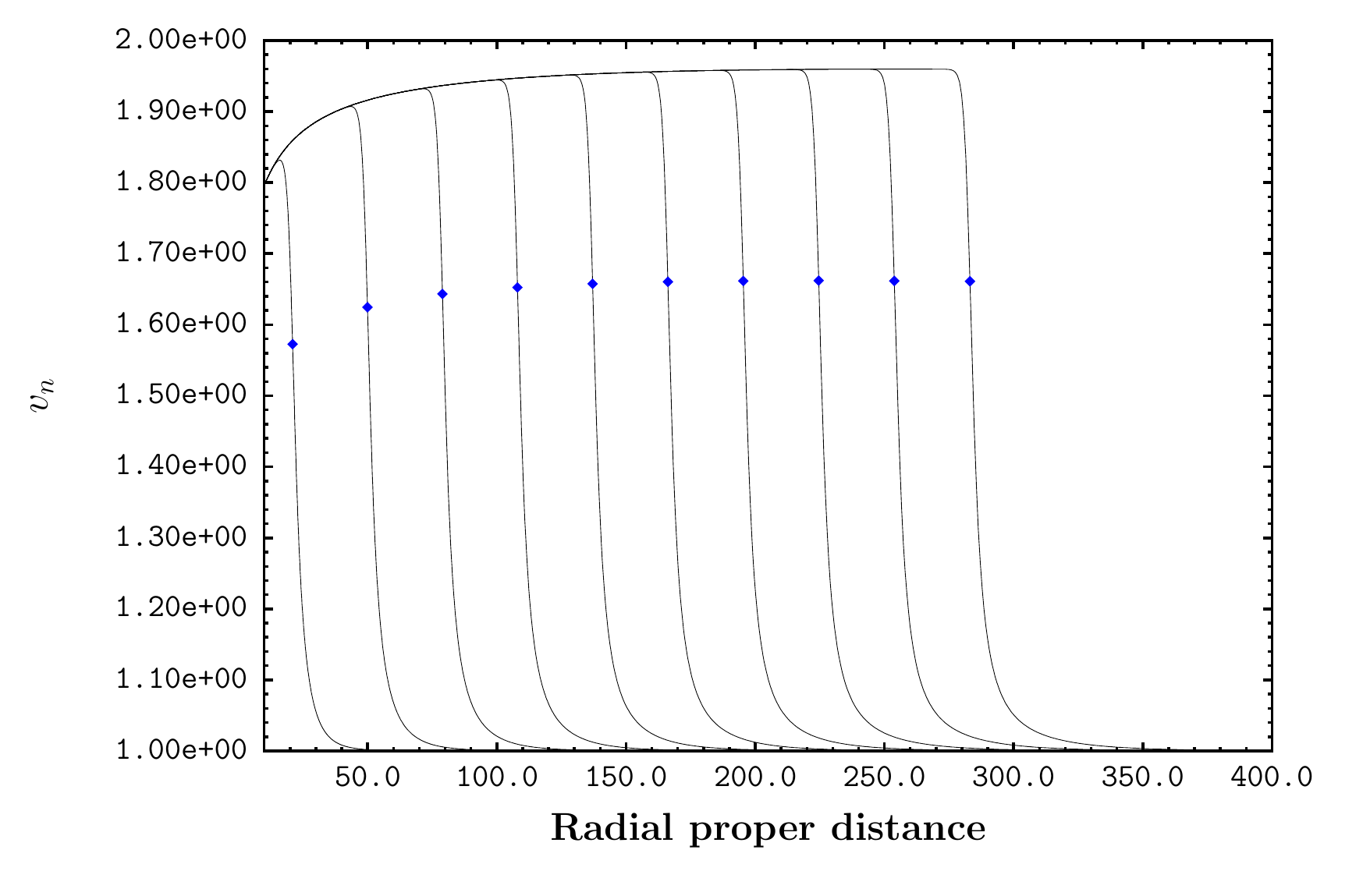}
\caption{\normalfont The particle velocity $\Vn$.}
\label{Max:Vn}
\end{figure}

\clearpage

\begin{figure}[t]
\FigOne{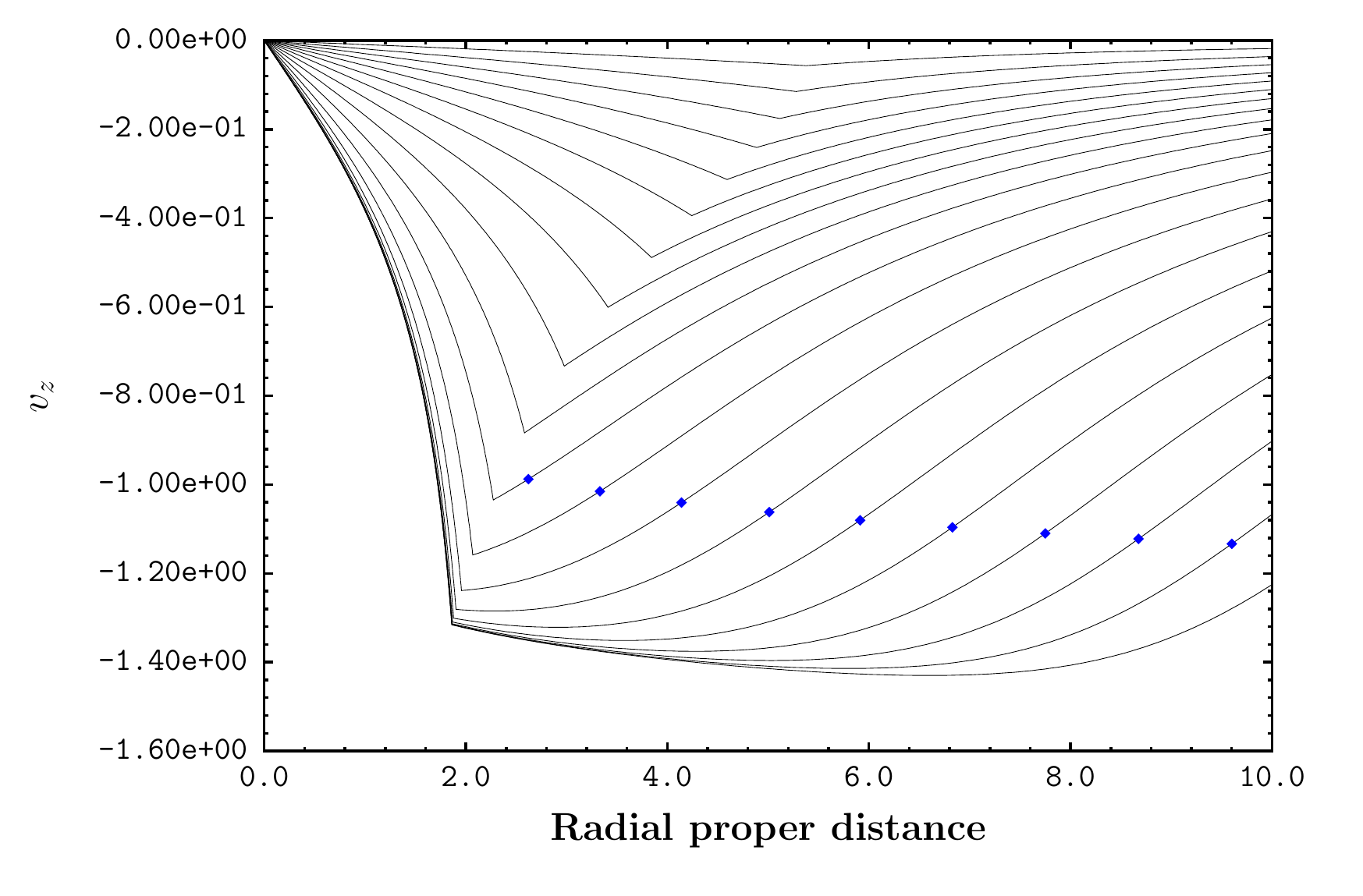}
\FigOne{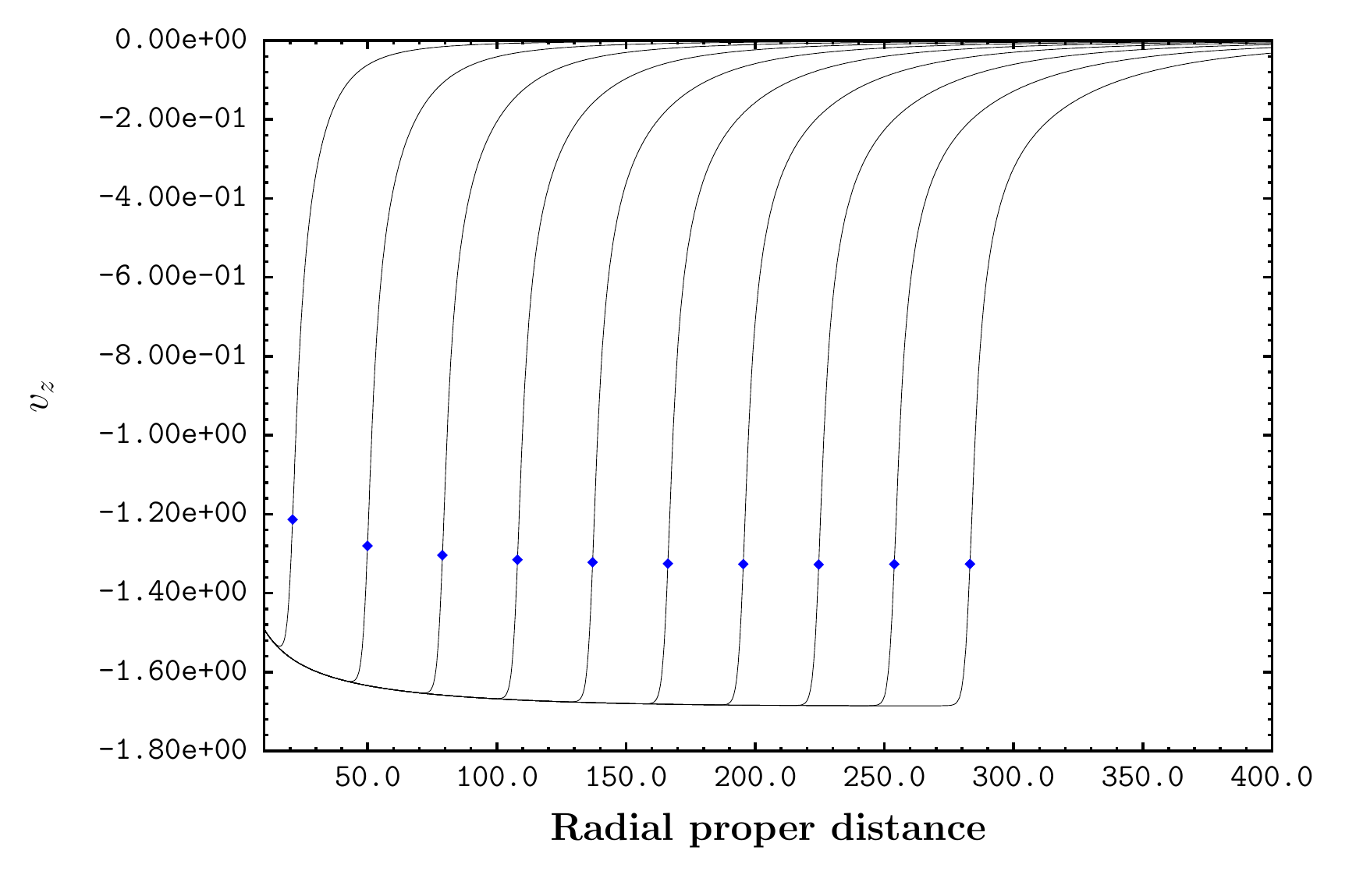}
\caption{\normalfont The particle velocity $\Vz$.}
\label{Max:Vz}
\end{figure}

\clearpage

\begin{figure}[t]
\FigOne{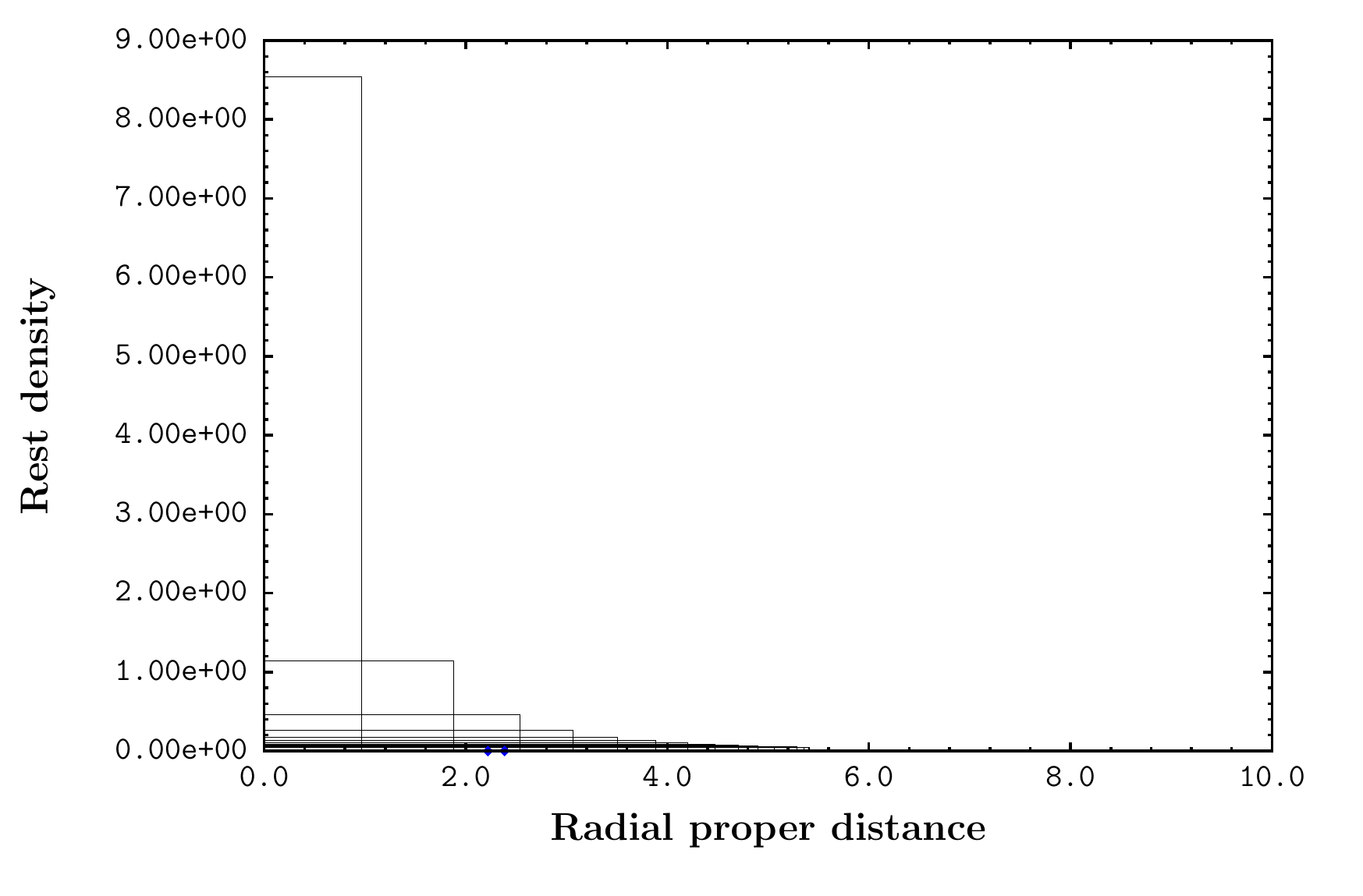}
\FigOne{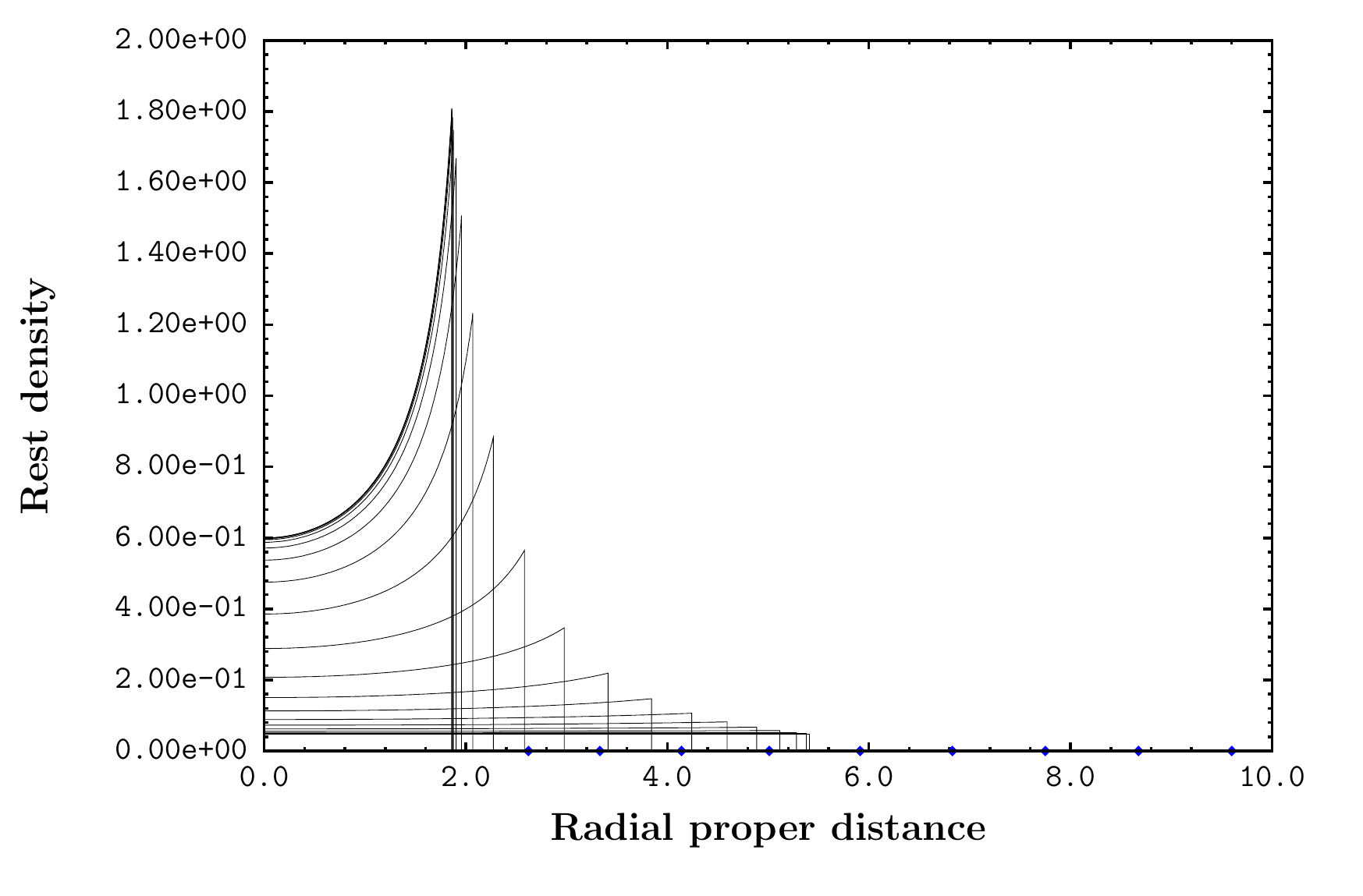}
\caption{\normalfont The The rest density for geodesic slicing (top) and maximal slicing (bottom).}
\label{fig:GeoMaxRho}
\end{figure}

\clearpage

% === convergence =============================================================

\begin{figure}[t]
\FigOne{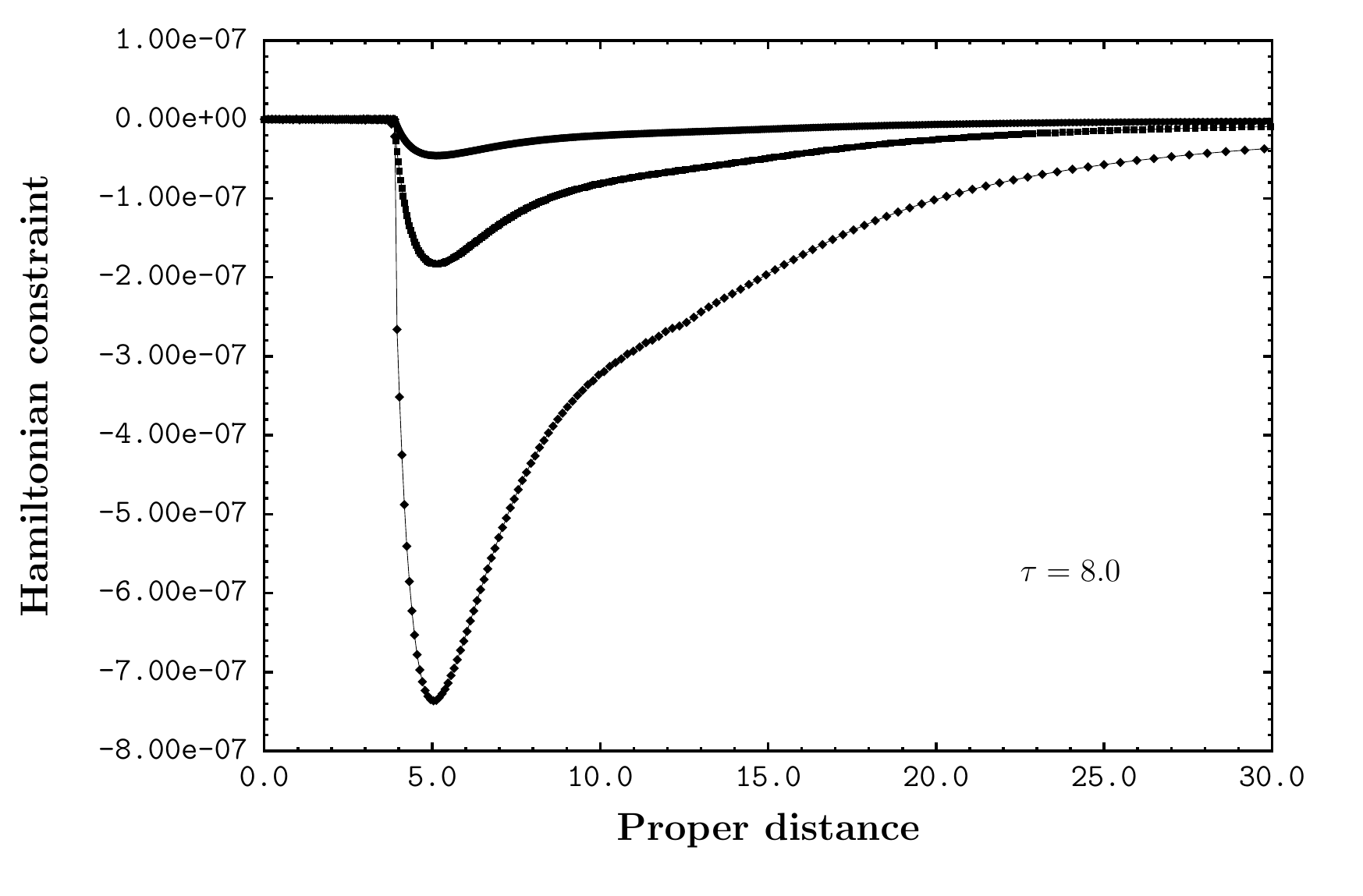}
\FigOne{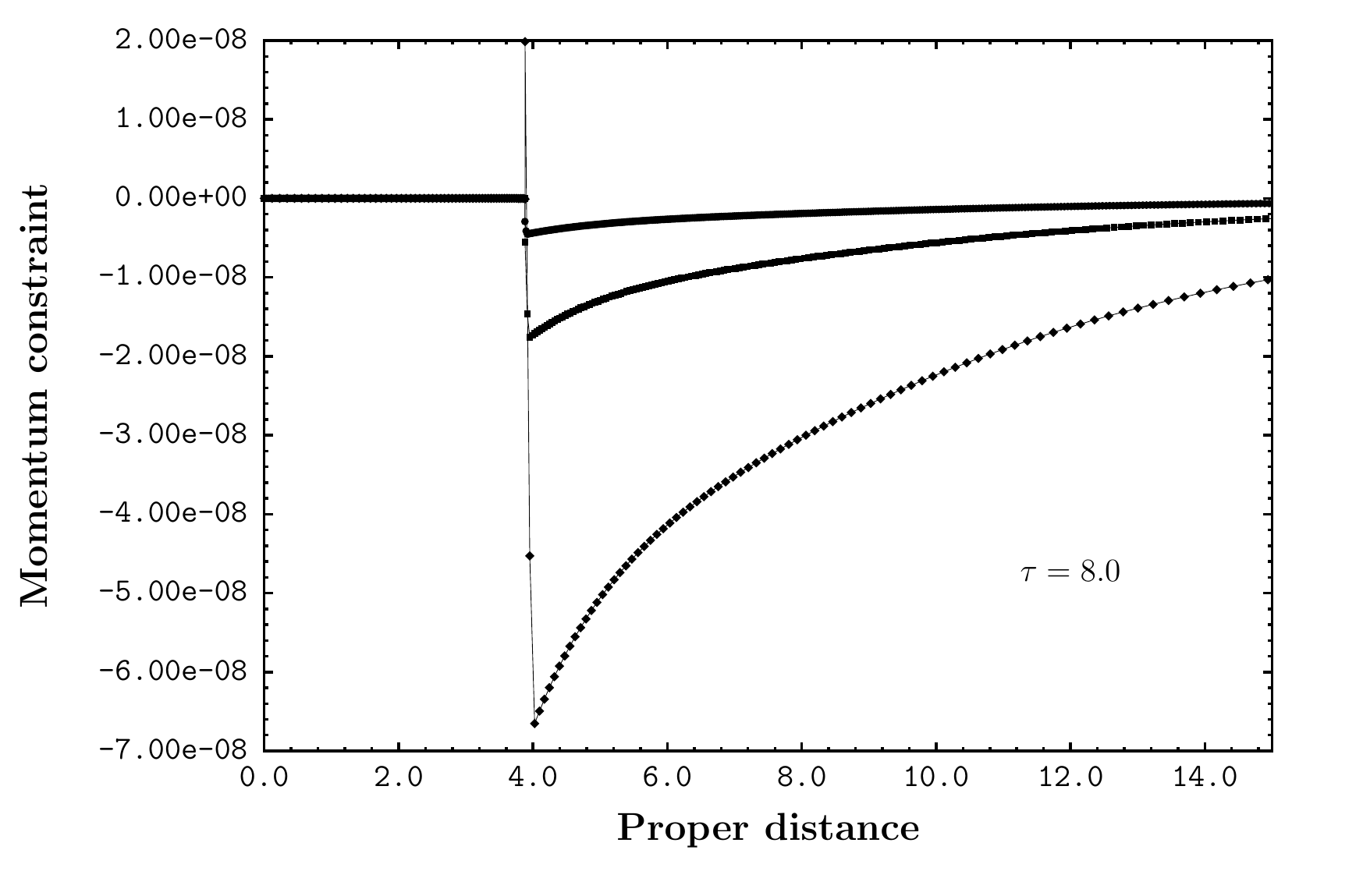}
\caption{\normalfont This is a snapshot of the constraints across the lattice at a fixed time in
geodesic slicing. The three curves correspond to the three models described in the text. Note that
the peaks decrease rapidly as the number of lattice nodes is increased. The horizontal axes have
been truncated to give a better view of the data.}
\label{Geo:CnvrgHamMom}
\end{figure}

\clearpage

\begin{figure}[t]
\FigOne{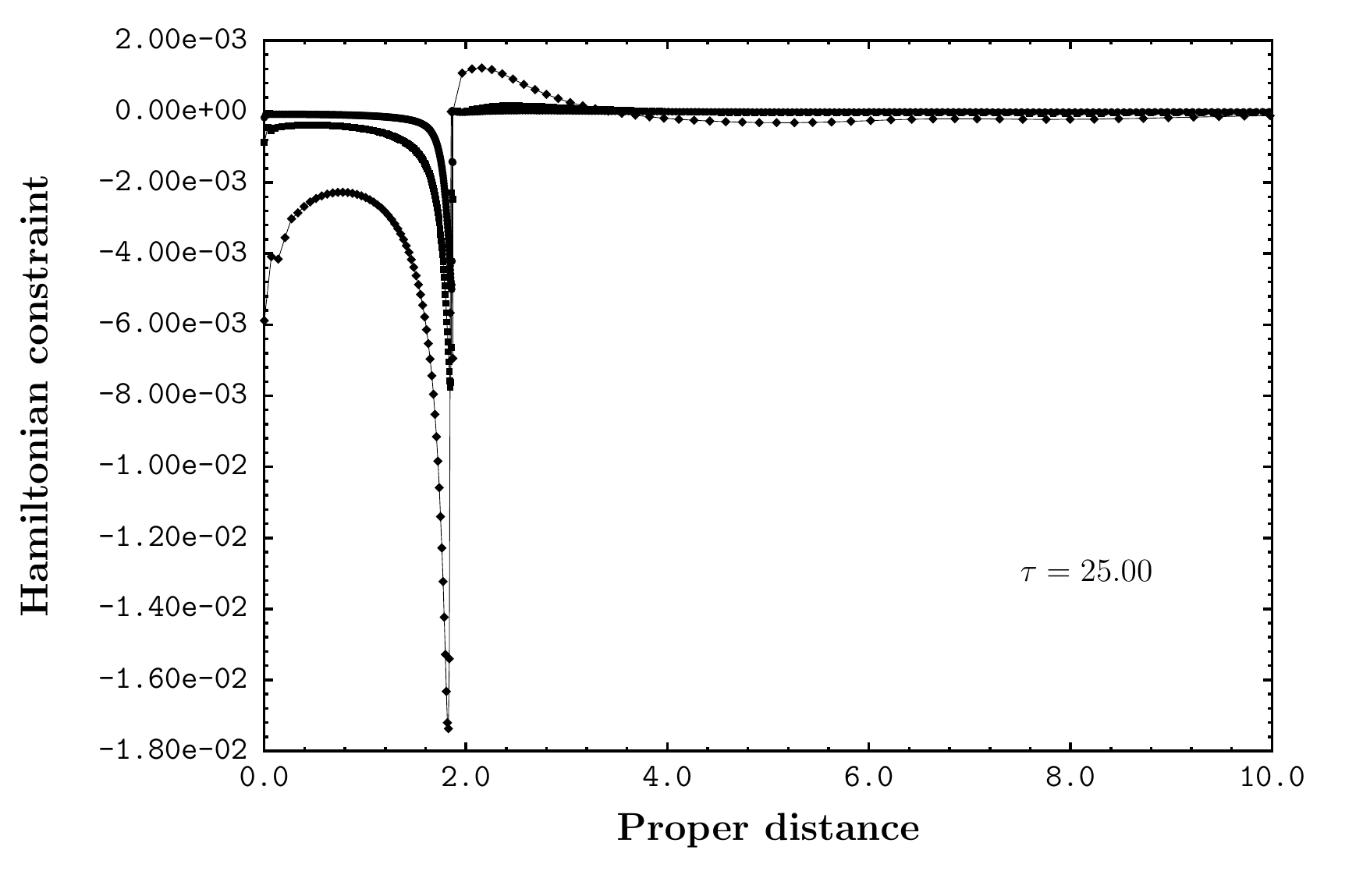}
\FigOne{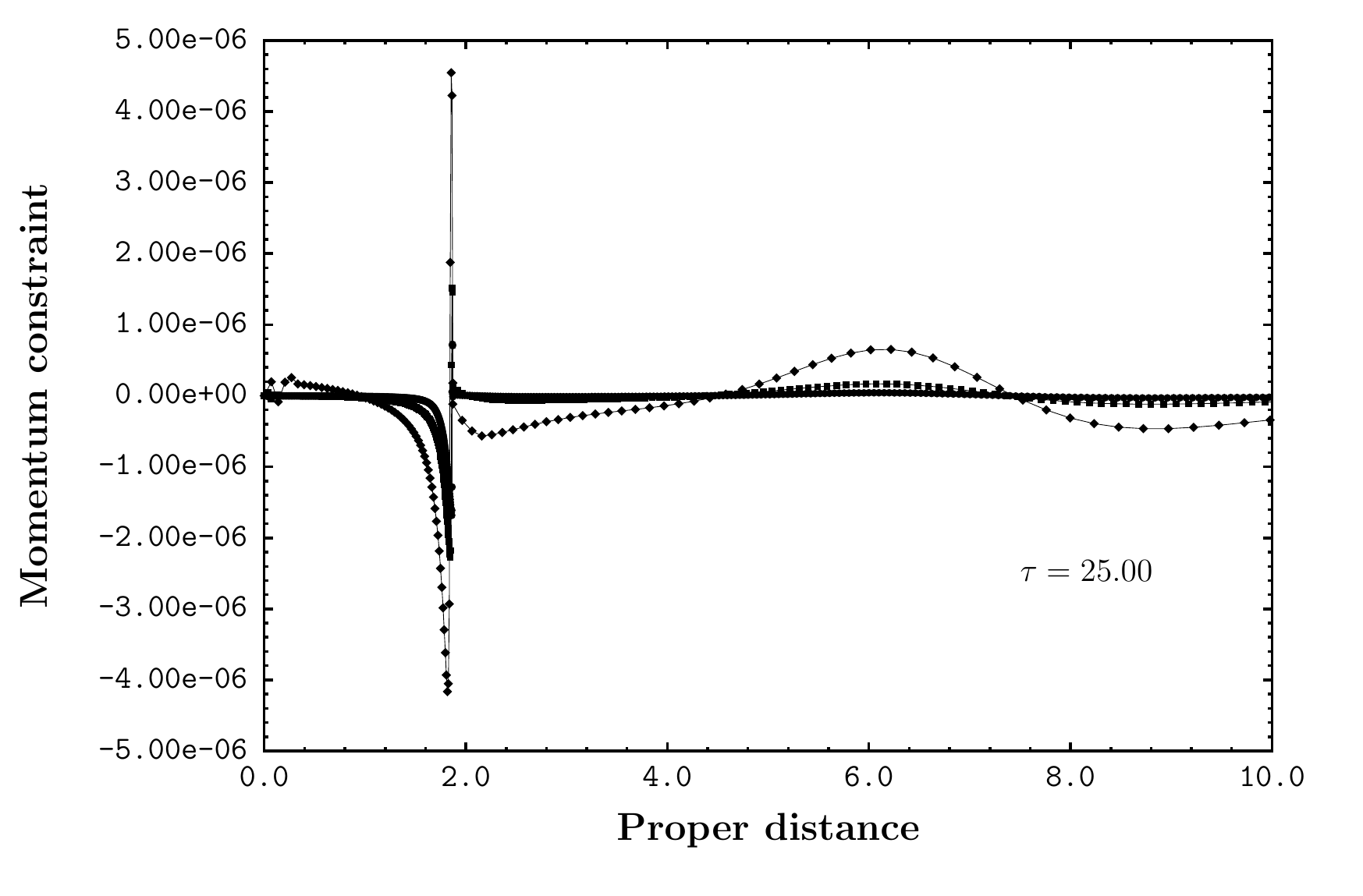}
\caption{\normalfont The Hamiltonian and momentum constraints across the lattice at a fixed time
with maximal slicing. The peak occurs near the junction and appears to vary as $1/\nvac$.}
\label{Max:CnvrgHamMom}
\end{figure}

\clearpage

\begin{figure}[t]
\FigOne{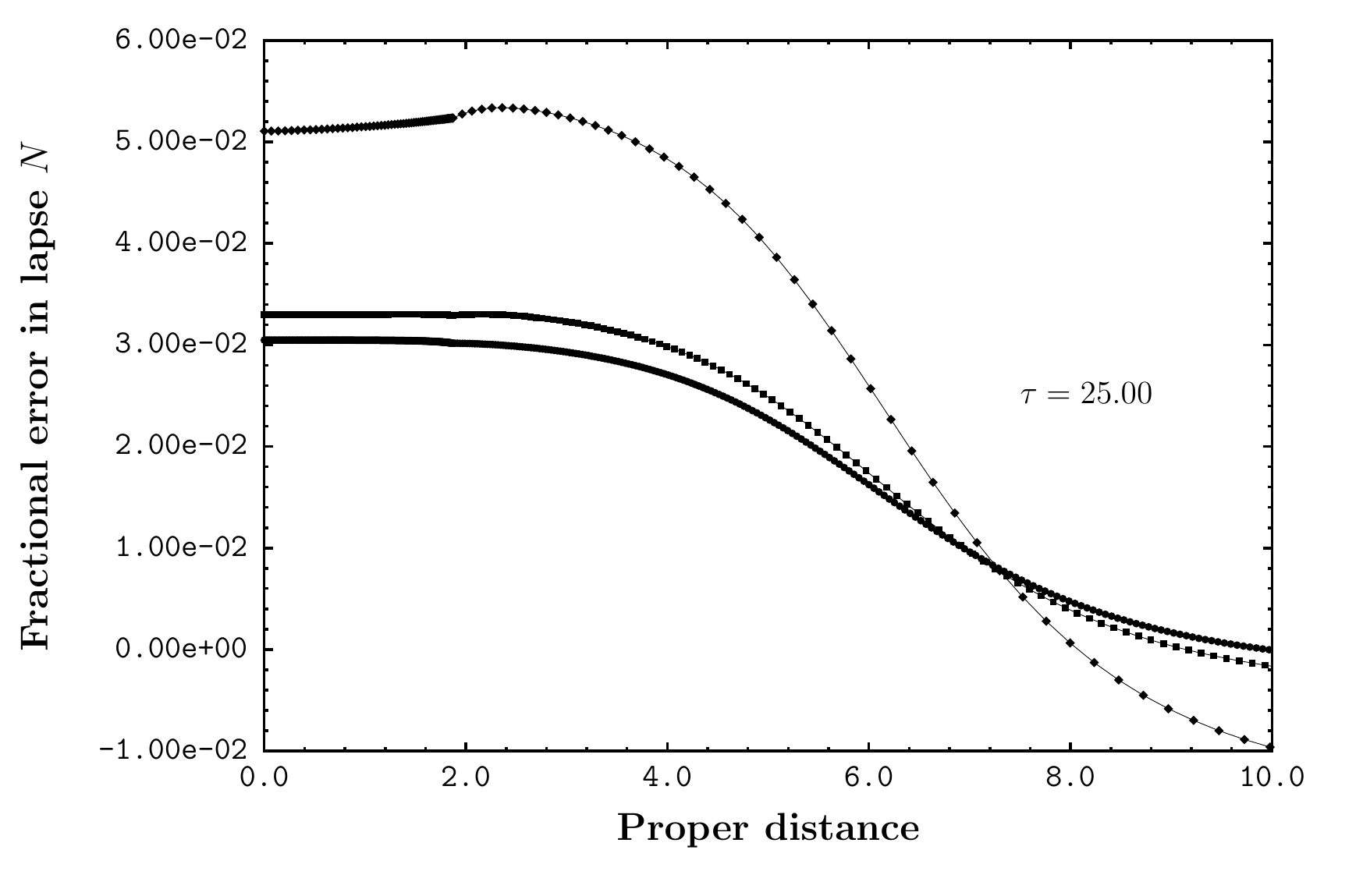}
\FigOne{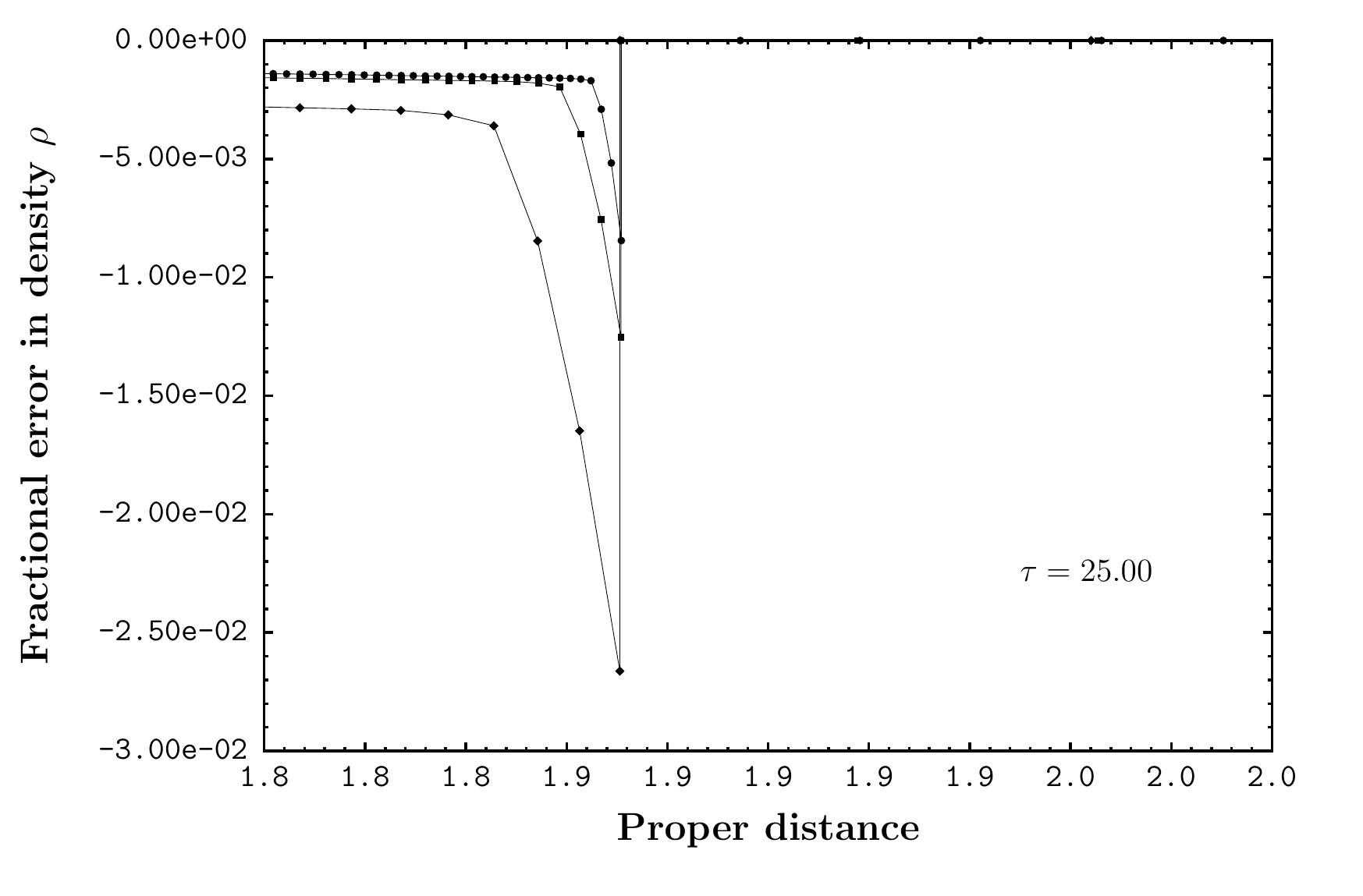}
\caption{\normalfont The fractional errors for the lapse and the rest density across the lattice
with maximal slicing. The large error in the coarsest lattice is probably due to having two few
nodes. The finer lattice show much better errors but note that the lapse appears not to converge at
the origin. This is due to the use of a finite outer boundary for the lapse.}
\label{Max:CnvrgNRho}
\end{figure}

\clearpage

\begin{figure}[t]
\FigOne{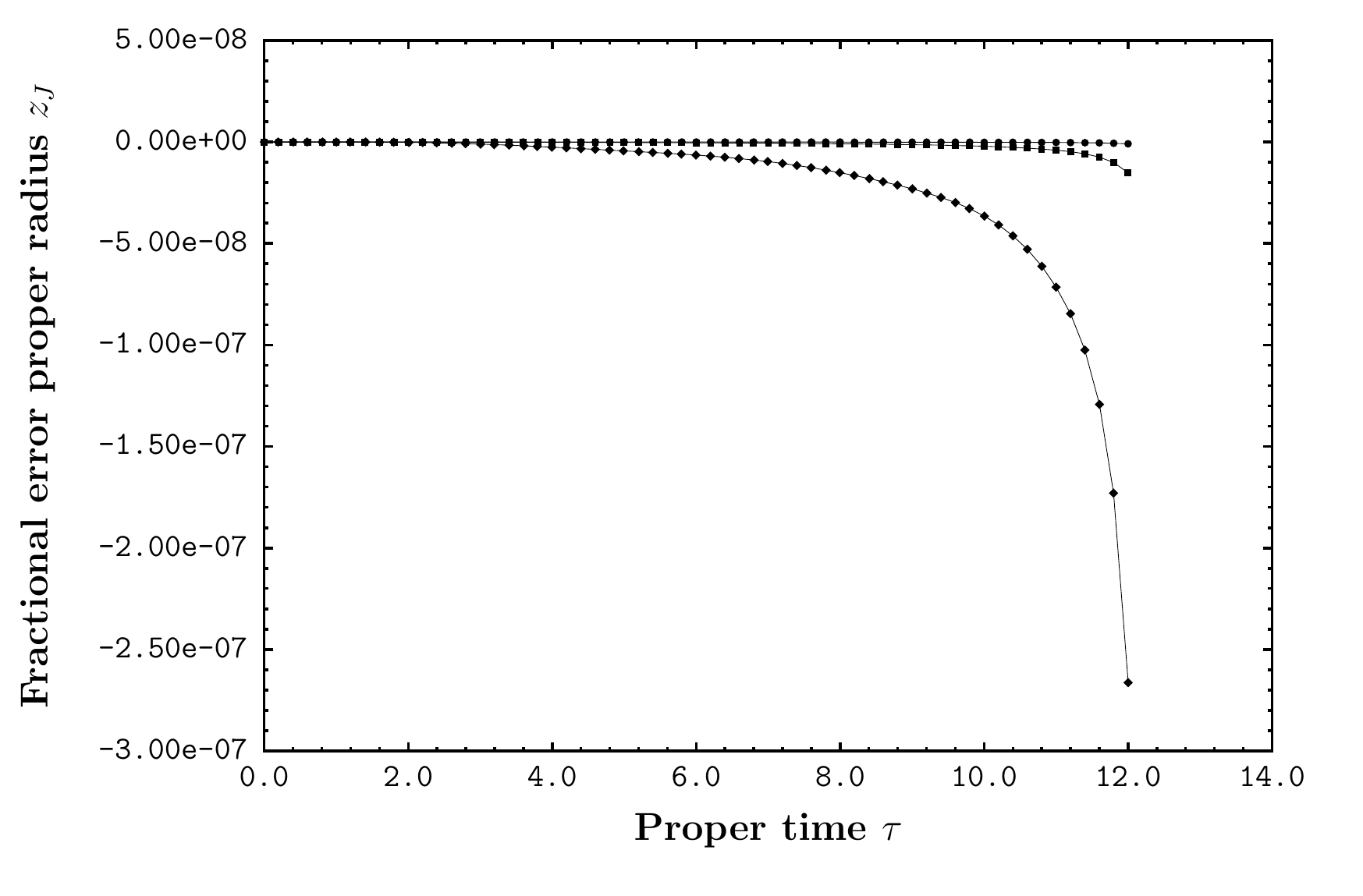}
\FigOne{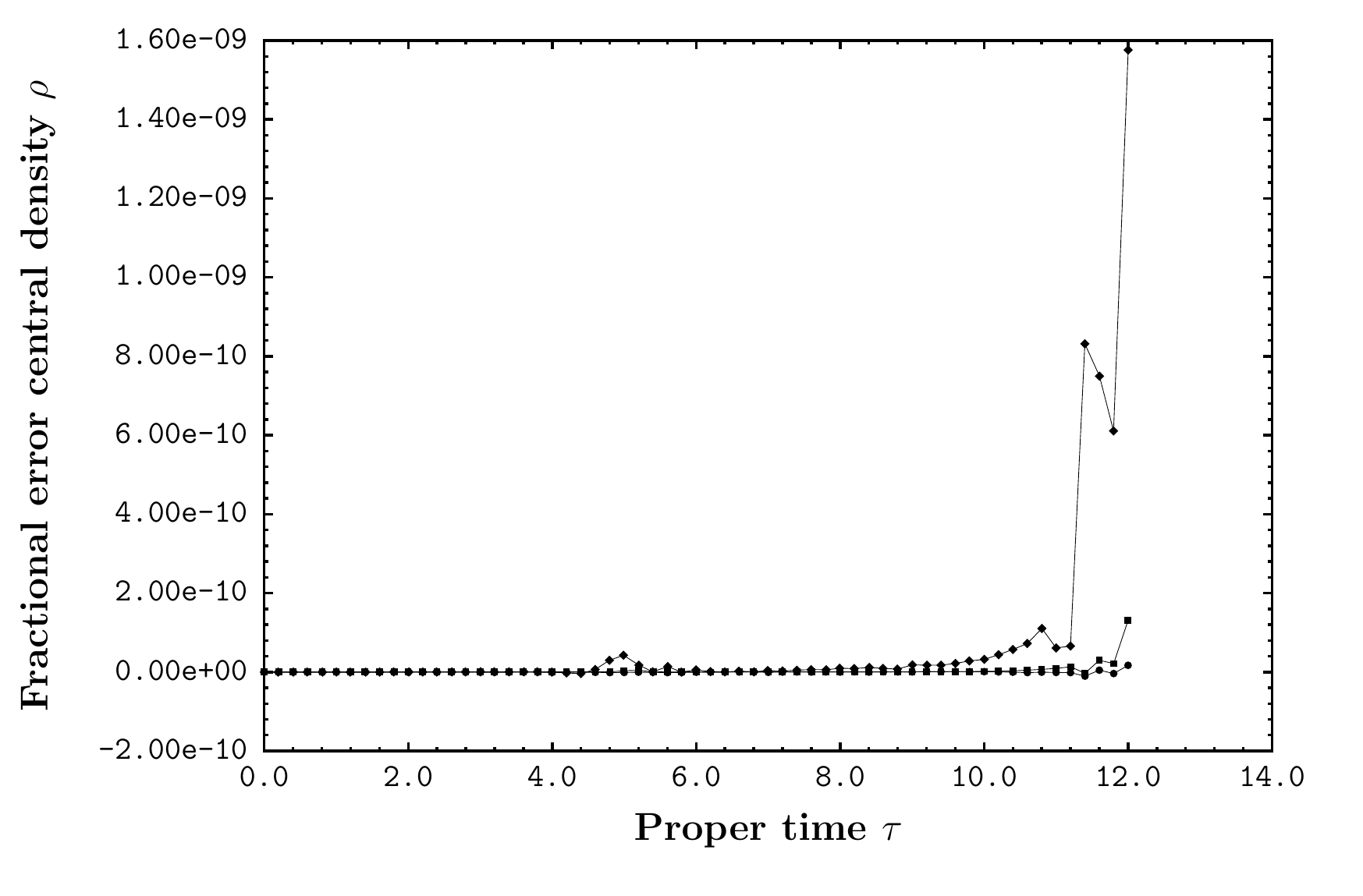}
\caption{\normalfont The fractional errors for the radius and the central density for geodesic
slicing. The convergence is clear and it is rapid (we make no attempt to estimate the order of the
convergence).}
\label{Geo:CnvrgRadRho}
\end{figure}

\clearpage

\begin{figure}[t]
\FigOne{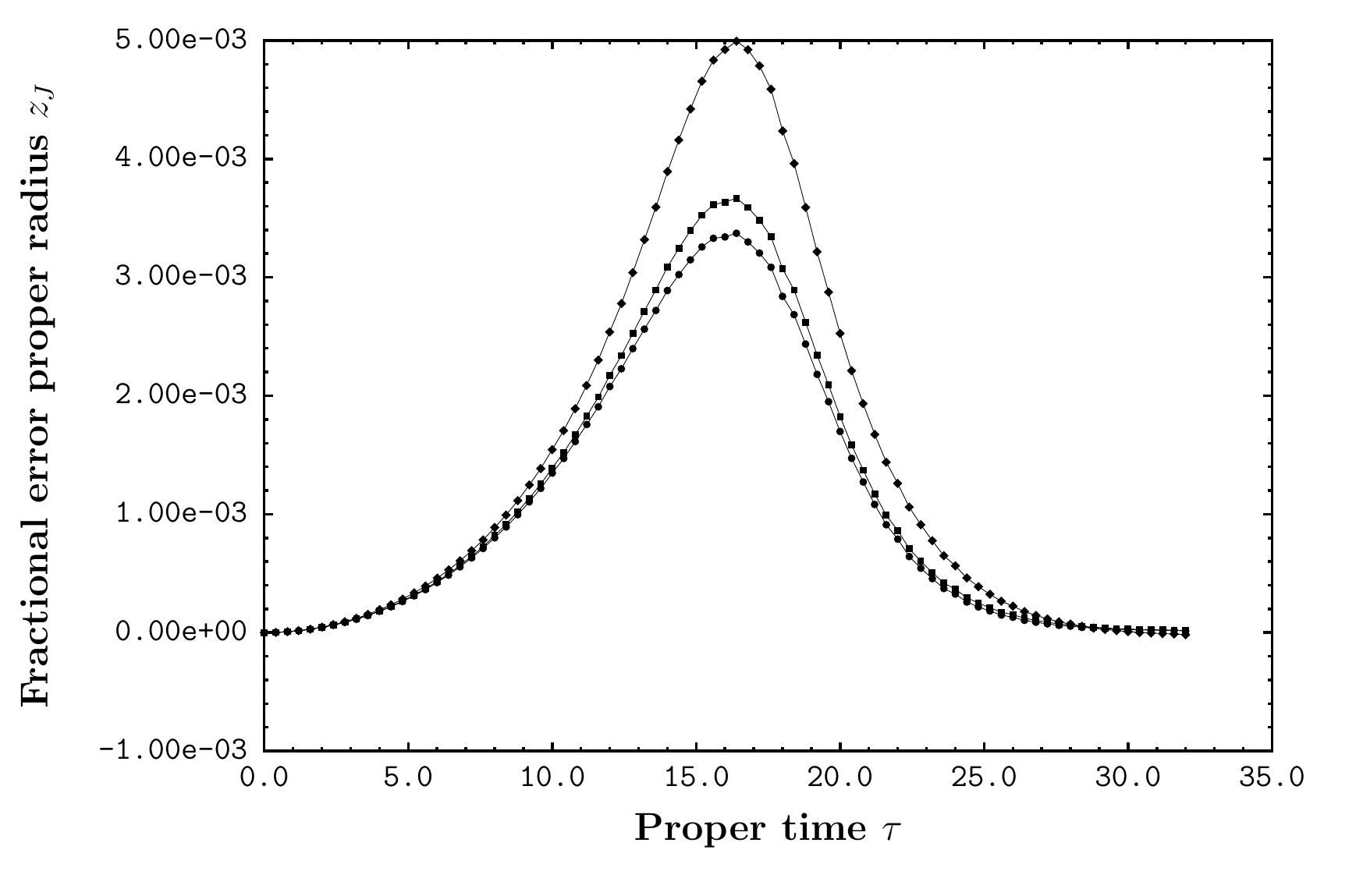}
\FigOne{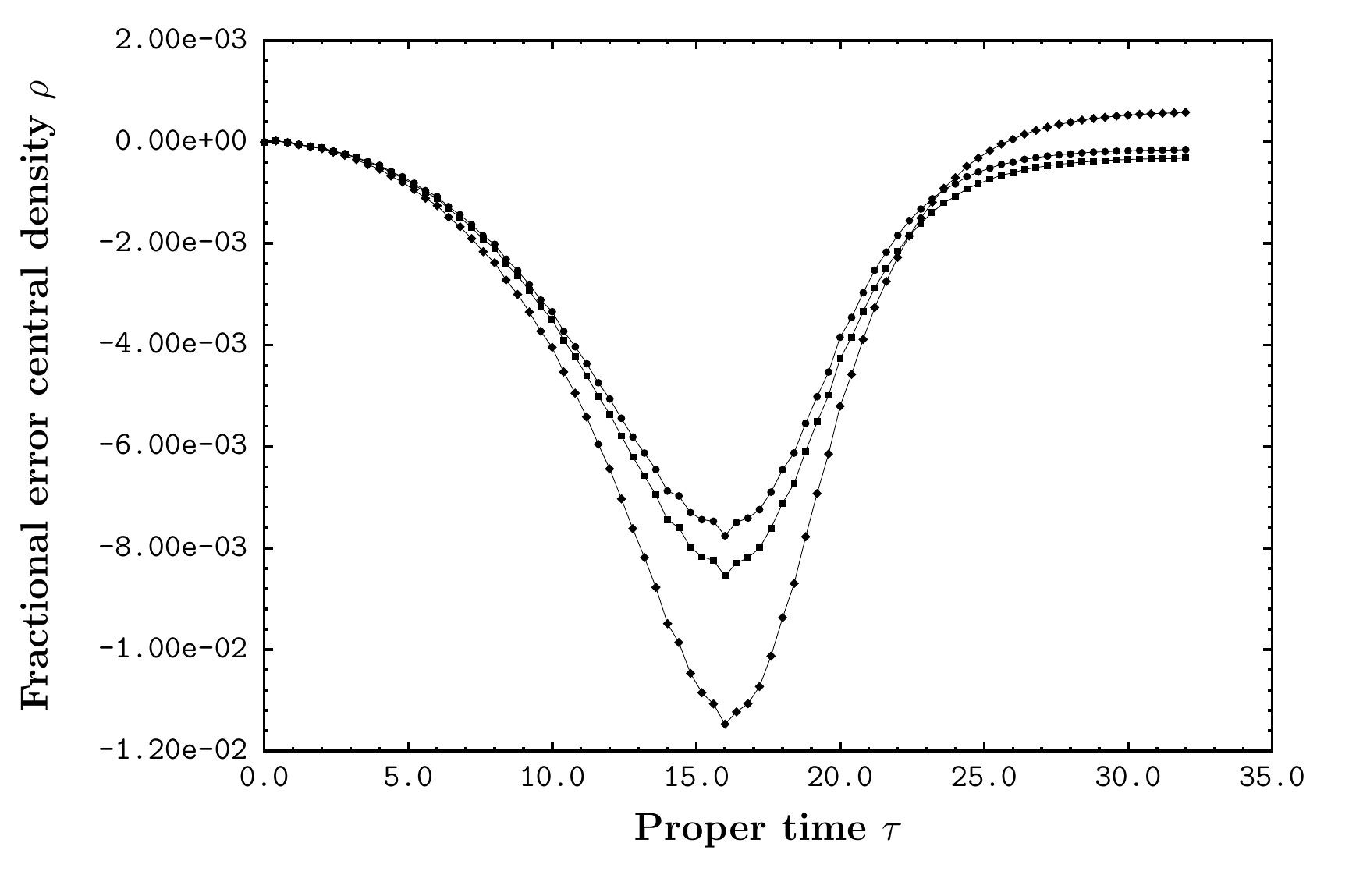}
\caption{\normalfont The fractional errors for the radius and the central density for maximal
slicing. The peak occurs around the time when the apparent horizon forms. The height of the peak
for the finest resolution is limited by the location of the outer boundary. Doubling $\zvac$ halves
the height of the peak.}
\label{Max:CnvrgRadRho}
\end{figure}

\clearpage

\begin{figure}[t]
\FigOne{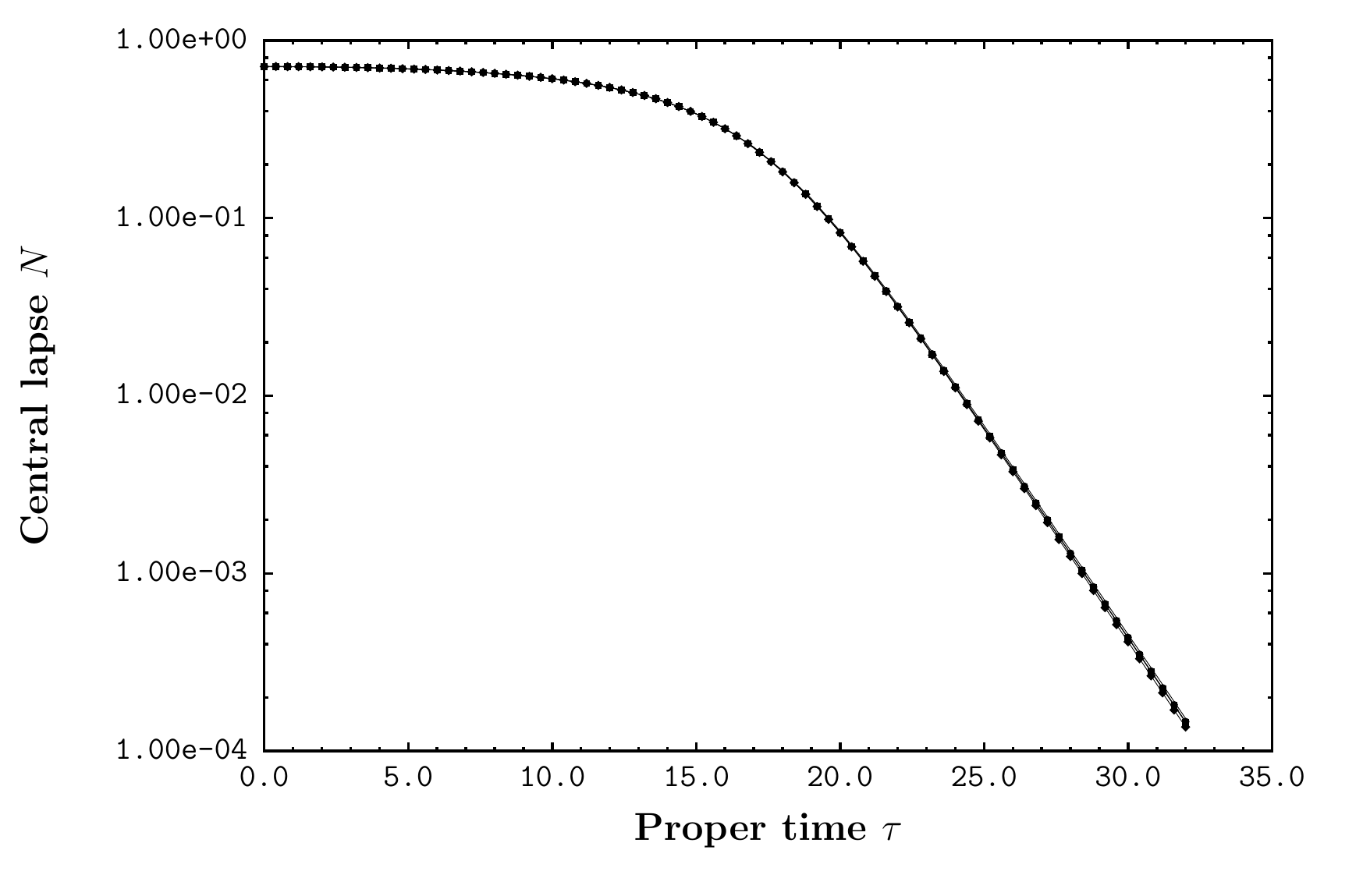}
\FigOne{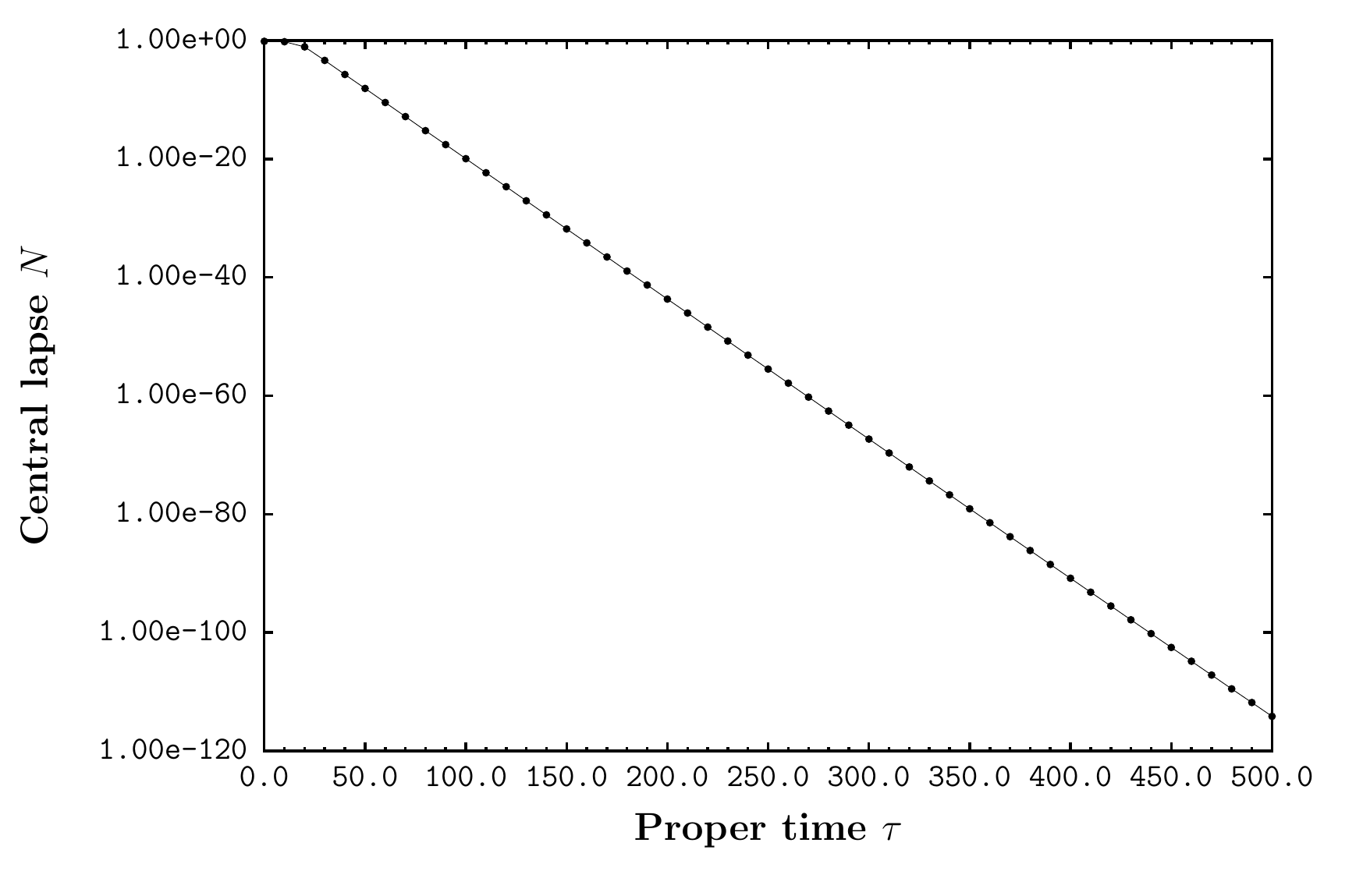}
\caption{\normalfont The lapse at the origin for the three models superimposed on the exact data of
Petrich \etal (top) and for the single long term model (bottom). This shows clearly that the lapse
collapses exponentially.}
\label{Max:ExpoLapse}
\end{figure}

\clearpage

\begin{figure}[t]
\FigOne{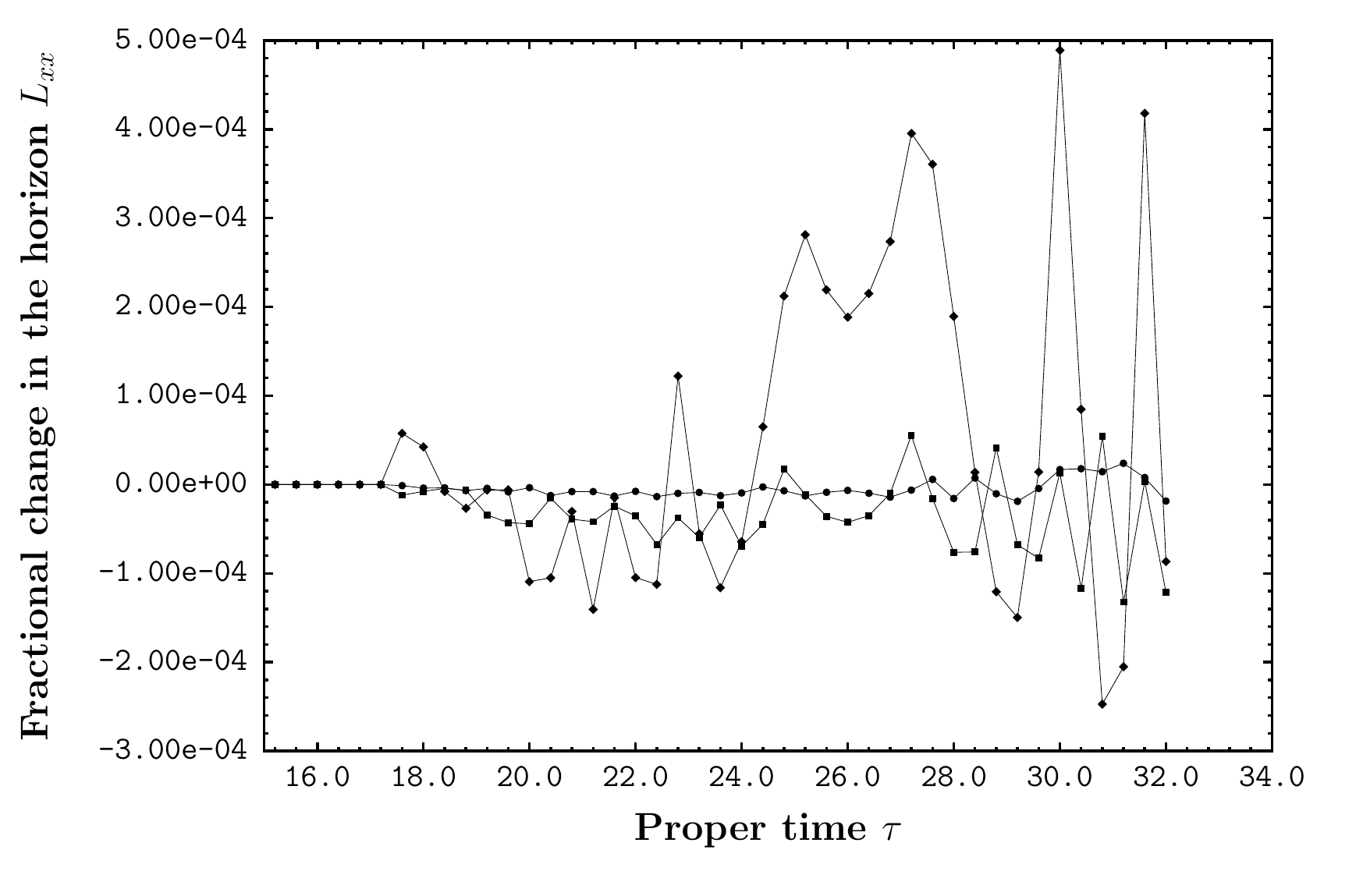}
\FigOne{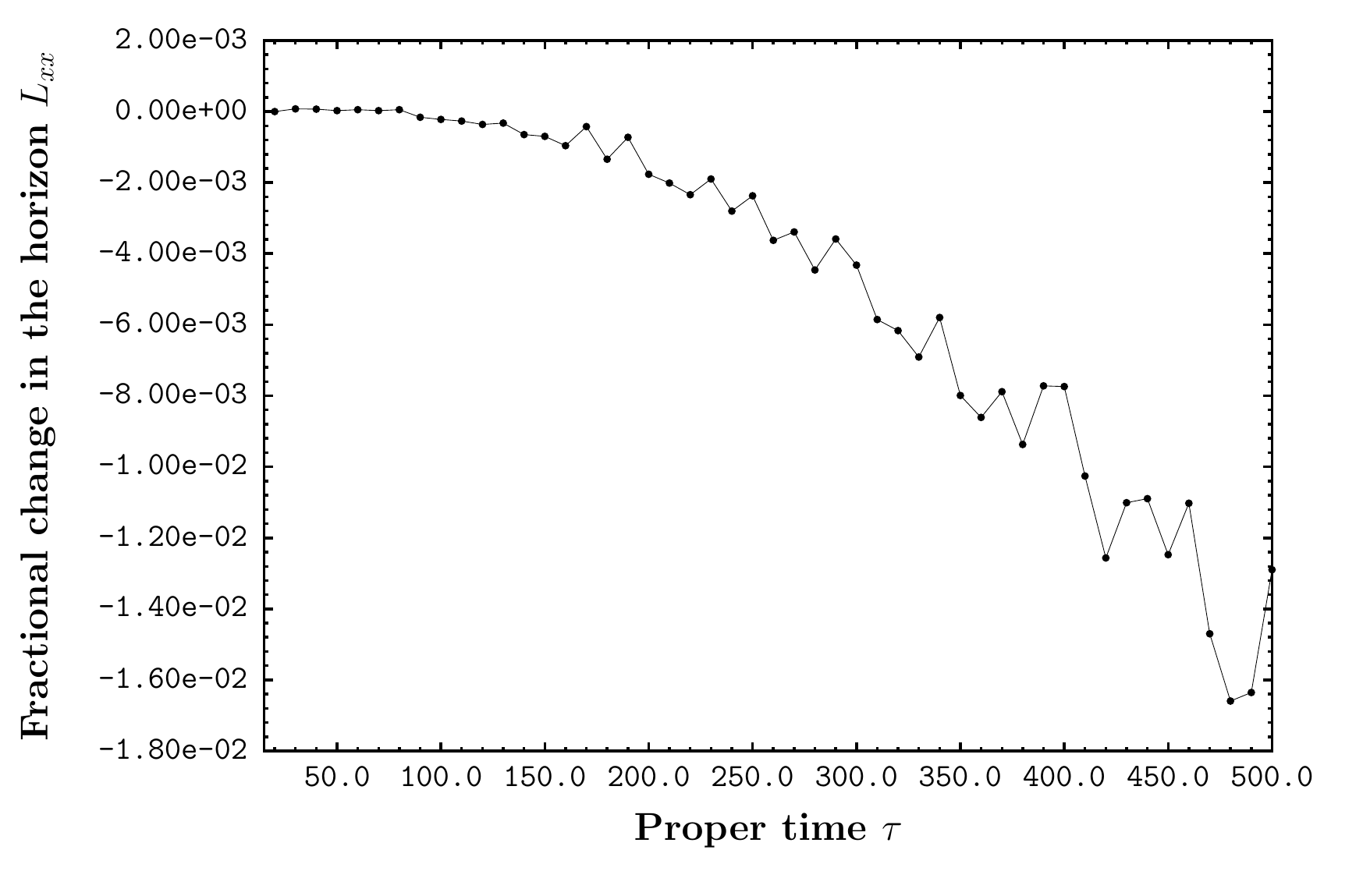}
\caption{\normalfont The fractional change in the $\Lxx$ on the horizon for the three models (top)
and for the long-term integration (bottom) in maximal slicing). These errors should be zero by
Hawking area theorem.}
\label{Max:ErrHorizon}
\end{figure}

\clearpage

% ============================================================================================
% \bibliographystyle{brewin}
% \bibliography{brewin}

\providecommand{\href}[2]{#2}\begingroup\raggedright\endgroup

% --- replace the entries in paper.bbl after bibtex, to create links for Paper 1 and 2 ---
%
% \bibitem{brewin:2002-01}
% L.~Brewin, {\hypertarget{paper1}{{\bf (Paper 1)}} Long term stable integration of a maximally sliced Schwarzschild
%   black hole using a smooth lattice method}, {\em Classical and Quantum
%   Gravity} {\bf 19} (2002)  429--455.
% 
% \bibitem{brewin:2009-04}
% L.~Brewin, {\hypertarget{paper2}{{\bf (Paper 2)}} Deriving the ADM 3+1 evolution equations from the second variation
%   of arc length}. In preparation, 2009.

\end{document}